\documentclass[11pt]{article}
\usepackage[utf8]{inputenc}
\usepackage{arydshln}
\usepackage[]{bm}
\usepackage[]{amsmath}
\usepackage{rotating}
\usepackage[]{verbatim}
\usepackage{amsthm}
\usepackage{afterpage}
\usepackage{enumerate}
\usepackage[flushleft]{threeparttable}
\usepackage{mathtools}
\usepackage[mathscr]{euscript}

\DeclareMathOperator*{\argmin}{arg\,min}
\usepackage[titletoc]{appendix}
\usepackage{epstopdf} 
\usepackage{graphicx}
\usepackage{lscape}
\usepackage{amssymb}
\usepackage[capposition=top]{floatrow}
\usepackage[left=0.9in, right=0.9in, top=1.1in, bottom=1.1in]{geometry}
\usepackage[normalem]{ulem}

\usepackage[linesnumbered,ruled,vlined]{algorithm2e}
\newtheorem{definition}{Definition}
\newtheorem{proposition}{Proposition}

\newtheorem{theorem}{Theorem}
\newtheorem{remark}{Remark}
\newtheorem{lemma}{Lemma}
\usepackage{tabularx,booktabs}
\usepackage{longtable}
\usepackage{booktabs}
\usepackage{multirow}
\usepackage{adjustbox}
\usepackage{array}
\newcolumntype{H}{>{\setbox0=\hbox\bgroup}c<{\egroup}@{}}
\usepackage{tikz}
\usetikzlibrary{shapes.geometric, arrows, chains}
\usetikzlibrary{calc, shapes, positioning}
\usepackage{subcaption}
\usepackage{authblk}

\usepackage{setspace}
\usepackage{xr}
\usepackage{bbm}
\usepackage{csquotes}
\usepackage{algpseudocode}
\allowdisplaybreaks
\algblock{Input}{EndInput}
\algnotext{EndInput}
\algblock{Output}{EndOutput}
\algnotext{EndOutput}

\usepackage[inline]{enumitem}
\usepackage[colorlinks=true, linkcolor=blue, citecolor=blue, urlcolor=blue]{hyperref}
\usepackage[justification=centering]{caption}

\usepackage{indentfirst} 
\usepackage{titlesec}
\titlelabel{\thetitle.\hspace{.5em}} 

\tikzset{
  startstop/.style={
    rectangle, 
    rounded corners,
    minimum width=3cm, 
    minimum height=1cm,
    align=center, 
    draw=black, 
    fill=red!30
    },
  process/.style={
    rectangle, 
    minimum width=3cm, 
    minimum height=1cm, 
    align=center, 
    draw=black, 
    fill=blue!30
    },
  decision/.style={
    rectangle, 
    minimum width=3cm, 
    minimum height=1cm, align=center, 
    draw=black, 
    fill=green!30
    },
  arrow/.style={thick,->,>=stealth},
  dec/.style={
    ellipse, 
    align=center, 
    draw=black,
    fill=green!30
    },
}

\usepackage{natbib}
\begin{document}

\setlength{\abovedisplayskip}{5pt}
\setlength{\belowdisplayskip}{5pt}
\begin{spacing}{1.2}
\title{Robust Inference Methods for Latent Group Panel Models under Possible Group Non-Separation\thanks{Okui acknowledges financial support from JSPS KAKENHI Grant number 23K25501 and the NOMURA Foundation. We thank the participants of the 30th International Panel Data Conference (June 30 - July 1, 2025), the 2025 Symposium for High Dimensional Econometrics and Machine Learning at Tsinghua University (August 24, 2025), and the LMU-Todai Econometrics Workshop (September 25-26, 2025), Singapore Workshop on Econometrics \& Data Sciences at Nanyang Technological University (November 22, 2025), Kansai Econometrics Study Group (January 10-11, 2026), and Econometrics Forum 2026 (June 6-7, 2026). All errors are our own. E-Mail: oguzhan.akgun@u-bourgogne.fr (O. Akgün), okuiryo@e.u-tokyo.ac.jp (R. Okui).}
}

\author[1]{Oğuzhan Akgün}
\affil[1]{Université Bourgogne Europe, LEDi UR 7467, 21000 Dijon, France}

\author[2]{Ryo Okui}
\affil[2]{University of Tokyo, Japan}

\date{\today}

\maketitle

\begin{abstract}
We develop robust inference methods for general linear hypotheses in linear panel data models with latent group structure in the coefficients. We employ a selective conditional inference approach based on the conditional distribution of coefficient estimates given the group structure estimated from the data. The resulting inference procedures remain valid even when group separation fails (i.e., when the distributional properties of the group-specific coefficients are not established) and, because they account for uncertainty in estimating the group structure, they also improve on conventional asymptotic procedures in finite samples when separation does hold. Our tests are exactly valid under Gaussian errors with known variances and asymptotically valid under general error distributions. Unlike much of the post-clustering inference literature, which focuses on testing group homogeneity, our framework accommodates arbitrary linear restrictions on the group-specific coefficients. Inverting the conditional tests yields selective confidence sets with valid coverage conditional on the estimated group structure. We illustrate the methods through Monte Carlo simulations and an application to growth convergence clubs. Simulations demonstrate accurate size control and good power in finite samples, including in the presence of serial correlation and cross-sectional dependence. The applications show sharp differences between the traditional inference methods and robust methods proposed in this paper, illustrating the importance of taking the estimated group structure into account.
\end{abstract}


\end{spacing}

\newpage

\section{Introduction}\label{sec:intro}

Latent group structure has recently become a popular approach for analyzing panel data. Recent contributions include \cite{lin_estimation_2012}, \cite{bonhomme_grouped_2015}, \cite{su_identifying_2016}, \cite{lumsdaine_estimation_2023}. This specification assumes that the coefficients vary across individual units but are homogeneous within each group. Consequently, it provides a flexible yet manageable way to model various forms of heterogeneity. Since a group contains multiple units, the parameters of the model can be estimated with high precision.

This paper addresses an important yet under-explored issue: conducting valid inference for general linear hypotheses on group-specific slope parameters when the groups are not well separated in the population, that is, when their parameters are not sufficiently different from each other. This arises in many applications. Testing the homogeneity of the entire slope parameter vector serves as a general test of group separation, allowing one to determine whether two or more groups are identical and can be combined. In other applications, it is of interest to test the homogeneity of only a subset of parameters, for example to identify which aspects of the parameter vectors contribute to group heterogeneity, or whether any specific group pairs share coefficient values. Since group separation is not guaranteed and the groups may not be well-defined, it is desirable to have a testing procedure that does not require it, even when the interest lies in the value of a particular parameter. Section~\ref{sec:settings} presents more concrete empirical examples that illustrate the importance of these considerations.

The central difficulty of this testing problem is that the groups are estimated from the same data used to estimate the coefficients, so valid inference must account for the resulting uncertainty. When group separation holds, this is straightforward in large samples: the group-structure estimator is super-consistent, so the estimated structure coincides with the true one with probability approaching one, and the standard Wald test is asymptotically valid. In finite samples, however, the entire structure is rarely recovered exactly, and this pointwise asymptotic argument can poorly approximate the behavior of the Wald statistic. When group separation fails, the difficulty is more fundamental: the asymptotic distribution of the coefficient estimators is generally unavailable in closed form, and only limited results exist \citep[e.g., ][S3.1]{bonhomme_grouped_2015}.

Our approach utilizes selective conditional inference. Specifically, we employ a method following \cite{lee_exact_2016}, later adapted to the clustering problem by \cite{gao24} and \cite{chen23}, which in our case involves calculating the distribution of statistics conditional on the estimated group structure. The key point is that, once we condition on the estimated group structure and nuisance terms, the exact distribution of the Wald test statistic for the linear hypothesis can be derived and computed analytically, and is a truncated $\chi^2$ distribution. The main challenge is characterizing the estimated group structure as a subset of the data space, and we provide analytical formulas to compute the conditioning set for two estimation algorithms for the group structure.

Our testing procedure achieves the desired size conditional on the estimated groups. We first develop the theory under the assumptions of Gaussian errors and homoskedasticity, so that the results are exact and valid in finite samples. We then extend the results to settings where the key unit-specific statistics are asymptotically normal and show that our methods are asymptotically valid.

To assess the quality of the asymptotic approximation and the robustness of our procedure to potential violations of these assumptions, we conduct Monte Carlo simulations. Our findings indicate that the method performs well, particularly when the sample size is large, even in the presence of serial or cross-sectional correlations. Our method offers advantages even when group separation holds, and the group structure can be estimated consistently. The estimator is then super-consistent, but this is a pointwise property that fixes the data-generating process and need not hold uniformly, so the associated asymptotic distribution of the coefficient estimator can approximate the finite-sample distribution poorly when separation is weak.\footnote{The importance of uniformity and the potential danger of relying on pointwise asymptotic results are discussed in \citet{leeb_model_2005}. \citet{Andrews2020GenericResults} provide theoretical discussions on establishing uniform asymptotic results.} Conventional asymptotic procedures ignore this source of uncertainty. In contrast, conditioning on the estimated group structure accounts for it directly. Our simulation results confirm better size control without a loss of power.

We present three empirical applications to illustrate our testing procedures. Our main application revisits the growth convergence club hypothesis in the spirit of \cite{canova2004testing} and \cite{phillips2007transition}, using data from the Penn World Table. Two further applications, reported in the appendix, examine heterogeneity in firm-level R\&D investment dynamics \citep{aguilar-loyo_grouped_2024}, and the relationship between income and democracy \citep{bonhomme_grouped_2015}. In all three cases, naive post-clustering inference points to pervasive heterogeneity, whereas our procedures deliver more disciplined conclusions, underscoring the importance of accounting for the uncertainty in group estimation.

\subsection*{Relation to the literature}

Our study falls into two strands of the literature: latent group structure and selective conditional inference. We discuss how our study contributes to each strand.

\paragraph{Latent group structure.}
Models with latent group structures have long been studied as clustering problems in statistics. For a single cross-section ($T=1$), the regression model with group-specific coefficients is referred to as a clusterwise regression model by \cite{spath79}, who developed an algorithm similar to the KMeans algorithm \citep{macqueen67} that applies the idea of grouping similar units to a regression model with general regressors. A large number of studies then proposed different estimation methods and algorithms to estimate the parameters of the model, and the research area in statistics and operations research is still active \citep[see, for instance,][]{carbonneau_globally_2011,carbonneau_extensions_2012,carbonneau_globally_2014,ingrassia_model-based_2014,bagirov_algorithm_2015,park_algorithms_2017,bagirov_nonsmooth_2018,karmitsa_missing_2022}.

The emergence of similar literature in econometrics is more recent. A seminal contribution in this field is \cite{lin_estimation_2012}, which examines two types of estimators that we consider in this paper. The first estimator simultaneously estimates both the clustering structure and the parameters, extending the KMeans algorithm. \cite{bonhomme_grouped_2015} extend it further to estimate models with group-specific but time-varying intercepts, called grouped fixed effects (GFE), which is particularly useful when the parameters are time-varying, making unit-by-unit estimation impossible \citep{okui_heterogeneous_2021}. We refer to it as Panel Clusterwise Regression (PCR).\footnote{In the literature, this estimator may be called the GFE estimator. Our primary model does not include GFE, and the coefficients exhibit a grouped structure of heterogeneity. We use a different term to avoid confusion.} The second estimator uses a clustering algorithm on unit-specific time-series parameter estimates: we estimate the coefficients for each unit individually and then apply the KMeans algorithm, which we refer to as the Two-Step KMeans (TSK) estimator. The literature extending these frameworks is now extensive \citep[see, for example,][]{ke_homogeneity_2015,ke_structure_2016,qian_shrinkage_2016,su_identifying_2016,ando_panel_2016,su_identifying_2018,wang_homogeneity_2018,zhang_quantile-regression-based_2019,su_sieve_2019,cytrynbaum_blocked_2020,huang_identifying_2020,okui_heterogeneous_2021,wang_identifying_2021,bonhomme2022discretizing, chetverikov_spectral_2022,cheng2023clustering,su_identifying_2023,lu_uniform_2023,lumsdaine_estimation_2023,yu_spectral_2024,wang_panel_2024,aguilar-loyo_grouped_2024}.

This paper contributes to the literature on panel data models with latent group structures by developing methods that are robust to violations of the group separation assumption. The literature has traditionally relied on this assumption, but its validity is not always guaranteed, making it essential to establish inference procedures that are resilient to potential group non-separation.

Our study is somewhat related to the question of how to test the number of groups, since the group separation assumption is violated by construction when the number of groups is overspecified. In the context of sequential testing, \cite{lin_estimation_2012} and \cite{wang_panel_2024} advocate the use of homogeneity tests such as \cite{hashem_pesaran_testing_2008} and \cite{ando_simple_2015}, while \cite{lu_determining_2017} propose a residual-based LM test to be used at each iteration. \cite{raiola2025testing} also argue that homogeneity testing is useful for examining the existence of latent group structure. \cite{patton23} employ the sample splitting technique to establish valid procedures for homogeneity testing; because of the nature of sample splitting, their procedure cannot use the entire sample for the estimation and inference, which may result in a loss of statistical power. Other papers in the literature rely on information criteria (e.g., \cite{lin_estimation_2012}, \cite{bonhomme_grouped_2015}, \cite{su_identifying_2016}, \cite{yu_spectral_2024}, \cite{wang_panel_2024}, \cite{aguilar-loyo_grouped_2024}), and \cite{lu_determining_2017} provide some comparison between the two approaches. Our focus differs: they concentrate on the number of groups, whereas we examine whether (a subset of) parameters exhibit commonality within (a subset of) groups. Although these questions are closely related, they are distinct.

Another related but distinct question is how to perform statistical inferences on common parameters when group-specific nuisance parameters are present, whereas our focus is on group-specific coefficients that are directly influenced by the estimation.

\paragraph{Selective conditional inference.}
Selective conditional inference is a topic of growing interest in recent statistics literature \citep[see][]{taylor2015statistical,benjamini2020selective,kuchibhotla22}, and is gaining attention in econometrics as well \citep[see][]{andrews2021inference,andrews2024inference,akgun24}. This paper extends the scope of this selective inference method. The main issue addressed in this literature is that using a sample to choose a null hypothesis and consecutively testing it with the same sample conflicts with the mathematical foundations of the classical tests, such as Wald, LM, or LR, an issue sometimes referred to as \textit{double dipping} \citep{kriegeskorte09}. \cite{kuchibhotla22} reviews these efforts with a particular emphasis on inference following model selection. The main problem is that testing a null hypothesis of model fit, using a procedure that best fits the sample at hand, yields highly anti-conservative test statistics \citep[see][for the consequences of homogeneity testing post-clustering]{chen23,gao24}. Two of the most popular solutions for double dipping are sample splitting and selective inference. As mentioned above, \cite{patton23} proposed a split-sample method; although simple to implement, their methodology has important drawbacks, such as the question of how to split the sample, invalidity under general forms of time series dependence, and loss of power due to the reduction of the number of observations \citep[see][for a more detailed discussion on sample splitting]{kuchibhotla22}.

This paper employs the conditional selective inference method, which has been used to test group separation in simple clustering settings by \cite{gao24, chen23, chen2024}. We extend this approach to panel data models. Our second proposed test is based on the KMeans estimates of the group structure using unit-specific time-series estimates. In this case, we derive a novel decomposition of the vector of unit-specific estimates, expressing it as a function of the constrained least-squares estimator and the deviation of the estimator under the alternative hypothesis from the constrained estimates. In addition to being based on the unit-specific estimators of a panel data model, our decomposition is the most general in the literature. It includes the special cases of \cite{chen23}, \cite{gao24}, \cite{chen2024}, and \cite{yun2024selective}.

Our first test, which is a novel contribution to the literature, is based on the PCR estimator. Determining the conditioning set for selective inference in this case presents several technical challenges, since the PCR estimator simultaneously estimates the coefficient parameters and the group memberships without relying on individual coefficient estimators. This makes it challenging to compute the conditioning set, particularly when deriving the vector orthogonal to the parameter of interest in the null hypothesis. Nonetheless, we have successfully derived an appropriate conditioning set, thereby making the procedure feasible. This contribution and its insights are expected to broaden the application of the selective inference approach.

While preparing this paper, we became aware of the contemporaneous and independent work by \cite{wan2025conditional}, which investigates a related problem. We note several key distinctions between our approaches. First, our framework addresses general linear hypotheses, whereas \cite{wan2025conditional} focuses specifically on the homogeneity between two given groups. Second, our procedure can accommodate parameters that cannot be estimated through unit-by-unit procedures, such as group fixed effects (GFE). Third, we employ Wald statistics, yielding a testing procedure based on a truncated $\chi^2$ distribution, which is much simpler to implement. These differences make our methodology applicable to a broader class of inference problems in panel data models.

The paper is organized as follows. Section \ref{sec:settings} presents the settings and discusses empirical examples in which our testing may be relevant. Section \ref{sec:estimation} introduces two estimators for the group structure: the PCR and TSK estimators. Section \ref{sec:tests} develops our testing procedures, generalizing the framework of \citet{chen23} to panel data settings, and establishes their finite-sample and asymptotic validity. Section \ref{sec:ci} constructs selective confidence intervals. Section \ref{sec:extensions} discusses potential extensions of the main analysis. These include models with unit-specific heterogeneity (for instance, Extension 1 of \citealp{bonhomme_grouped_2015} and Example 1 of \citealp{su_identifying_2016}) and models with time-varying GFE \citep{bonhomme_grouped_2015,aguilar-loyo_grouped_2024}. The results of Monte Carlo simulations are included in Section \ref{sec:motivating_simulations}. Empirical illustrations are given in Section \ref{sec:applications}. Section \ref{sec:otherquestions} concludes the paper with directions to future research.

\textit{Notation:} Random variables are denoted by uppercase letters, and their realizations by the corresponding lowercase letters.
Further, $\| \cdot \|$ denotes Euclidean norm, $\mathbf{1} \{ \cdot \}$ is the indicator function, $\mathrm{bdiag}( \cdot )$ forms a block-diagonal matrix by given elements,
$\otimes$ denotes the Kronecker product.
For any vector $v$, $\mathrm{dir}(v)$ stands for its direction (i.e., $\mathrm{dir}(v) = v/\lVert v \rVert$).
{Additionally, $[v]_s^h$ represents the vector formed by the rows from $(1+h(s-1))$ to $hs$ of the vector $v$.}

\section{Settings} \label{sec:settings}

This section introduces panel data models with latent group structure and defines the null hypotheses of our testing problem. It also lists empirical examples where our testing procedure is relevant.

\subsection{Panel data models with latent group structure}

We consider panel data models with a grouped pattern of heterogeneity. Our primary focus is testing general linear hypotheses about the group-specific slope coefficients, and the key issue is how to account for the fact that the unknown group structure must be estimated from the data.

Suppose we observe a panel data set $D = \{ (Y_{it}, X_{it}); i=1, \dots, N, t=1,\dots, T \}$ where the subscripts $i$ and $t$ denote the observational unit and the time period, respectively, $Y_{it}$ represents a scalar dependent variable for unit $i$ at time $t$, and $X_{it}= (X_{1, it},\dots, X_{K, it})'$ is a vector of explanatory variables for unit $i$ at time $t$ with $X_{k, it}$ being the scalar $k$-th explanatory variable. Consider the following panel data model with heterogeneous coefficients:
\begin{equation}\label{eq:main_model}
    Y_{it} = X_{it}' B_i + \varepsilon_{it}, \; i=1,\dots,N, \; t=1,\dots,T, \\
\end{equation}
where $B_i = (B_{1,i},\dots,B_{K,i})'$ is a $K \times 1$ vector of unknown coefficients, and $\varepsilon_{it}$ is the error term. Note that $B_i$ depends on $i$, meaning that the coefficients are potentially heterogeneous.

We model the parameter heterogeneity using group structure. Each unit $i$ is assigned to exactly one of $G$ groups, and the membership indicator $g^0_i \in \{ 1,\dots, G \}$ signifies the true group assignment for unit $i$. The group membership variables $g^0_i$, $i=1,\dots, N$, are unobserved and need to be estimated from data; we collect them into $\gamma^0 = (g^0_1,\dots,g^0_N)'$. Units within a group share the same coefficient values, while units in different groups may have distinct ones. Specifically, we set $B_i = \theta^0_g$ if $g^0_i = g$ with $\theta^0_g = (\theta^0_{1,g},\dots,\theta^0_{K,g})'$ representing the underlying group-specific coefficients.

The primary objective of this paper is to develop tests for general linear hypotheses that account for estimation uncertainty in the group assignments. However, stating the null hypothesis in this context is nontrivial because the group membership is unknown. Suppose that we would like to test the null hypothesis $H_0: R\theta^0 = r$, where $R$ is a $q \times GK$ nonrandom matrix with $\mathrm{rank}(R) = q$, $\theta^0 = (\theta^{0'}_1,\dots,\theta^{0'}_G)'$, and $r$ is a $q \times 1$ nonrandom vector. Despite appearing to be a standard testing problem, testing $H_0$ is nonstandard because the group structure $\gamma^{0}$ is unknown and needs to be estimated.

Instead, we consider null hypotheses given a group membership structure. For any group membership structure $\gamma = (g_1,\dots,g_N)'$, we define the following pseudo true parameters:
\begin{align*} 
    \theta_{\textrm{PCR}, g} (\gamma) &= \left( \sum_{i=1}^N \sum_{t=1}^T X_{it} X_{it}' \cdot \mathbf{1}\{ g_i = g \} \right)^{-1}  \sum_{i=1}^N \sum_{t=1}^T X_{it} X_{it}' B_i \cdot \mathbf{1}\{ g_i = g \}, \\
    \theta_{\textrm{TSK}, g} (\gamma) &= \left( \sum_{i=1}^N \mathbf{1}\{ g_i = g \} \right)^{-1} \sum_{i=1}^N B_i \cdot \mathbf{1}\{ g_i = g \},
\end{align*}
where PCR and TSK correspond to clustering procedures we introduce below. We then stack the group-specific coefficients so that $\theta_{v} (\gamma) = (\theta'_{v,1} (\gamma),\dots,\theta'_{v,G}(\gamma))'$ where $v \in \{PCR, TSK\}$. We emphasize that the pseudo true parameters under the null hypothesis depend on the clustering procedure we use. When the group structure holds, and the groups are assigned according to the true group membership structure, that is $\gamma = \gamma^0$, then $\theta_{v}(\gamma^0) = \theta^0$ for the true group-specific coefficients for $v \in \{PCR, TSK\}$. However, the groups do not need to be assigned according to the true group membership structure, and indeed the group structure need not hold at all in our framework. This feature makes our procedures robust to the misspecification of the number of groups and violations of group separation.

We hence state the null hypotheses based on the estimated group membership structure. Let $\gamma_{D} = (g_{1,D},\dots,g_{N,D})'$ be an estimator of $\gamma^0$ based on the data set $D$. Our null hypothesis is
\begin{equation}\label{eq:main_null}
    H_0: R\theta_{v} ( \gamma_{D}) = r, \quad v \in \{PCR, TSK\}.
\end{equation}
This null hypothesis depends on the estimated group structure and thus is data-dependent, and we solve this issue by conditioning on $\gamma_D$ as discussed in Section \ref{sec:preliminaries}. The usual Wald test may lose validity when the group structure is unobserved and estimated from the dataset prior to testing $H_0$. A particularly important example is the homogeneity hypothesis, which posits that all groups, or a subset of them, share the values of all or some of the coefficients.

We note that, in the rest of the paper, we write the null hypothesis as $H_0: R\theta = r$. That is, we drop the dependence of the pseudo-true coefficients on the estimated group structure and on the particular estimation method adopted. This choice is purely for notational simplicity and will not cause any confusion.

\subsection{Examples}\label{sec:examples}

In this section, we demonstrate the practical importance of having valid tests of the null hypothesis $H_0$ in three empirical examples, which are examined in Section \ref{sec:applications} and in Appendix \ref{sec:add_applications}.

\paragraph{Example 1: Growth convergence.}
The literature on economic growth suggests that it may exhibit convergence properties, and that convergence patterns may exhibit heterogeneity across countries, which is attributed to convergence clubs. Various methods for testing the hypothesis of convergence clubs are proposed by \cite{canova2004testing} and \cite{phillips2007transition}, among others. Let the first difference of the natural logarithm of per capita GDP be given by
\begin{equation*}
    \Delta \log GDP_{it} = \theta_{1,g_i} \log GDP_{i,t-1} + Z'_{it} \theta_{-1,g_i} + \varepsilon_{it}
\end{equation*}
where $Z_{it}$ is a vector of conditioning variables, $t$ typically represents years, and $i$ may denote countries or regions.\footnote{All three examples include a lagged dependent variable and are therefore dynamic, so that $X_{it}$ is predetermined rather than strictly exogenous. Our procedures accommodate such models as long as the relevant unit-specific statistics are asymptotically Gaussian as $T\to \infty$.}
Our model is derived by setting $Y_{it}=\Delta \log GDP_{it}$, $X_{it} = (\log GDP_{i,t-1},Z'_{it})'$ and $\theta_g = (\theta_{1, g},\theta'_{-1,g})'$ in \eqref{eq:main_model}. For example, assume that $K=3$ and $G=2$. The null hypothesis of a single $\beta$-convergence club is then the single restriction $\theta_{1,1} = \theta_{1,2}$, obtained by setting $r = 0$ and $R = (1,0,0,-1,0,0)$ in \eqref{eq:main_null}, where the non-zero entries of $R$ correspond to the regressors of interest and the zeros indicate the control variables.

\paragraph{Example 2: Firm level R\&D investment and the business cycle.}
The literature on the relationship between output and R\&D investment provides mixed evidence concerning the effect of output on R\&D, and as the evidence in \cite{bachmann2014investment} suggests, a particular question pertains to its heterogeneity. A version of the model estimated by \cite{aguilar-loyo_grouped_2024} is
\begin{equation*}
    \Delta \log RD_{it} = \theta_{1,g_i} \Delta \log X_{st} + Z'_{it} \theta_{-1,g_i} + \varepsilon_{it},
\end{equation*}
where $RD_{it}$ is the R\&D investment of firm $i$ in year $t$ and $X_{st}$ is the output of industry $s$. Setting $Y_{it} = \Delta \log RD_{it}$, $X_{it} = (\Delta \log X_{st},Z'_{it})'$ and $\theta_g = (\theta_{1,g},\theta'_{-1,g})'$ corresponds to the model in \eqref{eq:main_model}. \cite{aguilar-loyo_grouped_2024} argues that group heterogeneity in the effect of output on the R\&D investment may not be substantial, and proposes an estimation method that incorporates information about variance heterogeneity, but does not formally assess the extent of coefficient heterogeneity across groups. Supposing for simplicity that $K=3$ and $G=2$, this claim can be represented as $\theta_{1,1} = \theta_{1,2}$ such that $r = 0$ and $R = (1,0,0,-1,0,0)$ in \eqref{eq:main_null} as in the previous example. Note that the model estimated by \cite{aguilar-loyo_grouped_2024} incorporates time-varying GFE, an extension of the main model discussed in Section \ref{sec:gfe}.

\paragraph{Example 3: Income and democracy.} A typical application of panel models with latent group structure is the estimation of the relationship between income and democracy. \cite{bonhomme_grouped_2015} \citep[see also][]{okui_heterogeneous_2021} estimate a model similar to
\begin{equation*} 
    DEM_{it} = \theta_{1,g_i} DEM_{i,t-1} + \theta_{2,g_i} \log GDP_{i,t-1} + \varepsilon_{it}
\end{equation*} 
where $DEM_{it}$ is the Freedom House indicator of democracy and $GDP_{it}$ is the per capita GDP of country $i$ at year $t$. The model is obtained by setting $Y_{it} = DEM_{it}$, $X_{it} = (DEM_{i,t-1},\log GDP_{i,t-1})'$ and $\theta_g = (\theta_{1,g},\theta_{2,g})'$ in \eqref{eq:main_model}. Here, $K=2$ and the authors choose $G=4$. In their exercise with heterogeneous coefficients, they estimate two models: first, $\theta_{1, g_i}$ is heterogeneous, while $\theta_{2, g_i}=\theta$ for all $i$; second, both are heterogeneous. To determine whether the coefficient of lagged democracy is equal across all groups, one may set $R$ to the $3 \times 8$ matrix with rows $(1,0,0,0,0,0,-1,0)$, $(0,0,1,0,0,0,-1,0)$, and $(0,0,0,0,1,0,-1,0)$, together with $r = (0,0,0)'$, whereas to test whether the coefficient of lagged income is equal for Group 3 and Group 4, we can select $R = (0,0,0,0,0,1,0,-1)$ and $r = 0$.

\section{Estimators}\label{sec:estimation}

In this section, we discuss the estimation of the group membership variables $g_i$, $i=1,\dots, N$.
Section \ref{sec:panel_KMeans} introduces the PCR estimator that simultaneously estimates the coefficient parameters and the group memberships.
The TSK estimator defined in Section \ref{sec:simple_KMeans} utilizes unit-specific estimates of model parameters to cluster the units into $G$ groups.

\subsection{Panel Clusterwise Regression estimator}\label{sec:panel_KMeans}

The panel clusterwise regression (PCR) estimator, a panel data analog of clusterwise regression methods \citep{spath79}, gained prominence in econometrics following the work of \citet{bonhomme_grouped_2015}. The main advantage of this estimator, relative to the alternative clustering approach introduced below, namely TSK, is that it can accommodate a broader class of models. In particular, it can be extended to incorporate important generalizations of \eqref{eq:main_model}, such as GFE \citep{bonhomme_grouped_2015} or group-specific structural breaks \citep{okui_heterogeneous_2021}. Moreover, unlike related two-step procedures based on unit-by-unit estimation, such as \citet{wan2025conditional}, it does not require the number of regressors to be smaller than the time dimension.

Consider the following least squares estimator of the group membership variables and the group-specific parameters:
\begin{equation}\label{eq:panelKMeans_est}
\big(\hat{\theta}_{1,D},\dots,\hat{\theta}_{G,D},\hat{g}_{1,D},\dots,\hat{g}_{N,D}\big) = \argmin_{\left(\theta_1,\dots,\theta_G,g_1,\dots,g_N\right)} \sum_{i=1}^N \sum_{t=1}^T ( Y_{it} - X_{it}'\theta_{g_i} )^2.
\end{equation}
The estimator and its numerical implementations are well studied in the literature. We rely on the following algorithm, which is a panel data generalization of the clusterwise regression algorithm \citep{spath79}. It iterates between two optimizations: for any given $\gamma$, we have $(\hat{\theta}_{1,D},\dots , \hat{\theta}_{G,D}) = \argmin_{\left(\theta_1,\dots,\theta_G\right)} \sum_{i=1}^N \sum_{t=1}^T ( Y_{it} - X_{it}'\theta_{g_i} )^2$, and for any given $\{\theta_{g}, g=1,\dots,G \}$, $\hat{\gamma}_{D} = \left(\hat{g}_{1,D},\dots,\hat{g}_{N,D}\right) = \argmin_{\left(g_1,\dots,g_N\right)} \sum_{i=1}^N \sum_{t=1}^T ( Y_{it} - X_{it}'\theta_{g_i} )^2$. Apart from the absence of GFE and the presence of group-specific slope coefficients in our main model, Algorithm \ref{algo:panel_KMeans}, which takes a realization of the original data set $D$ as input, is equivalent to the main iterative algorithm of \cite{bonhomme_grouped_2015}.
In Section \ref{sec:gfe}, we discuss the extension to the models with GFE.

\begin{algorithm}[ht]
\caption{Panel Clusterwise Regression}\label{algo:panel_KMeans}
\SetAlgoLined
\DontPrintSemicolon

\KwIn{Realization $d = \{ (y_{it}, x_{it}); i=1, \dots, N, t=1,\dots, T \}$ of $D$; number of groups $G$}
\KwOut{Group assignment vector $\hat{\gamma}_{d}$; group-specific coefficients $\hat{\theta}_{g,d}$; number of iterations $M$}

Initialize $\hat{\theta}_{g,d}^{(0)}$ for all $g = 1, \dots, G$; set $m = 0$\;

\Repeat{$\hat{g}_{i,d}^{(m+1)} = \hat{g}_{i,d}^{(m)}$ for all $i = 1, \dots, N$}{
    \For{$i = 1$ \KwTo $N$}{
        Assign group:
        \[
        \hat{g}_{i,d}^{(m+1)} = \argmin_{g \in \{1, \dots, G\}} \sum_{t=1}^T \big(y_{it} - x_{it}' \hat{\theta}_{g,d}^{(m)}\big)^2
        \]
    }

    \For{$g = 1$ \KwTo $G$}{
        Update group parameters:
        \[
        \hat{\theta}_{g,d}^{(m+1)} =
        \left( \sum_{i=1}^N \sum_{t=1}^T x_{it} x_{it}' \cdot \mathbf{1}\big\{ \hat{g}_{i,d}^{(m+1)} = g \big\} \right)^{-1}
        \sum_{i=1}^N \sum_{t=1}^T x_{it} y_{it} \cdot \mathbf{1}\big\{ \hat{g}_{i,d}^{(m+1)} = g \big\}
        \]
    }

    Update $m \gets m+1$
}

Set $M = m$; $\hat{\theta}_{g,d}  = \hat{\theta}_{g,d}^{(M)}$; $\hat{\gamma}_{d} = (\hat{g}_{1,d}^{(M)},\dots,\hat{g}_{N,d}^{(M)})'$

\end{algorithm}

\subsection{Two-Step KMeans estimator}\label{sec:simple_KMeans}

Next, we discuss an alternative estimator. A straightforward method for estimating group memberships and group-specific slopes is to apply the standard KMeans algorithm to unit-specific estimates of the slope parameters. Several studies in the literature employ this strategy. For instance, \cite{wang_panel_2024} use unit-specific slope parameter estimates to cluster the units into different groups using KMeans. However, this method has shortcomings: it cannot accommodate important generalizations of the model in \eqref{eq:main_model}, such as GFE \citep{bonhomme_grouped_2015} or group-specific structural breaks {\citep{okui_heterogeneous_2021}.} In addition, it is infeasible when $T < K$. Nevertheless, we study this estimator as it is popular in empirical work and easy to implement.

The TSK estimator of the group membership variables $g_i$, $i=1,\dots,N$, and the group-specific slopes $\theta_g$, is defined as
\begin{equation}\label{eq:KMeans_ghat}
\big(\tilde{\theta}_{1,D},\dots,\tilde{\theta}_{G,D},\tilde{g}_{1,D},\dots,\tilde{g}_{N,D}\big) = \argmin_{\left(\theta_1,\dots,\theta_G,g_1,\dots,g_N\right)} \sum_{i=1}^N \big\lVert \hat{B}_{i} - \theta_{g_i} \big\rVert^2
\end{equation}
where $\hat{B}_{i} = \big( \sum_{t=1}^T X_{it} X'_{it} \big)^{-1} \sum_{t=1}^T X_{it}Y_{it}$ represents the unit-by-unit OLS estimator of $B_i$. For any given $\gamma = (g_1,\dots,g_N)'$, this definition implies $\tilde{\theta}_{g,D} = \argmin_{\left(\theta_1,\dots,\theta_G\right)} \sum_{i=1}^N \lVert \hat{B}_{i} - \theta_{g_i} \rVert^2$, and for any given $\{ \theta_{g}, g=1,\dots,G \}$, $\tilde{\gamma}_{D} = \left(\tilde{g}_{1,D},\dots,\tilde{g}_{N,D}\right)' = \argmin_{\left(g_1,\dots,g_N\right)} \sum_{i=1}^N \lVert \hat{B}_{i} - \theta_{g_i} \rVert^2$. Algorithm \ref{algo:simple_KMeans}, which takes a realization of $\{ \hat{B}_{i}; i=1, \dots, N \}$ as input, summarizes the steps of the standard iterative KMeans algorithm associated with the above estimator.

\begin{algorithm}[ht]
\caption{Two-Step KMeans}\label{algo:simple_KMeans}
\SetAlgoLined
\DontPrintSemicolon

\KwIn{{Realization $\{ \hat{b}_{i}; i=1, \dots, N \}$ of $\{ \hat{B}_{i}; i=1, \dots, N \}$; number of groups $G$}}
\KwOut{Group assignment vector $\tilde{\gamma}_{d}$; group means $\tilde{\theta}_{g,d}$; number of iterations $M$}

Initialize $\tilde{\theta}_{g,d}^{(0)}$ for all $g = 1, \dots, G$; set $m = 0$\;

\Repeat{$\tilde{g}_{i,d}^{(m+1)} = \tilde{g}_{i,d}^{(m)}$ for all $i = 1, \dots, N$}{
    \For{$i = 1$ \KwTo $N$}{
        Assign cluster:
        $\tilde{g}_{i,d}^{(m+1)} = \argmin_{g \in \{1, \dots, G\}} \big\| \hat{b}_{i} - \tilde{\theta}_{g,d}^{(m)} \big\|^2$
    }
    
    \For{$g = 1$ \KwTo $G$}{
        Compute:
        \[
        \tilde{\theta}_{g,d}^{(m+1)} =
        \frac{1}{\tilde{n}_{g,d}^{(m+1)}} \sum_{i=1}^N \hat{b}_{i} \cdot \mathbf{1}\big\{ \tilde{g}_{i,d}^{(m+1)} = g \big\}
        \]
        where $\tilde{n}_{g,d}^{(m+1)} = \sum_{i=1}^N \mathbf{1}\big\{ \tilde{g}_{i,d}^{(m+1)} = g \big\}$
    }
    
    Update $m \gets m+1$
}

Set $M = m$; $\tilde{\theta}_{g,d}  = \tilde{\theta}_{g,d}^{(M)}$; $\tilde{\gamma}_{d} = (\tilde{g}_{1,d}^{(M)},\dots,\tilde{g}_{N,d}^{(M)})'$

\end{algorithm}
We see that $\tilde{\theta}_{g,D} = \tilde{n}_{g,D}^{-1} \sum_{i=1}^N \hat{B}_{i} \cdot \mathbf{1} \{\tilde{g}_{i,D} = g\}$ where $\tilde{n}_{g,D} = \sum_{i=1}^N  \mathbf{1} \{ \tilde{g}_{i,D} = g \}$. While $\tilde{\theta}_{g, D}$ is not the least squares estimator using observations in group $g$, as for the PCR estimator, it can be shown that $\tilde{\theta}_{g, D} \overset{p}{\longrightarrow} \theta_g$ as $(T, N) \rightarrow \infty$. This estimator is the Mean Group estimator applied to each group. The advantages of the Mean Group estimator for short dynamic panels with heterogeneous coefficients are well documented in the literature \citep{pesaran1995estimating}.

Algorithm \ref{algo:simple_KMeans} is easy to implement. It is the KMeans clustering, available in most statistical software, using data obtained by least squares. To examine basic subcases of the null hypothesis $H_0$ regarding parameter homogeneity across groups, a simple adjustment to the computational procedures outlined by \cite{chen23} and \cite{chen2024} can be made. However, these studies focus on a narrow null hypothesis of equality between two cluster centers and operate under restrictive assumptions that may not hold in typical economic panel data settings. Our work expands the scope of the test by introducing a more realistic set of assumptions for economic applications.

\begin{remark}
\label{rem:weighted}
The clustering step can be extended from the Euclidean criterion to a weighted quadratic criterion of the form:
\[
\sum_{i=1}^N (\hat{B}_i - \theta_{g_i})' \Phi_i (\hat{B}_i - \theta_{g_i}),
\]
where each $\Phi_i$ is a symmetric positive definite weight matrix, the Euclidean case in the main text corresponding to $\Phi_i = I_K$ for all $i$. Note that when all of $\hat{B}_i$ can be computed, the PCR procedure coincides with the weighted version of the TSK procedure by choosing $\Phi_i = X_i'X_i$, an insight also used in \citet{wan2025conditional}. However, this correspondence fails when some or all of $\hat{B}_i$ cannot be computed, such as cases with GFE or, in general, cases with limited time series variations in $X_{it}$. The extension to weighted versions would be useful in the presence of heteroskedasticity, but would require a careful treatment of how the weights are constructed, which we discuss in the theory section below.
\end{remark}

\section{Test Statistics}\label{sec:tests}

In this section, we propose two main test statistics, each corresponding to an estimator discussed in the previous section.
In Section \ref{sec:preliminaries}, we introduce the selective Type I error rate and other preliminary definitions.
In Section \ref{sec:panel_KMeans_tests}, the tests based on the PCR estimator are derived, and Section \ref{sec:simple_KMeans_tests} is concerned with the derivation of the TSK estimator-based tests.

\subsection{Conditional inference in panel models}\label{sec:preliminaries}

In this section, we establish the testing situation and define the Type I error rate we aim to control.
The construction of the null hypothesis \eqref{eq:main_null} indicates that the group structure is estimated from the data and may differ from the population's group structure.
Under \eqref{eq:main_null}, the group separation may fail, and the number of groups may be overspecified.

The null \eqref{eq:main_null} is random because $\gamma_{D}$, which in turn defines the vector $\theta_{v}$, $v \in \{PCR, TSK\}$, is computed from the data.
The matrix $R$ and the vector $r$ in \eqref{eq:main_null} are nonrandom for given $G$.\footnote{If $G$ is estimated, these terms would potentially depend on the data as well, through their dimensions being related to the number of clusters estimated. We assume throughout the paper that $G$ is prespecified.}
Thus, the coefficients being tested are fixed for any \emph{given} $\gamma$.
However, when the groups are unknown and must be estimated from the data, the coefficients are random because they depend on $D$.
Acknowledging this randomness is important for handling situations in which the groups are not well-defined under the null hypothesis.
Failing to do so invalidates the conventional testing procedures of $H_0$ in \eqref{eq:main_null}, such as the use of a $\chi^2$ distribution to compute the critical value for the Wald test.

\begin{figure}[!htbp]
\centering
\captionsetup[subfigure]{position=top}

\setlength{\fboxrule}{1pt}
\setlength{\fboxsep}{0pt}

\begin{subfigure}[t]{0.32\textwidth}
    \centering
    \caption{Spurious PCR groups}
    \vspace{0.5em}
    \includegraphics[width=\textwidth]{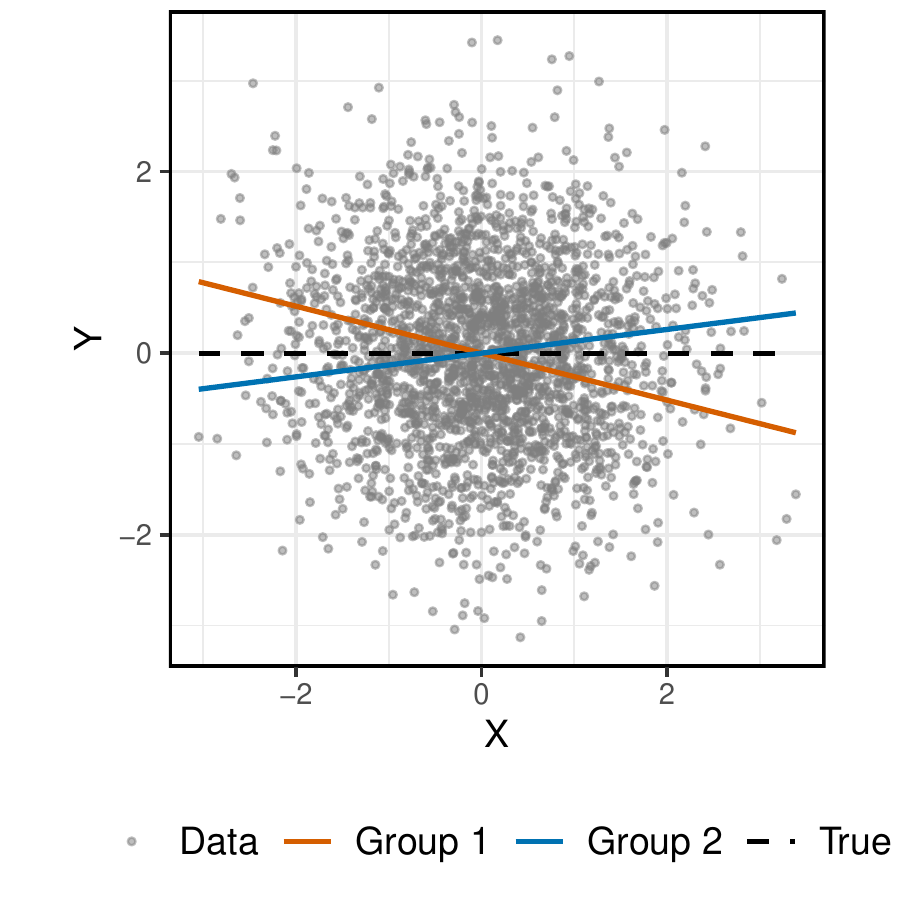}
\end{subfigure}
\hfill
\begin{subfigure}[t]{0.32\textwidth}
    \centering
    \caption{Q-Q plot under $H_0:\theta_{1,1}=\theta_{1,2}$}
    \vspace{0.5em}
    \includegraphics[width=\textwidth]{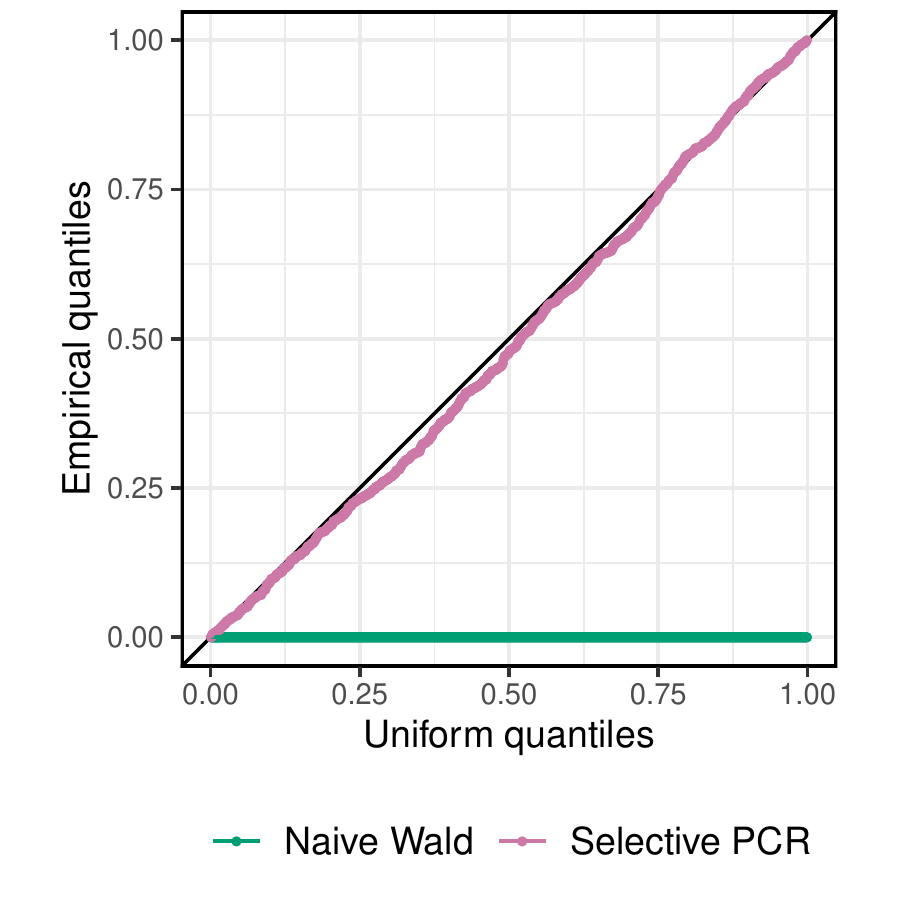}
\end{subfigure}
\hfill
\begin{subfigure}[t]{0.32\textwidth}
    \centering
    \caption{Q-Q plot under $H_0:\theta_{1,1}=0$}
    \vspace{0.5em}
    \includegraphics[width=\textwidth]{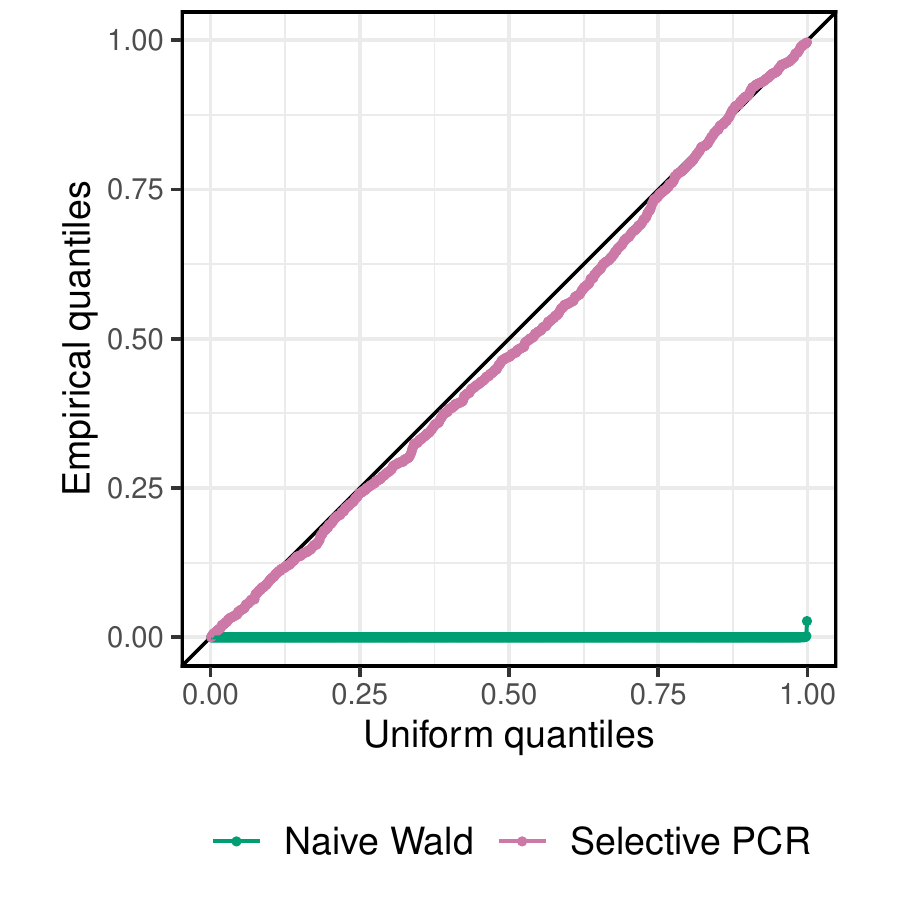}
\end{subfigure}

\vspace{1em}

\caption{PCR clustering and inference under the null of parameter homogeneity}
\label{fig:pcr_null_homogeneity}

\vspace{0.5em}

\begin{minipage}{0.96\textwidth}
\small
\textit{Note.} The data are generated from the homogeneous panel model $Y_{it}=0 \cdot X_{it}+U_{it}$, $i=1,\dots,N $, $t=1,\dots,T$, $X_{it}\sim N(0,1)$, and $U_{it}\sim N(0,1)$, independently over $i$ and $t$. Thus, there is no latent group structure in the population. In the left panel, however, PCR is forced to estimate two groups. The dashed line represents the true population regression function, while the two solid lines are the regression functions associated with the two estimated PCR groups. The middle and right panels report Q-Q plots of Monte Carlo $p$-values for the naive Wald test and the selective PCR test under the null hypotheses $H_0:\theta_{1,1}=\theta_{1,2}$ and $H_0:\theta_{1,1}=0$, respectively. The test statistics are calculated under the iid assumption and with a degree-of-freedom correction to the variance estimator.
\end{minipage}
\end{figure}

Figure \ref{fig:pcr_null_homogeneity} illustrates the consequences of ignoring the randomness in the null hypothesis. In Panel (a), even though the data are generated from a fully homogeneous model, the PCR estimator mechanically partitions the sample into two groups and produces clearly distinct fitted regression lines, showing that estimated group separation may arise purely from sampling variation rather than genuine population heterogeneity. This artificial separation has important consequences for inference: the Q-Q plot in Panel (b) shows that the naive Wald test is severely distorted under a homogeneity null, with $p$-values concentrated near zero, whereas the selective PCR $p$-values are uniformly distributed. Ignoring the uncertainty inherent in data-driven clustering therefore leads to substantial overrejection, whereas the selective procedure corrects for this source of distortion.

Furthermore, the problem does not concern only the cases of testing homogeneity hypotheses of the form $H_0:\theta_{1,1}=\theta_{1,2}$ as indicated in the previous literature. Panel (c) shows that when group separation fails, even the simple significance Wald tests for hypotheses such as $H_0:\theta_{1,1}=0$ fail to control the Type I error. Hence, the problem of using the same data to select the groups and then testing for general linear hypotheses is more serious than previously thought.

We design tests that control the Type I error conditional on the estimated group structure.
This approach eliminates the randomness in the null hypothesis. 
Below, we define the concept of selective Type I error rate.

\begin{definition}\normalfont\label{definition:selectivepvalue}
{A test of $H_0$ controls the selective Type I error rate at level $\alpha \in (0,1)$ if}
\begin{equation}\label{eq:selective_pvalue_def}
    \Pr_{H_0} \left[ \text{Reject } H_0 \text{ at level } \alpha \;\middle|\; \gamma_D = \gamma_d \right] \leq \alpha, 
\end{equation}
where $\gamma_D$ is a data-dependent choice of $\gamma$, such as the output of Algorithm \ref{algo:simple_KMeans} or Algorithm \ref{algo:panel_KMeans}, and $\gamma_d$ is its realized value associated to the realization $d$ of $D$.
\end{definition}

The definition states that a test controls the selective Type I error rate if the probability of rejecting $H_0$ when it is true is at most $\alpha$ across all realizations of the data set $D$ that yield the same group membership estimates. Below, we demonstrate how to construct tests by conditioning on the estimated groups, thereby asymptotically controlling the selective Type I error rate.

\subsection{Tests based on the PCR estimator}\label{sec:panel_KMeans_tests}

We now discuss our testing procedures. First, we consider the circumstances under which the PCR estimator is used. Our test statistic is based on the conditional distribution of the quadratic form of the constrained estimator given the estimated group structure. The conditional distribution follows a truncated $\chi^2$ distribution.

The test statistic we consider is the Wald statistic based on the PCR estimator under the null hypothesis $H_0$, whose conditional distribution is used to test $ H_0$. It is given by
\begin{equation}\label{eq:pcr_test_stat}
    W_{\textrm{PCR}} = (R \hat{\theta}_D - r)'\widehat{\Omega}_{R,D}^{-1} (R \hat{\theta}_D - r),
\end{equation}
where $\hat{\theta}_{D} = (\hat{\theta}'_{1,D},\dots,\hat{\theta}'_{G,D})'$, $\widehat{\Omega}_{R,D} = R \widehat{\Omega}_D R'$ with $\widehat{\Omega}_D$ being an estimator of $\Omega_{\textrm{PCR}} = \mathrm{V} ( \hat{\theta}_D )$. To provide the explicit formula of $\Omega_{\textrm{PCR}}$, we introduce the following notation. Let $\mathbb{X}$ be the $NT \times NK$ block-diagonal matrix whose $i$-th block is $X_i = (X_{i1},\dots,X_{iT})'$, and define $\widehat{\mathbb{H}} = \widehat{H} \otimes I_K$ where $\widehat{H}$ is the $N \times G$ matrix of group dummies based on $\hat{\gamma}_D$. The group-stacked regressor matrix $X_{\hat{\gamma}} = \mathbb{X}\widehat{\mathbb{H}}$ is $NT \times GK$. Let $\sigma^2 = \mathrm{V} ( \varepsilon_{it} )$. Using this notation, $\hat{\theta}_D = (X_{\hat{\gamma}}'X_{\hat{\gamma}})^{-1}X_{\hat{\gamma}}'Y$ and $\Omega_{\textrm{PCR}} = \sigma^2(X_{\hat{\gamma}}'X_{\hat{\gamma}})^{-1}$. 

We examine the conditional distribution of $W_{\textrm{PCR}}$ under the assumption that the group structure estimator is $\hat{\gamma}_D$. However, obtaining this distribution is quite challenging. Therefore, we condition on additional random variables that are independent of $W_{\textrm{PCR}}$. To this end, we derive a decomposition of $Y$ into a component that determines $W_{\textrm{PCR}}$ and a nuisance component that is independent of it.

We start from the identity $Y = X_{\hat{\gamma}}\hat{\theta}_{R,D} + X_{\hat{\gamma}}(\hat{\theta}_D - \hat{\theta}_{R,D}) + \hat{\varepsilon}$, where $\hat{\varepsilon} = Y - X_{\hat{\gamma}}\hat{\theta}_D$ and the constrained PCR estimator $\hat{\theta}_{R,D}$ is given by
$
\hat{\theta}_{R,D} = \hat{\theta}_D - (X_{\hat{\gamma}}'X_{\hat{\gamma}})^{-1} R' [R(X_{\hat{\gamma}}'X_{\hat{\gamma}})^{-1} R']^{-1} (R \hat{\theta}_D - r),
$,
which follows from the standard textbook formula of the constrained least squares estimator \citep[see][p.122, for instance]{greene2011econometric}.
Since $\hat{\theta}_D - \hat{\theta}_{R,D} = (X_{\hat{\gamma}}'X_{\hat{\gamma}})^{-1}R'[R(X_{\hat{\gamma}}'X_{\hat{\gamma}})^{-1}R']^{-1}(R\hat{\theta}_D - r)$, we obtain the decomposition
\begin{equation}\label{eq:decomposition_pcr_Y}
Y = X_{\hat{\gamma}}(X_{\hat{\gamma}}'X_{\hat{\gamma}})^{-1}R' [R(X_{\hat{\gamma}}'X_{\hat{\gamma}})^{-1}R']^{-1}(R\hat{\theta}_D - r) + \hat{V},
\end{equation}
where $\hat{V} = X_{\hat{\gamma}}\hat{\theta}_{R,D} + \hat{\varepsilon}$.
Define
$
\mathbb{Q}_{\textrm{PCR}} = X_{\hat{\gamma}}(X_{\hat{\gamma}}'X_{\hat{\gamma}})^{-1}R' [  R(X_{\hat{\gamma}}'X_{\hat{\gamma}})^{-1}R']^{-1/2} \sigma
$,
which is $NT \times q$. Let $\mathcal{L}_D = [\sigma^2 R(X_{\hat{\gamma}}'X_{\hat{\gamma}})^{-1}R']^{-1/2}(R\hat{\theta}_D - r)$, so that $W_{\textrm{PCR}}^{1/2} = \lVert \mathcal{L}_D \rVert$ and $\hat{J} = \mathrm{dir}(\mathcal{L}_D)$. Applying the norm-direction decomposition to $\mathcal{L}_D$, we obtain
\begin{equation}\label{eq:decomposition_pcr_Y2}
Y = W_{\textrm{PCR}}^{1/2}\, \mathbb{Q}_{\textrm{PCR}} \hat{J} + \hat{V}.
\end{equation}

In the above formulation, the constrained estimator is needed because we consider a general linear hypothesis rather than simple homogeneity restrictions between cluster centers, as in much of the existing post-clustering inference literature \citep{cheng2023clustering,gao24,wan2025conditional}. The constrained estimator allows us to separate the component of the data that determines the test statistic from the remaining nuisance component. This decomposition is essential for extending selective inference to general linear hypotheses.

Premultiplying \eqref{eq:decomposition_pcr_Y2} by $\mathbb{X}'$ we obtain the following alternative expression
\begin{equation}\label{eq:decomposition_pcr_S}
S
=
W_{\textrm{PCR}}^{1/2}\mathbb{P}_{\textrm{PCR}}\hat{J}
+
\hat{U},
\end{equation}
where $S=\mathbb{X}'Y$, $\hat{U}=\mathbb{X}'\hat{V}$ and $\mathbb{P}_{\textrm{PCR}}=\mathbb{X}'\mathbb{Q}_{\textrm{PCR}}$ is an $NK\times q$ matrix. This alternative decomposition is important because the PCR clustering procedure depends on the data through the statistic $S=\mathbb{X}'Y$, which is $NK\times 1$, contrary to \eqref{eq:decomposition_pcr_Y2}, which operates on a $NT\times 1$ vector. We develop this idea in the following remark.

\begin{remark}
\label{rem:measurability}
All quantities entering our analysis depend on the data only through $S = \mathbb{X}'Y$, including the group assignments produced by Algorithm~\ref{algo:panel_KMeans}. At any iteration, the PCR rule allocates unit $i$ to the group minimizing
$
\lVert Y_i - X_i \theta_g \rVert^2
= \lVert Y_i \rVert^2 - 2\, \theta_g' S_i + \theta_g' \Sigma_i \theta_g
$
where $S_i = X_i'Y_i$ and $\Sigma_i = \sum_{t=1}^T X_{it}X_{it}'$. The first term does not depend on $g$ and cancels in every pairwise comparison, while $\Sigma_i$ is a function of the fixed regressors alone. The assignment of unit $i$ is therefore a function of $S_i$ and the regressors only, for arbitrary unit-specific designs $\Sigma_i$. Iterating, the entire assignment path, and hence the selection event $\bigcap_{m=0}^{M}\{\hat\gamma_D^{(m)} = \hat\gamma_d^{(m)}\}$, is measurable with respect to $S$. This justifies conducting the conditional analysis using $S$ and conditioning on $\hat U = \mathbb{X}'\hat V$ rather than on the higher-dimensional $\hat V$.
\end{remark}

This reduction of dimension will be useful in the proof of Theorem \ref{thm:pcr} to obtain the necessary independence results of the nuisance terms from the main conditioning event, which we will present in what follows. We consider the following conditioning set:
\begin{equation*}
\mathcal{C}_{\textrm{PCR}} = \left\lbrace \bigcap_{m=0}^M \lbrace
        \hat{\gamma}^{(m)}_D =
        \hat{\gamma}^{(m)}_d
    \rbrace,\; \hat{J} = \hat{j},\; \hat{U} = \hat{u} \right\rbrace
\end{equation*}
where $\hat{u}$ is the realization of $\hat{U}$ associated with the realization $d$ of $D$, $\hat{j}$ is that of $\hat{J}$, and $\widehat{\gamma}^{(m)}_d = (\hat{g}^{(m)}_{1,d},\dots,\hat{g}^{(m)}_{N,d})'$, namely the group membership structure in the $m$-th step of Algorithm ~\ref{algo:panel_KMeans}.
The conditioning set $\mathcal{C}_{\textrm{PCR}}$ contains more information than the condition in Definition~\ref{definition:selectivepvalue}. The first term implies that we focus on realizations of $D$ that yield the same group membership estimates at each iteration of Algorithm~\ref{algo:panel_KMeans} as those of the current realization. Although this is more restrictive than conditioning solely on the final group membership estimates, it greatly simplifies the analytical formulas for the truncation set \citep{chen23}. Moreover, to obtain a nuisance-free conditional distribution of the test statistic, we need the second and third terms \citep[see also a related discussion in][]{gao24}.
By the law of iterated expectations, conditioning on these additional terms does not affect the Type I error rate \citep[see also][]{taylor2015statistical}.

The following result summarizes how to construct a test statistic that successfully controls the Type I error rate.
\begin{theorem}\normalfont
\label{thm:pcr}
Suppose that $X_{it}$ is nonrandom,
$S \sim N(\mathbb{X}'\mathbb{X} B , \sigma^2 \mathbb{X}'\mathbb{X})$ and $\sum_{t=1}^T X_{it} X_{it}' = \Sigma$ for any $i$, where $\Sigma$ is nonsingular.
Then, under $H_0$,
$W_{\textrm{PCR}} \; | \; \mathcal{C}_{\textrm{PCR}} \sim \chi^2_{q}|_{\mathcal{T}_{\textrm{PCR}}}$,
with $\chi^2_{q}|_{\mathcal{T}_{\textrm{PCR}}}$ a $\chi^2_{q}$ random variable truncated to the set $\mathcal{T}_{\textrm{PCR}}$, which is given by
\begin{equation}\label{eq:truncationset_pcr}
\mathcal{T}_{\textrm{PCR}} = \left\lbrace \phi^2 \in \mathbbm{R}_{\geq 0} : \bigcap_{m=0}^M \lbrace \widehat{\gamma}^{(m)}_{d(\phi)} = \widehat{\gamma}^{(m)}_{d} \rbrace \right\rbrace,
\end{equation}
where $d(\phi) = \{[y_{it}(\phi), X_{it}]; i=1,\dots,N,\, t=1,\dots,T\}$ with $y_{it}(\phi)$ being the $(i,t)$-th element of the perturbed outcome vector
\begin{equation}\label{eq:perturbation_pcr}
y(\phi) = \phi \cdot \mathbb{Q}_{\textrm{PCR}}\hat{j} + \hat{v},
\end{equation}
and the corresponding perturbed sufficient statistic is $s(\phi) = \mathbb{X}'y(\phi) = \phi \cdot \mathbb{P}_{\textrm{PCR}}\hat{j} + \hat{u}$.
\end{theorem}

Equation \eqref{eq:perturbation_pcr} defines a perturbation of the outcome vector $Y$ as a function of the scalar $\phi$, holding the regressors $X_{it}$ fixed. Let $w_{\textrm{PCR}}$ be the realization of $W_{\textrm{PCR}}$ associated with the data $d$. When $\phi = w_{\textrm{PCR}}^{1/2}$, the original data $y$ is recovered. If $\phi > w_{\textrm{PCR}}^{1/2}$, the outcomes are shifted so that the estimated group-specific slopes move further apart, increasing the test statistic. Conversely, if $\phi < w_{\textrm{PCR}}^{1/2}$, the outcomes are shifted to bring the estimated slopes closer together, reducing the test statistic. The truncation set $\mathcal{T}_{\textrm{PCR}}$ consists of the values of $\phi$ for which Algorithm~\ref{algo:panel_KMeans}, applied to the perturbed data $\{[y_{it}(\phi), X_{it}]\}$, yields the same group assignment as with the original data. The analytical formulas for computing $\mathcal{T}_{\textrm{PCR}}$ are derived in Appendix~\ref{sec:truncationset}.

\begin{remark}
\label{rem:pcr-unequal-design}
The theorem assumes normally distributed, homoskedastic errors as well as the normalization $\sum_{t=1}^T X_{it}X_{it}' = \Sigma$ for all $i$. A representative example is $\Sigma = I_K$. It is possible to relax the restriction on $\sum_{t=1}^T X_{it}X_{it}'$ which is used in Lemma~\ref{lemma-wind-pcr} only to obtain the compact block forms $\mathbb{X}'\mathbb{X} = I_N \otimes \Sigma$ and $\widehat{\mathbb{H}}'\mathbb{X}'\mathbb{X}\widehat{\mathbb{H}} = \operatorname{diag}(n_1, \dots, n_G) \otimes \Sigma$. It is not essential to the independence result. Retaining general designs, $\mathbb{X}'\mathbb{X} = \operatorname{bdiag}(\Sigma_1, \dots, \Sigma_N)$ is an arbitrary block-diagonal SPD matrix, and $\hat{\theta}_D$, $\hat{\theta}_{R,D}$, and $\hat{U}$ are all defined directly in terms of $X_{\hat{\gamma}} = \mathbb{X}\widehat{\mathbb{H}}$ and $X_{\hat{\gamma}}'X_{\hat{\gamma}}$. The cross-covariance computation in the proof of Lemma~\ref{lemma-wind-pcr} then goes through verbatim with $X_{\hat{\gamma}}'X_{\hat{\gamma}}$ in place of $I_N \otimes \Sigma$, since the cancellation relies only on $X_{\hat{\gamma}} = \mathbb{X}\widehat{\mathbb{H}}$ and the idempotence of $\mathbb{X}(\mathbb{X}'\mathbb{X})^{-1}\mathbb{X}'$, not on the Kronecker structure. With Remark~\ref{rem:measurability}, exact finite-sample validity of the PCR test therefore does not require a common design second moment, provided the errors remain homoskedastic normal.

This is a statement about unequal designs, not heteroskedastic errors. When all $\hat{B}_i$ can be computed, Euclidean PCR coincides with weighted TSK with $\Phi_i = X_i'X_i$ (Remark~\ref{rem:weighted}); since $\hat{B}_i \sim N(B_i, \sigma^2 \Sigma_i^{-1})$, this is exactly the inverse-variance weight of Lemma~\ref{lem:tsk-weighted-independence} up to the scalar $\sigma^2$. The robustness of PCR to unequal designs is thus an instance of that lemma rather than a construction-specific advantage. The genuine advantage of PCR lies elsewhere: as discussed in Sections~\ref{sec:panel_KMeans_tests} and~\ref{sec:gfe}, it remains applicable when the $\hat{B}_i$ cannot be formed at all, such as with grouped fixed effects or limited time-series variation in $X_{it}$, where the weighted-TSK correspondence is unavailable.
\end{remark}

\subsection{Tests based on the TSK estimator}\label{sec:simple_KMeans_tests}

Next, we examine the case in which the parameters are estimated using the TSK estimator. The process of constructing the test is similar to that of the PCR estimator, but with subtle differences. Specifically, a different decomposition is needed to derive the conditional distribution of the estimator. 

In this case, the test statistic is given by
\begin{equation}\label{eq:twostep_KMeans_test_stat}
    W_{\textrm{TSK}} = (R \tilde{\theta}_{D} - r)'\tilde{\Omega}_{R,D}^{-1} (R \tilde{\theta}_{D} - r)
\end{equation}
where $\tilde{\theta}_{D} = (\tilde{\theta}'_{1,D},\dots,\tilde{\theta}'_{G,D})'$, $\tilde{\Omega}_{R,D} = R \tilde{\Omega}_D R'$ with $\tilde{\Omega}_D$ being an estimator of
the variance of $\tilde{\theta}_{D}$. 

We assume that $\hat{B} \sim N(B, \sigma^2 I_N \otimes \Sigma^{-1})$ where $\hat{B} = (\hat{B}'_1,\dots,\hat{B}'_N)'$ is the $NK \times 1$ vector of estimated slope coefficients, $B$ is an $NK \times 1$ vector of the true values of the coefficients
and $\Sigma$ is an invertible $K\times K$ matrix. 
Here, we implicitly assume that the regression error is normally distributed and homoskedastic, and that $\sum_{t=1}^T X_{it} X_{it}' = \Sigma$ for all $i$. Note that we make similar assumptions for the PCR procedure.
In this setting, we have $\tilde{\Omega}_{R,D} = \sigma^2 R \widetilde{\mathcal{N}}_{\Sigma}^{-1}R'$, where $\widetilde{\mathcal{N}}_{\Sigma} = \mathrm{diag}(\tilde{n}_{1},\dots,\tilde{n}_{G}) \otimes \Sigma$.

Our derivation of the conditional distribution is based on the deviations from the constrained estimator of \(\theta\). Describing the constrained estimator, denoted as $\tilde{\theta}_{R, D}$, requires newly introduced notation.
Define $\widetilde{\mathbb{H}} =  \tilde{H} \otimes I_K $ where $\tilde{H}$ is the $N \times G$ matrix of group dummies based on the TSK estimator $\tilde{\gamma}_D$ and $I_K$ is the $K$-dimensional identity matrix. This definition differs from that for the PCR estimator only in the use of $\tilde{\gamma}_D$ instead of $\hat{\gamma}_D$.
The TSK estimator of the group-specific slope coefficients can be written as $\tilde{\theta}_{D} = (\widetilde{\mathbb{H}}'\widetilde{\mathbb{H}})^{-1}\widetilde{\mathbb{H}}'\hat{B}$.
The constrained estimator is then given by
\begin{equation}\label{eq:constrained_tsk}
\tilde{\theta}_{R,D} = \tilde{\theta}_{D} - \widetilde{\mathcal{N}}_{\Sigma}^{-1} R' (R \widetilde{\mathcal{N}}_{\Sigma}^{-1} R')^{-1} (R \tilde{\theta}_{D} - r).
\end{equation}

We now obtain the following decomposition:
\begin{equation}\label{eq:decomposition_KMeans1}
    \hat{B} = \widetilde{\mathbb{H}} \tilde{\theta}_{R,D} + \mathbb{Q}_{\textrm{TSK}} \tilde{\Omega}_{R,D}^{-1/2} (R \tilde{\theta}_{D} - r) + \mathbb{M}\hat{B}
\end{equation}
where
$
\mathbb{Q}_{\textrm{TSK}} = \widetilde{\mathbb{H}} \widetilde{\mathcal{N}}_{\Sigma}^{-1} R' (R \widetilde{\mathcal{N}}_{\Sigma}^{-1} R')^{-1/2} \sigma
$ and $ \mathbb{M} = I_{NK} - \widetilde{\mathbb{H}}(\widetilde{\mathbb{H}}'\widetilde{\mathbb{H}})^{-1} \widetilde{\mathbb{H}}' $. The first term in \eqref{eq:decomposition_KMeans1} contains the vector of constrained estimates of the group means under $H_0$. The second term is the deviations of the unconstrained grouped means from the constrained estimates. The third term $\mathbb{M}\hat{B}$ is the residual of $\hat{B}$ from its unweighted group means.\footnote{Under the inverse-variance weight of Remark \ref{rem:wkmeans}, $\mathbb{M}$ generalizes to the oblique projector $I_{NK} - \widetilde{\mathbb{H}}(\widetilde{\mathbb{H}}'\Phi \widetilde{\mathbb{H}})^{-1}\widetilde{\mathbb{H}}' \Phi$. The two coincide here because the clustering metric is Euclidean and $\sum_{t=1}^T X_{it} X_{it}'$ is common across units.}
The decomposition in Equation \eqref{eq:decomposition_KMeans1} is important because it allows us to measure the distance of the unconstrained estimates from the constrained estimates of the group means through the second term of the right-hand side.
The test statistic $W_{\textrm{TSK}}$ can easily be expressed in the second term of \eqref{eq:decomposition_KMeans1}.
To see this, first note that $W_{\textrm{TSK}} = \big\lVert \tilde{\Omega}_{R,D}^{-1/2} (R \tilde{\theta}_{D} - r) \big\rVert^2$.
Since for any vector $v$ with $\lVert v \rVert \neq 0$ we have $v = \lVert v \rVert \mathrm{dir}(v)$, applying this identity to $W_{\textrm{TSK}}$, we obtain
\begin{equation}\label{eq:perturbation_KMeans_ts}
    \hat{B} =
    W_{\textrm{TSK}}^{1/2} \mathbb{Q}_{\textrm{TSK}}  \tilde{J} +
    \tilde{U}
\end{equation}
where
$
\tilde{J} = \mathrm{dir} ( \tilde{\Omega}_{R,D}^{-1/2} ( R \tilde{\theta}_{D} - r ) )
$
and
$
\tilde{U} = \widetilde{\mathbb{H}} \tilde{\theta}_{R,D} + \mathbb{M}\hat{B}
$.
To the best of our knowledge, the current paper is the first in the literature that considers a decomposition at this level of generality. For instance, \cite{chen23} consider the equality of the centers of only two clusters, \cite{yun2024selective} focused on the equality of all cluster centers, and \cite{chen2024} developed a test for the equality of the mean of a single feature.
Our decomposition enables consideration of general linear constraints by selecting different matrices $R$.
Furthermore, all the decompositions proposed in the aforementioned literature are special cases of \eqref{eq:perturbation_KMeans_ts}.

\begin{remark}
A feature that distinguishes our construction from the post-clustering inference literature \citep[e.g.,][]{chen23,gao24,chen2024} is the metric used in the orthogonalization for the construction of the nuisance component $\tilde{U}$. In the existing literature, the observations are spherical Gaussian, so the residual can be the ordinary Euclidean projection of the data onto the tested contrast, whereas in our panel setting the unit-specific estimators satisfy $\hat{B} \sim N(B, \sigma^2 I_N \otimes \Sigma^{-1})$, which is generally anisotropic. The metric enters not through the residual $\mathbb{M}\hat{B}$, which is the ordinary projection residual, but through the constrained estimator $\tilde{\theta}_{R, D}$ and the whitening $\tilde{\Omega}_{R, D}^{-1/2}$, both taken in the variance metric $\tilde{\mathcal{N}}_{\Sigma}^{-1}$ of $\tilde{\theta}_D$. This delivers $R\tilde{\theta}_D \perp \tilde{U}$ established in Lemma~\ref{lemma-wind}, which fails if the contrast is removed in the Euclidean metric.
\end{remark}

Lastly, given the estimated group structure, we derive the conditional distribution of the quadratic form defined above.
Let
$\tilde{\Omega}_{R,d}$ be the realization of $\tilde{\Omega}_{R,D}$ associated with the realization $d$ of $D$,
$\tilde{\theta}_d$ that of $\tilde{\theta}_D$, 
and $\tilde{u}$ that of $\tilde{U}$.
Define the conditioning set:
\begin{equation*}
\mathcal{C}_{\textrm{TSK}} = \left\lbrace \bigcap_{m=0}^M \lbrace 
        \tilde{\gamma}^{(m)}_D = 
        \tilde{\gamma}^{(m)}_d 
    \rbrace, \tilde{J} = \tilde{j}, \tilde{U} = \tilde{u} \right\rbrace
\end{equation*}
where $\tilde{\gamma}^{(m)}_d = (\tilde{g}^{(m)}_{1,d},\dots,\tilde{g}^{(m)}_{N,d})'$ and $\tilde{\gamma}^{(m)}_D$ is its population counterpart.

Now we can state the following result, which shows that the exact conditional distribution of the test statistic $W_{\textrm{TSK}}$ is truncated $\chi^2_{q}$.
\begin{theorem}\normalfont
\label{thm:tsk}
Suppose that $\hat{B} \sim N(B, \sigma^2 I_N \otimes \Sigma^{-1})$, $\sigma^2 >0$, and $\Sigma$ is nonsingular.
Then, under $H_0$,
$W_{\textrm{TSK}} \; | \; \mathcal{C}_{\textrm{TSK}} \sim \chi^2_{q}|_{\mathcal{T}_{\textrm{TSK}}}$
with $\chi^2_{q}|_{\mathcal{T}_{\textrm{TSK}}}$ a $\chi^2_{q}$ random variable truncated to the set $\mathcal{T}_{\textrm{TSK}}$ which is given by
\begin{equation}\label{eq:truncationset_tsk}
\mathcal{T}_{\textrm{TSK}} = \left\lbrace \phi^2 \in \mathbbm{R}_{\geq 0} : \bigcap_{m=0}^M \lbrace \tilde{\gamma}^{(m)}_{d(\phi)} = \tilde{\gamma}^{(m)}_{d} \rbrace \right\rbrace
\end{equation}
where $d(\phi) = \{ \hat{b}_{i}(\phi); i=1, \dots, N \}$ with $\hat{b}_{i}(\phi) = [\hat{b}(\phi)]_i^K$ and
\begin{equation}\label{eq:perturbation_simple_KMeans}
    \hat{b}(\phi) =
    \phi \cdot \mathbb{Q}_{\textrm{TSK}} \tilde{j} +
    \tilde{u}
\end{equation}
\end{theorem}

\begin{remark}
\label{rem:wkmeans}
  The same results can be established for the weighted procedure discussed in Remark~\ref{rem:weighted}. Under weighting, the group estimator becomes $\tilde{\theta}_{g,D} = \bigl( \sum_{i \in g} \Phi_i \bigr)^{-1} \sum_{i \in g} \Phi_i \hat{B}_i$, which is linear in $\hat{B}$ and therefore remains Gaussian; the entire decomposition discussed above carries over with $\widetilde{\mathbb{H}}'\widetilde{\mathbb{H}}$ replaced by $\widetilde{\mathbb{H}}'\Phi \widetilde{\mathbb{H}}$ throughout, where $\Phi = \operatorname{bdiag}(\Phi_1, \dots, \Phi_N)$. This extension relaxes the homoskedasticity assumption of the exact theory. Suppose the unit-specific estimators are heteroskedastic with possibly unequal designs, $\hat{B}_i \sim N(B_i, V_i)$ with $V_i = \sigma_i^2 \Sigma_i^{-1}$ and $\Sigma_i = \sum_{t=1}^T X_{it} X_{it}'$ varying across $i$. Choosing the inverse-variance weights $\Phi_i = V_i^{-1}$ makes $\tilde{\theta}_{g,D}$ the GLS group average, the independence $R\tilde{\theta}_D \perp \tilde{U}$ continues to hold, as we establish in Lemma~\ref{lem:tsk-weighted-independence}, and the conditional distribution of the Wald statistic is exactly truncated $\chi^2_q$ as before, now without requiring a common error variance or a common design across units. Note that this exact finite-sample result requires that the clustering weight coincide with the inverse variance, $\Phi_i = V_i^{-1}$. For a generic positive-definite $\Phi_i$, the selection event is still characterized by quadratic inequalities in the data and the algorithm is well-defined, but the independence in Lemma~\ref{lem:tsk-weighted-independence} breaks down, so the exact truncation result no longer applies and the procedure is justified only asymptotically under consistency of $\hat{\Phi}_i$. In the main text, we adopt the Euclidean criterion for simplicity, as it isolates the role of group-structure uncertainty, our primary focus.
\end{remark}

\subsection{Asymptotic analysis}\label{sec:asymptotics}

We examine the asymptotic properties of our testing procedures. The previous discussions assume that the statistics $S$ or $\hat B$ follow a Gaussian distribution in finite samples. However, this assumption can be restrictive in real-world applications. Therefore, we consider scenarios in which these unit-specific statistics vectors asymptotically follow a Gaussian distribution.

For the PCR procedure, we assume that $S$ is asymptotically Gaussian. Namely, instead of assuming that $S\sim N(\mathbb{X}'\mathbb{X} B, \sigma^2 \mathbb{X}'\mathbb{X})$, we assume that as $T \to \infty$, $S \to_d S^* \sim N(\mathbb{X}'\mathbb{X} B, \sigma^2 \mathbb{X}'\mathbb{X})$, where $N$ is held fixed. 
For the TSK procedure, we assume that $\hat B$ is asymptotically Gaussian, namely $\hat B\to_d B^* \sim N(B, \sigma^2 I_N \otimes \Sigma^{-1})$.

We have two remarks regarding this assumption. First, this setup examines a situation in which the time series within each unit is not sufficiently informative. In this case, $S / T$ does not converge in probability but has an asymptotic distribution. If we consider a scenario in which $ S/T$ converges, this raises several theoretical challenges. When group separation is maintained, the group structure is consistent, yielding a degenerate test statistic and a conditioning probability that converges to 0 or 1.
Our setup effectively mimics situations in which there is statistical uncertainty in unit-specific coefficients and in the group structure. Theoretically, this assumption could be justified in certain contexts, such as:
\begin{itemize*}
\item[(i)] $\sum_{t=1}^T X_{it} X_{it}' \to \Sigma$ and $\mathrm{V}(\varepsilon_{it}) = \sigma^2$, or
\item[(ii)] $\frac{1}{T}\sum_{t=1}^T X_{it} X_{it}' \to \Sigma$ and $\mathrm{V}(\varepsilon_{it}) = T\sigma^2$.
\end{itemize*}

Second, we examine the scenario where $T$ approaches infinity while $N$ remains constant. Specifically, we focus on cases where the number of cross-sectional units is limited but the time series is sufficiently long to enable a Gaussian approximation (even though it may not provide sufficient information for consistent estimation).

The following theorem demonstrates that our PCR procedure asymptotically controls the size.
\begin{theorem}
\label{thm:pcr_asymptotic} 
    Suppose that $S \to_d N(\mathbb{X}'\mathbb{X}B, \sigma^2 \mathbb{X}'\mathbb{X})$ and $\sum_{t=1}^T X_{it} X_{it}' = \Sigma$ for any $i$, where $\Sigma$ is nonsingular. 
     Let $F_{\textrm{PCR}}$ be the cumulative distribution function of $\chi^2_{q}|_{\mathcal{T}_{\textrm{PCR}}}$.
Then, under $H_0$, as $T\to \infty$,
\begin{align*}
    \Pr\left(F_{\textrm{PCR}} (W_{\textrm{PCR}}) \geq 1- \alpha 
    \; \middle | \;
    \bigcap_{m=0}^M \lbrace \hat{\gamma}^{(m)}_D = \hat{\gamma}^{(m)}_d  \rbrace \right)
    \to \alpha.
\end{align*}
\end{theorem}

We reject $H_0$ if $F_{\textrm{PCR}}(W_{\textrm{PCR}}) \geq 1-\alpha$. Thus, the theorem above shows that, conditional on the group assignment algorithm’s realized sequence, the probability of rejecting the null hypothesis converges to the intended size $\alpha$. The proof is in the appendix. It uses the continuity of $F_{\textrm{PCR}}$, allowing the continuous mapping theorem, and that the probability $\Pr \left( \bigcap_{m=0}^{M} \{ \hat{\gamma}^{(m)}_D = \hat{\gamma}^{(m)}_d \} \right)$ is asymptotically non-zero in our setting.

A similar result holds for the TSK procedure. In this case, instead of assuming $\hat{B}\sim N(B, \sigma^2 I_N \otimes \Sigma^{-1})$, we assume that as $T \to \infty$, $\hat{B} \to_d N(B, \sigma^2 I_{N}\otimes \Sigma^{-1})$, where $N$ is held fixed.
\begin{theorem}
\label{thm:tsk_asymptotic} 
Suppose that $\hat{B}\to_d N( B, \sigma^2 I_N \otimes \Sigma^{-1})$, where $\sigma^2>0$ and $\Sigma$ is non-singular. Let $F_{\textrm{TSK}}$ be the cumulative distribution function of $\chi^2_{q}|_{\mathcal{T}_{\textrm{TSK}}}$.
Then, under $H_0$, as $T\to \infty$,
\begin{align*}
    \Pr\left(F_{\textrm{TSK}} (W_{\textrm{TSK}}) \geq 1- \alpha 
    \; \middle | \;
    \bigcap_{m=0}^M \lbrace \tilde{\gamma}^{(m)}_D = \tilde{\gamma}^{(m)}_d  \rbrace \right)
    \to \alpha.
\end{align*}
\end{theorem}

\begin{remark}
    The extension to cases with $N, T\to \infty$ raises technical challenges that the current literature has not been able to accommodate. When the dimension of $S$ or $\hat B$ grows, we need to employ a high-dimensional CLT. However, the current versions of the high-dimensional CLT require that the event whose probability we would like to evaluate be a convex set or exhibit properties similar to those of a convex set \citep[see, for instance][]{chang2024central}. This requirement causes a problem in our setting, because, for example, the event $F_{PCR} (W_{PCR} )\geq 1- \alpha$ is not a convex set in the space of $S$.
\end{remark}

\subsection{Variance estimation}\label{sec:variance}

We now describe the variance estimators used for the PCR and TSK procedures. The theoretical results above assume known variances, but in practice these are unknown and need to be estimated; in the simulations presented later, we examine the performance of our procedures with estimated variances. Although the two estimators aggregate the panel differently, both admit, conditional on the estimated group structure, an exact representation of their estimation error as the time average of a group-level score process. We exploit this common structure and estimate both variances within a single framework: we construct estimator-specific scores and apply to them a kernel-based long-run variance estimator in the spirit of \citet{driscoll98}. Because the score aggregates over the cross-section before outer products are taken, the resulting estimator is robust to arbitrary cross-sectional dependence (CSD) as well as to serial correlation of unknown form.

Two ingredients are common to both constructions. First, let
\[
\hat{Q}_i = \frac{1}{T}\sum_{t=1}^{T} X_{it}X_{it}',
\quad
\hat{\varepsilon}_{it} = y_{it} - X_{it}'\hat{B}_i,
\]
denote the unit-level second-moment matrix and the residuals from the unit-level time-series regressions; the scores below are evaluated at these unit-level residuals, a choice we return to at the end of the subsection. Second, define the Bartlett weights
\[
w_{ts}
  = k_T\!\left( \frac{|t-s|}{L_T + 1} \right),
\quad
k_T(x)
  = \begin{cases}
      1 - x, & 0 \le x \le 1,\\[4pt]
      0, & \text{otherwise},
    \end{cases}
\]
where $L_T$ is a bandwidth parameter with default value $L_T = \lfloor T^{1/3} \rfloor$ in our implementation.

Consider first the PCR estimator, which relies on a pooled regression within each selected group. Writing
\[
\hat{Q}_{g,D}
  = \frac{1}{\hat{n}_{g,D} T}
    \sum_{i=1}^{N} \sum_{t=1}^{T}
      X_{it} X_{it}'
      \mathbf{1}\!\left\{ \hat{g}_{i,D} = g \right\},
\quad
\hat{\zeta}_{g,t,D}^{\mathrm{PCR}}
  = \hat{Q}_{g,D}^{-1}
    \frac{1}{\hat{n}_{g,D}} \sum_{i=1}^{N}
      \mathbf{1}\!\left\{ \hat{g}_{i,D} = g \right\}
      X_{it} \hat{\varepsilon}_{it},
\]
the estimation error of $\hat{\theta}_{g,D}$ around its target conditional on the estimated group memberships admits the representation as a time average of $\hat{\zeta}_{g,t,D}^{\mathrm{PCR}}$. The variance estimator takes the form
\begin{equation}\label{eq:pcr-dk}
\widehat{\Omega}_{g,D}
  = \frac{1}{T^{2}} \sum_{t=1}^{T} \sum_{s=1}^{T}
      w_{ts}
      \big( \hat{\zeta}_{g,t,D}^{\mathrm{PCR}} - \bar{\zeta}_{g,D}^{\mathrm{PCR}} \big)
      \big( \hat{\zeta}_{g,s,D}^{\mathrm{PCR}} - \bar{\zeta}_{g,D}^{\mathrm{PCR}} \big)'.
\end{equation}
Expanding the products shows that, up to the centering of the scores, \eqref{eq:pcr-dk} coincides with the familiar Driscoll--Kraay sandwich form:
\[
\widehat{\Omega}_{g,D}
  = \hat{Q}_{g,D}^{-1}
    \left[
      \frac{1}{T^{2}}
      \sum_{i,j=1}^{N} \sum_{t,s=1}^{T}
        \frac{w_{ts}}{\hat{n}_{g,D}^{\,2}}
        X_{it} X_{js}'
        \hat{\varepsilon}_{it} \hat{\varepsilon}_{js}
        \mathbf{1}\!\left\{ \hat{g}_{i,D} = g \right\}
        \mathbf{1}\!\left\{ \hat{g}_{j,D} = g \right\}
    \right]
    \hat{Q}_{g,D}^{-1}.
\]

We next turn to TSK and build a robust estimator using the same score-based long-run variance idea. Since $\tilde{\theta}_{g, D}$ is the average of the unit-level slope estimates over the selected group, its estimation error satisfies the exact identity
\begin{equation}\label{eq:tsk-score-identity}
\tilde{\theta}_{g,D}
  - \frac{1}{\tilde{n}_{g,D}} \sum_{i=1}^{N} \mathbf{1}\!\left\{ \tilde{g}_{i,D} = g \right\} B_i
  = \frac{1}{T} \sum_{t=1}^{T} \tilde{\zeta}_{g,t,D}^{\mathrm{TSK}},
\quad
\tilde{\zeta}_{g,t,D}^{\mathrm{TSK}}
  = \frac{1}{\tilde{n}_{g,D}} \sum_{i=1}^{N}
      \mathbf{1}\!\left\{ \tilde{g}_{i,D} = g \right\}
      \hat{Q}_i^{-1} X_{it} \varepsilon_{it},
\end{equation}
so that the sampling variation of $\tilde{\theta}_{g,D}$ around its target conditional on the estimated group memberships is governed by the long-run variance of the cross-sectional average of the unit-level influence functions. Let $\hat{\zeta}$ denote the feasible score obtained by replacing $\varepsilon_{it}$ with $\hat{\varepsilon}_{it}$ in \eqref{eq:tsk-score-identity}, and let $\bar{\zeta}_{g,D}^{\mathrm{TSK}} = T^{-1}\sum_{t=1}^{T} \hat{\zeta}_{g,t,D}^{\mathrm{TSK}}$ be its time average. The variance estimator is
\begin{equation}\label{eq:tsk-dk}
\tilde{\Omega}_{g,D}
  = \frac{1}{T^{2}} \sum_{t=1}^{T} \sum_{s=1}^{T}
      w_{ts}
      \big( \hat{\zeta}_{g,t,D}^{\mathrm{TSK}} - \bar{\zeta}_{g,D}^{\mathrm{TSK}} \big)
      \big( \hat{\zeta}_{g,s,D}^{\mathrm{TSK}} - \bar{\zeta}_{g,D}^{\mathrm{TSK}} \big)'.
\end{equation}
A natural alternative would be the non-parametric estimator of \citet{pesaran2006estimation}, which scales the cross-sectional dispersion of the unit-level slope estimates within the group. That estimator is valid when the $\hat{B}_i$ are approximately uncorrelated across units, but under CSD in $\varepsilon_{it}$ the dispersion of the $\hat{B}_i$ no longer identifies the variance of their average. The resulting tests can be severely undersized or oversized. The score-based construction in \eqref{eq:tsk-dk} sums over units within the group before forming products, so contemporaneous cross-unit covariances enter the estimator automatically, exactly as in the pooled Driscoll--Kraay case. The two procedures therefore differ only in how the cross-section is weighted inside the score: TSK applies the unit-specific whitening $\hat{Q}_i^{-1}$ before averaging, whereas PCR averages first and applies the pooled $\hat{Q}_{g,D}^{-1}$.

A word on the choice of residuals is in order. Both estimators evaluate the score at the unit-level residuals $\hat{\varepsilon}_{it}$ rather than at the residuals of the fitted group model, $\hat{\varepsilon} = Y - X_{\hat{\gamma}}\hat{\theta}_D$. Under possible non-separation, the group-level residuals contain the within-group heterogeneity term $X_{it}'(B_i - \theta_{g})$, which would contaminate the long-run variance of the score and inflate the estimator. In contrast, the unit-level residuals remain valid estimates of $\varepsilon_{it}$ regardless of the group structure, which is precisely the regime our procedures are designed for. This choice also makes explicit that $\tilde{\Omega}_{g, D}$ and $\widehat{\Omega}_{g, D}$ estimate the variance of the group estimators around their targets conditional on the estimated group memberships, in line with the conditional inference framework of this paper.

Finally, since the estimators in \eqref{eq:pcr-dk} and \eqref{eq:tsk-dk} treat the selected groups as mutually independent clusters, the overall covariance matrix is obtained by block-diagonalizing the group-wise estimators,
\[
\tilde{\Omega}_{D}
  = \mathrm{bdiag} \left[
      \tilde{\Omega}_{1,D},
      \dots,
      \tilde{\Omega}_{G,D} \right],
\quad
\widehat{\Omega}_{D}
  = \mathrm{bdiag} \left[
      \widehat{\Omega}_{1,D},
      \dots,
      \widehat{\Omega}_{G,D} \right].
\]
Our implementation also provides two variants: a restricted version that imposes cross-sectional independence across units and no serial correlation, obtained by setting $L_T = 0$ and summing unit-wise products; and an unrestricted version that applies the kernel estimator to the stacked $GK$-dimensional score, thereby allowing dependence across groups, in which case $\tilde{\Omega}_{D}$ and $\widehat{\Omega}_{D}$ are no longer block-diagonal.

\begin{remark}
The exact finite-sample results of Theorems~\ref{thm:pcr} and~\ref{thm:tsk} are derived under homoskedasticity, $V(S)=\sigma^{2}\mathbb{X}'\mathbb{X}$ and $V(\hat{B})=\sigma^{2}(I_{N}\otimes\Sigma^{-1})$, which is what makes the nuisance $\hat{U}$ ($\tilde{U}$) independent of $R\hat{\theta}_{D}$ ($R\tilde{\theta}_{D}$) in Lemmas~\ref{lemma-wind-pcr} and \ref{lemma-wind}. Under serial and CSD, this independence requires that the orthogonalization be taken in the relevant variance metric rather than the Euclidean one, exactly as in the inverse-variance construction of Remarks~\ref{rem:weighted} and~\ref{rem:wkmeans}. Plugging the robust estimator into the whitening $\hat{\Omega}_{R, D}^{-1/2}$ and the perturbation direction implements this metric, so that the conditional distribution remains truncated $\chi^{2}_{q}$ asymptotically, under consistency of the long-run variance estimator. The exact truncation result is thus a finite-sample statement for spherical errors; under dependence, the procedure is justified asymptotically, in the same sense as the generic-weight case of Remark~\ref{rem:wkmeans}.
\end{remark}

\section{Selective Confidence Sets}\label{sec:ci}

\subsection{Definition}

We now construct confidence sets for scalar linear contrasts of the group-specific parameters, valid conditional on the clustering outcome, following the inversion principle applied to the selective tests developed above. We consider the case $q=1$, that is, $R$ is a $1\times GK$ row vector and $r$ is a scalar. Let $\check{\theta}_D$ denote the post-clustering estimator, either $\hat{\theta}_D$ for PCR or $\tilde{\theta}_D$ for TSK, and let $\check{\Omega}_{R,D}$ be the corresponding estimator of the conditional variance of $R\check{\theta}_D$, constructed as in Section~\ref{sec:variance}. The scalar parameter of interest is $\kappa = R\theta$, and for a candidate value $\kappa_0$, the Wald statistic associated with the null hypothesis $H_0:\kappa=\kappa_0$ is
\[
W_v(\kappa_0)
=
\frac{(R\check{\theta}_D-\kappa_0)^2}{\check{\Omega}_{R,D}},
\]
where $v\in\{\mathrm{PCR},\mathrm{TSK}\}$ denotes the estimator used.

For each candidate value $\kappa_0$, the selective decomposition is formed under $H_0:\kappa=\kappa_0$, so the one-dimensional representation of the selection event may depend on $\kappa_0$. We therefore write the corresponding truncation set in the scale of the Wald statistic as $\mathcal T_{v, D}(\kappa_0)\subseteq\mathbb{R}_{\geq 0}$. Conditional on the clustering event and on the nuisance statistics used in the decomposition, $W_v(\kappa_0)$ has a truncated $\chi^2_1$ distribution under $H_0:\kappa=\kappa_0$, with truncation set $\mathcal T_{v, D}(\kappa_0)$. Note that the observed statistic always satisfies $W_v(\kappa_0)\in\mathcal T_{v, D}(\kappa_0)$, since the observed data trivially satisfy the observed selection event. Since the clustering inequalities are quadratic in the one-dimensional perturbation parameter, $\mathcal T_{v, D}(\kappa_0)$ need not be connected and is in general a finite union of intervals, which explains why we construct ``selective confidence sets'' for scalar contrasts rather than ``selective confidence intervals.''

For an observed value $w_v(\kappa_0)$, the selective $p$-value for testing $H_0:\kappa=\kappa_0$ is
\[
p_v(\kappa_0)
=
\Pr\left(
\chi^2_1 \geq w_v(\kappa_0)
\mid
\chi^2_1\in\mathcal T_{v,D}(\kappa_0)
\right).
\]
The exact selective confidence set with nominal level $1-\alpha$ is obtained by inverting these selective tests:
\[
\mathcal C_{v,1-\alpha}
=
\left\{
\kappa_0\in\mathbb R:
p_v(\kappa_0)\geq\alpha
\right\}.
\]
This definition is the direct confidence-set analog of the selective tests, with the truncation set recomputed for each candidate value $\kappa_0$, because the null value enters the decomposition used to express the clustering event along the scalar selective direction.

Equivalently, for each $\kappa_0$, let $c_{v,\alpha,D}(\kappa_0)$ be the critical value satisfying
\[
\Pr\left(
\chi^2_1 \geq c_{v,\alpha,D}(\kappa_0)
\mid
\chi^2_1\in\mathcal T_{v,D}(\kappa_0)
\right)
=
\alpha,
\]
taking the largest such value if it is not unique, which occurs only when $\alpha$ coincides with the conditional probability of a gap in $\mathcal T_{v, D}(\kappa_0)$. The acceptance region in the scale of the Wald statistic is
$
\mathcal A_{v,D}(\kappa_0)
=
\mathcal T_{v,D}(\kappa_0)\cap[0,c_{v,\alpha,D}(\kappa_0)]
$,
so the exact selective confidence set can also be written as
\[
\mathcal C_{v,1-\alpha}
=
\left\{
\kappa_0\in\mathbb R:
\frac{(R\check{\theta}_D-\kappa_0)^2}{\check{\Omega}_{R,D}}
\in
\mathcal A_{v,D}(\kappa_0)
\right\}.
\]
The confidence set is thus the inverse image of the selective acceptance region. In contrast with classical Wald intervals, this inverse image is not necessarily connected, for two reasons: the truncation set $\mathcal T_{v, D}(\kappa_0)$ can be a union of intervals, and it may itself vary with the candidate null value $\kappa_0$.

The construction has the same theoretical justification as the selective tests. Under $H_0:\kappa=\kappa_0$, the conditional distributional result established above implies that $p_v(\kappa_0)$ is uniform on $[0,1]$, conditional on the selected clustering outcome and on the nuisance statistics: exactly so with known variance, and asymptotically once $\check{\Omega}_{R,D}$ is estimated, as discussed in Section~\ref{sec:variance}. Therefore, inversion of the selective tests yields
\[
\Pr\left(
\kappa\in\mathcal C_{v,1-\alpha}
\mid
\text{selection event}
\right)
=
1-\alpha+o(1),
\]
and the same coverage statement holds unconditionally after averaging over the selection event. This argument relies on the full selective inversion: if only one connected component is retained, the resulting set is a subset of $\mathcal C_{v,1-\alpha}$ and may undercover unless the omitted components are asymptotically irrelevant.

\subsection{Implementation and a fixed-truncation approximation}

In practice, exact inversion can be computationally demanding because it requires recomputing the selective decomposition and the truncation set $\mathcal T_{v, D}(\kappa_0)$ for many candidate values of $\kappa_0$. A useful approximation fixes the truncation set at a reference value, typically the observed contrast $\hat{\kappa}_D=R\check{\theta}_D$, and then inverts the resulting fixed truncated distribution. Define
$
\mathcal T^{\mathrm{fix}}_{v,D}
=
\mathcal T_{v,D}(\hat{\kappa}_D)
$.
The fixed-truncation selective $p$-value is
$
p^{\mathrm{fix}}_v(\kappa_0)
=
\Pr\left(
\chi^2_1 \geq w_v(\kappa_0)
\mid
\chi^2_1\in\mathcal T^{\mathrm{fix}}_{v,D}
\right)
$,
and the corresponding confidence set is
\[
\mathcal C^{\mathrm{fix}}_{v,1-\alpha}
=
\left\{
\kappa_0\in\mathbb R:
p^{\mathrm{fix}}_v(\kappa_0)\geq\alpha
\right\}.
\]
Let $c^{\mathrm{fix}}_{v,\alpha,D}$ satisfy
$
\Pr\left(
\chi^2_1 \geq c^{\mathrm{fix}}_{v,\alpha,D}
\mid
\chi^2_1\in\mathcal T^{\mathrm{fix}}_{v,D}
\right)
=
\alpha
$,
again taking the largest such value if it is not unique. Because the fixed-truncation tail probability is monotone in the observed value of the Wald statistic, the condition
$
p^{\mathrm{fix}}_v(\kappa_0)\geq\alpha
$
is equivalent to
\[
\frac{(R\check{\theta}_D-\kappa_0)^2}{\check{\Omega}_{R,D}}
\leq
c^{\mathrm{fix}}_{v,\alpha,D},
\]
so that the fixed-truncation confidence set has the closed-form representation
\[
\mathcal C^{\mathrm{fix}}_{v,1-\alpha}
=
\left[
R\check{\theta}_D
-
\sqrt{c^{\mathrm{fix}}_{v,\alpha,D}\check{\Omega}_{R,D}},
R\check{\theta}_D
+
\sqrt{c^{\mathrm{fix}}_{v,\alpha,D}\check{\Omega}_{R,D}}
\right].
\]
This fixed-truncation set should be interpreted as a computational approximation to the exact selective confidence set: it preserves the selection adjustment through the critical value of the truncated $\chi^2_1$ distribution, but does not reproduce all possible disconnected components that may arise from exact candidate-specific inversion.

The fixed-truncation approximation is justified when the truncation set is locally stable over the range of null values that can enter the confidence set. Intuitively, only candidate values within a few standard errors of $\hat{\kappa}_D$ can survive the inversion, so the approximation is accurate whenever the truncation set changes little as $\kappa_0$ ranges over this neighborhood. A sufficient (though high-level) condition is that, for every $C>0$,
\[
\sup_{|\kappa_0-\hat{\kappa}_D|\leq C\sqrt{\check{\Omega}_{R,D}}}
\;
\sup_x
\left|
\Pr\left(
\chi^2_1 \geq x
\mid
\chi^2_1\in\mathcal T_{v,D}(\kappa_0)
\right)
-
\Pr\left(
\chi^2_1 \geq x
\mid
\chi^2_1\in\mathcal T^{\mathrm{fix}}_{v,D}
\right)
\right|
=
o_p(1).
\]
This condition holds when the roots defining the truncation intervals are locally stable functions of $\kappa_0$, the conditioning probability is bounded away from zero, and the data are asymptotically separated from degeneracies such as clustering ties, repeated roots, or changes in the topology of the truncation set. The second requirement also guarantees that the true $\kappa$ falls in the relevant neighborhood of $\hat{\kappa}_D$ with high probability, so the same condition delivers both the accuracy of the approximation and its coverage: under local stability, the fixed-truncation confidence set has the same asymptotic coverage as the exact inverted set. In the Monte Carlo simulations, we use this fixed-truncation construction for computational feasibility. In empirical applications, where only a small number of contrasts are reported, exact numerical inversion can be implemented by recomputing $\mathcal T_{v, D}(\kappa_0)$ over a grid of candidate values. For ease of presentation, we may additionally report the connected component containing $R\check{\theta}_D$ as a compact summary of the local uncertainty around the estimate.

\section{Extensions}\label{sec:extensions}

This section discusses how the proposed methods can be extended to more general models. We consider models with unit-specific intercepts and those with GFE.

\subsection{Unit-specific heterogeneity}\label{sec:fe}

We consider a model with additive unit-specific unobserved heterogeneity and group-specific slope parameters. Unit-specific effects, referred to as one-way fixed effects (FE), control for time-invariant unobserved heterogeneity that may be correlated with the regressors. Incorporating them is standard practice in panel data analysis, as it mitigates omitted variable bias arising from unobserved heterogeneity that does not vary over time. When the regressors are strictly exogenous, accounting for one-way FE is straightforward: we apply the unit within transformation and then apply the proposed methods to the transformed data.

The model with one-way FE is
\begin{equation}\label{eq:femodel}
    Y_{it} = X_{it}'\theta_{g_i} + \mu_i + \varepsilon_{it},
\end{equation}
where $\mu_i$ denotes the unobserved time-invariant unit-specific effect. The regressor vector $X_{it}$ is assumed to be strictly exogenous, that is, $\mathrm{E} (\varepsilon_{it} \mid X_{i1}, \dots, X_{iT}) = 0$ for all $i,t$, and must exhibit sufficient time variation to ensure identification after the within transformation.

To eliminate the fixed effects, we apply the one-way within transformation, which removes the unit means. For any variable $a_{it}$, let
$
\dot{a}_{it} = a_{it} - \bar{a}_{i\cdot}
$
and
$
\bar{a}_{i\cdot} = T^{-1}\sum_{s=1}^T a_{is}
$.
Applying this transformation to equation~\eqref{eq:femodel} yields
\begin{align}
    \dot{Y}_{it} = \dot{X}_{it}'\theta_{g_i} 
    + \dot{\varepsilon}_{it}.
    \label{eq:owfetransformedmodel}
\end{align}

We then apply the proposed methods to equation~\eqref{eq:owfetransformedmodel}. The inference procedures remain valid provided that the transformed variables satisfy the assumptions stated earlier, in particular strict exogeneity and sufficient time variation of $X_{it}$. Under strict exogeneity, the OLS estimator, applied to each unit for the TSK estimator or to each group for the PCR estimator, retains its desirable properties after the unit-within transformation. If, however, $X_{it}$ is only predetermined (i.e.\ $\mathrm{E}(\varepsilon_{it} \mid X_{i1},\dots,X_{it}) = 0$ but $\mathrm{E}(\varepsilon_{it} \mid X_{i1},\dots,X_{iT}) \neq 0$), as in the examples discussed in Sections \ref{sec:settings} and \ref{sec:applications}, the within transformation induces a bias of order $O(1/T)$ that vanishes in our $T\to \infty$ asymptotics, so Theorems \ref{thm:pcr_asymptotic} and \ref{thm:tsk_asymptotic} continue to apply. Bias correction may nonetheless be desirable when $T$ is small.\footnote{The bias is of the \citet{Nickell1981} type. For the TSK procedure, the unit-specific estimators $\hat B_i$ enter only as asymptotically Gaussian inputs and may be replaced by bias-corrected or IV counterparts; the PCR procedure does not admit such a substitution as directly.} Finally, the requirement that $X_{it}$ be time-varying ensures that the within-transformed regressors $\dot{X}_{it}$ are not constant, preventing collinearity after the one-way transformation.

The same discussion applies to specifications that include an additive time effect $\lambda_t$ common across all groups, as in the empirical applications below. Since $\lambda_t$ does not vary across units, it is removed by the two-way within transformation,
$
\ddot{a}_{it}
=
a_{it}
-
\bar{a}_{i\cdot}
-
\bar{a}_{\cdot t}
+
\bar{a}_{\cdot\cdot}
$.
Starting from the original latent group model, however, this transformation does not generally preserve the term $X_{it}'\theta_{g_i}$, because time demeaning mixes observations from units belonging to different groups and therefore introduces slope coefficients from the other groups. Rather than deriving the transformed model mechanically from the original specification, we may instead assume directly that the two-way transformed data admit the latent group representation:
$
\ddot{Y}_{it}
=
\ddot{X}_{it}'\theta_{g_i}
+
\ddot{\varepsilon}_{it}
$,
and apply the proposed methods to this transformed model, as we do in the applications of Section \ref{sec:applications}. Even if the original errors $\varepsilon_{it}$ are cross-sectionally independent, the transformed errors $\ddot{\varepsilon}_{it}$ are cross-sectionally dependent, of the type accommodated by the Driscoll--Kraay estimator introduced in Section~\ref{sec:variance}. Group-specific time-varying effects are instead handled by the GFE extension discussed in the following subsection.

\subsection{Grouped fixed effects}\label{sec:gfe}

We now extend the analysis to models with GFE and group-specific slope parameters.  
Whereas the two-way fixed effects model discussed in Section~\ref{sec:fe} controls for additive unit and time effects that are common across groups, it cannot capture unobserved shocks that vary over time in a group-specific way.  
The GFE specification relaxes this restriction by allowing the intercept to vary over time within each group while preserving the pattern of heterogeneity across groups.

The model with GFE is given by
\begin{equation}\label{eq:gfemodel}
    Y_{it} = X_{it}'\theta_{g_i} + \eta_{g_i t} + \varepsilon_{it},
\end{equation}
where $\eta_{g_i t}$ denotes the fixed effect specific to group $g_i$ at time $t$.  
Compared with the model in Section~\ref{sec:fe}, this specification accommodates time-varying unobserved confounders that may differ systematically across groups.  
The grouped structure ensures parsimony by restricting the heterogeneity to evolve along $G$ latent trajectories rather than across $N$ individual units.  
A similar model, though with homogeneous slopes, appears as Extension~2 in the supplementary material of \citet{bonhomme_grouped_2015}.

Because $\eta_{g_i t}$ varies over time, the TSK estimator is no longer applicable.  
We therefore extend the PCR estimator to estimate group-specific slopes and group-time fixed effects jointly.
This is done by applying Algorithm \ref{algo:panel_KMeans} by extending the explanatory variables vector $X_{it}$ with time dummies.
The rest of the procedure for GFE remains identical to PCR.

The variance estimator for the GFE procedure follows the same score-based construction, after eliminating the group-time effects by the Frisch--Waugh--Lovell transformation. Let $\hat{e}^{\mathrm{GFE}}_{it} = y_{it} - X_{it}'\hat{\theta}_{g,D} - \hat{\eta}_{gt}$ denote the GFE residuals and define the within-group cross-sectionally demeaned regressors:
\[
\breve{X}_{it}
  = X_{it}
    - \frac{1}{\hat{n}_{g,D}}
      \sum_{j=1}^{N}
        \mathbf{1}\!\left\{ \hat{g}_{j,D} = g \right\} X_{jt},
\quad
i \in \left\{ i : \hat{g}_{i,D} = g \right\},
\]
which purge the group-time effects period by period. Since $\sum_{i} \mathbf{1}\{\hat{g}_{i,D}=g\}\,\breve{X}_{it} = 0$ for every $t$, the group-time effects drop out of the score, and the estimation error of the GFE slope estimator around its target conditional on the estimated group memberships is the time average of
\[
\hat{\zeta}^{\mathrm{GFE}}_{g,t,D}
  = \breve{Q}_{g,D}^{-1}
    \frac{1}{\hat{n}_{g,D}} \sum_{i=1}^{N}
      \mathbf{1}\!\left\{ \hat{g}_{i,D} = g \right\}
      \breve{X}_{it} \hat{e}^{\mathrm{GFE}}_{it},
\quad
\breve{Q}_{g,D}
  = \frac{1}{\hat{n}_{g,D} T}
    \sum_{i=1}^{N} \sum_{t=1}^{T}
      \breve{X}_{it} \breve{X}_{it}'
      \mathbf{1}\!\left\{ \hat{g}_{i,D} = g \right\}.
\]
The group-wise covariance estimator is then, with the same Bartlett weights and centering as in Section~\ref{sec:variance},
\[
\widehat{\Omega}^{\mathrm{GFE}}_{g,D}
  = \frac{1}{T^{2}} \sum_{t=1}^{T} \sum_{s=1}^{T}
      w_{ts}
      \big( \hat{\zeta}^{\mathrm{GFE}}_{g,t,D} - \bar{\zeta}^{\mathrm{GFE}}_{g,D} \big)
      \big( \hat{\zeta}^{\mathrm{GFE}}_{g,s,D} - \bar{\zeta}^{\mathrm{GFE}}_{g,D} \big)',
\]
and the overall matrix is obtained by block-diagonalization across groups. Unlike the PCR and TSK cases, the score is evaluated at the GFE residuals: unit-level time-series regressions with unit-specific time effects are infeasible, so no unit-level residual variant exists here. The demeaning nonetheless provides the analogous protection, as the group-time effects are removed exactly rather than estimated within the score.

\section{Monte Carlo}\label{sec:motivating_simulations}

In this section, we examine the finite-sample performance of the proposed selective inference procedures and illustrate the impact of ignoring the estimation uncertainty of the group structure.  
As emphasized in \citet{chen23} and \citet{gao24}, post-clustering inference based on estimated groups may lead to substantial size distortions, particularly when group separation is weak. Another purpose of these simulation exercises is to examine the robustness of our procedures to potential violations of the assumptions underlying the theoretical justification. To investigate this potential issue, we conduct Monte Carlo simulations.

\subsection{Monte Carlo design}\label{sec:mc_design}

Our Monte Carlo designs are linear panel models with latent group structure. All experiments are based on \(N=120\) units and \(T\in\{20,50\}\) time periods, with 1000 Monte Carlo replications. The data are generated according to
\begin{equation}\label{eq:monte_model}
Y_{it}=X_{it}'\theta_{g_i}+\varepsilon_{it},
\quad i=1,\dots,N,\quad t=1,\dots,T,
\end{equation}
where the true group membership is \(g_i=1\) for \(i\leq 40\) and \(g_i=2\) for \(i>40\). Thus, the first group contains 40 units and the second group contains 80 units. The regressor vector is \(X_{it}=(X_{1,it},X_{2,it})'\) and is independent of the error process.

The main text focuses on the baseline specification with no unit or time fixed effects. This case isolates the effect of estimating the latent slope groups and directly targets the setting emphasized in the theory. Additional Monte Carlo results reported in Appendix~\ref{sec:additional_sim} consider specifications with unit and grouped fixed effects. The appendix also reports the corresponding results for the GFE estimator.

We consider two experiments which differ in the properties of the regressors and the errors.

\paragraph*{Experiment 1.}
In the first experiment, observations are independent across units and over time. The errors satisfy
$
\varepsilon_{it}\overset{\mathrm{iid}}{\sim}N(0,1)
$,
and the regressors are iid across \(i\) and \(t\), with
\[
X_{it}
\overset{\mathrm{iid}}{\sim}
N\left(
\begin{pmatrix}
0\\
0
\end{pmatrix},
\begin{pmatrix}
1 & 0.4\\
0.4 & 1
\end{pmatrix}
\right).
\]
Thus, the two regressors are contemporaneously correlated, but there is no serial or CSD. This design provides a benchmark corresponding to the homoskedastic Gaussian setting.

\paragraph*{Experiment 2.}
The second experiment allows for serial dependence, within-group CSD, and non-Gaussian errors. The errors follow stationary AR(1) processes,
\[
\varepsilon_{it}
=
\rho_{\varepsilon}\varepsilon_{i,t-1}
+
\sqrt{1-\rho_{\varepsilon}^{2}}\,\varsigma_{it},
\quad
\rho_{\varepsilon}=0.5.
\]
The innovation vector \(\varsigma_t=(\varsigma_{1t},\dots,\varsigma_{Nt})'\) is spatially correlated within each true group. Its covariance matrix is
\[
\Sigma_S
=
\begin{pmatrix}
\rho_S \exp(-\Xi_1/\tau)+(1-\rho_S)I_{40} & 0\\
0 & \rho_S \exp(-\Xi_2/\tau)+(1-\rho_S)I_{80}
\end{pmatrix},
\quad
\rho_S=0.2,
\quad
\tau=0.3,
\]
where \(\exp(\cdot)\) is applied entrywise and \(\Xi_g\) is the \(n_g\times n_g\) matrix of pairwise distances \((\Xi_g)_{jk}=|\ell_{g,j}-\ell_{g,k}|\) between the equally spaced locations \(\ell_{g,j}=(j-1)/(n_g-1)\), \(j=1,\dots,n_g\), on the unit interval, with \(n_1=40\) and \(n_2=80\). Hence, dependence is stronger for nearby units and weaker for distant units, while the block-diagonal structure rules out cross-group dependence.

The regressors also follow stationary AR(1) processes,
\[
X_{i,t,k}
=
\rho_X X_{i,t-1,k}
+
\sqrt{1-\rho_X^2}\,\nu_{i,t,k},
\quad
\rho_X=0.5,
\quad
k=1,2.
\]
The innovations are jointly Gaussian, satisfy \(\operatorname{Corr}(\nu_{i,t,1},\nu_{i,t,2})=0.4\), and have the same within-group spatial covariance structure \(\Sigma_S\). To introduce a controlled departure from Gaussianity, the error innovations are Gaussian in the first half of the sample and standardized Student-\(t\) with six degrees of freedom in the second half:
\[
\varsigma_t
\sim
\begin{cases}
N(0,\Sigma_S), & t\leq T/2,\\[4pt]
t_6(0,\Sigma_S), & t>T/2.
\end{cases}
\]
The scaling of the Student-\(t\) innovations ensures that the covariance matrix remains \(\Sigma_S\).

\paragraph*{Slope parameters.}
We consider three data-generating processes for the slope parameters:
\[
\begin{array}{ll}
\text{DGP1:} & \theta_1=\theta_2=(2,1)',\\
\text{DGP2:} & \theta_1=(2,1)',\quad \theta_2=(4,1)',\\
\text{DGP3:} & \theta_1=(2,1)',\quad \theta_2=(4,2)'.
\end{array}
\]
DGP1 is homogeneous and violates group separation. DGP2 has one heterogeneous coefficient and therefore exhibits partial separation. DGP3 has two heterogeneous coefficients and exhibits full separation.

\paragraph*{Null hypotheses.}
For each design, we test
\[
H_{0,1}:\theta_1-\theta_2=0,
\quad
H_{0,2}:\theta_{2,1}-\theta_{2,2}=0,
\quad
H_{0,3}:\theta_{1,1}=\theta_{1,2}=0.
\]
The first hypothesis tests equality of the full slope vector across groups. The second hypothesis tests equality of the second slope across groups. The third hypothesis is false under all three DGPs and is used to evaluate power. We compare tests based on the true group assignment, conventional post-clustering Wald tests that ignore group-estimation uncertainty, and the proposed conditional tests. The post-clustering procedures are implemented using PCR, TSK, and GFE with \(G=2\). Standard errors and Wald statistics are computed using a covariance estimator matched to the design: an independent-score covariance estimator in Experiment 1 and a Driscoll--Kraay-type covariance estimator in Experiment 2. The reported rejection frequencies are computed at the 5 percent level.

\begin{table}
\centering
\caption{Truth of Null Hypotheses under Different DGPs}
\label{tab:nulltruth}
\begin{tabular}{clccc}
\toprule
& & DGP1 & DGP2 & DGP3\\
Null & Constraint & No group separation & Partial group separation & Full group separation\\
\midrule
\(H_{0,1}\) & \(\theta_1=\theta_2\) & \(\checkmark\) & \(\times\) & \(\times\)\\
\(H_{0,2}\) & \(\theta_{2,1}=\theta_{2,2}\) & \(\checkmark\) & \(\checkmark\) & \(\times\)\\
\(H_{0,3}\) & \(\theta_{1,1}=\theta_{1,2}=0\) & \(\times\) & \(\times\) & \(\times\)\\
\bottomrule
\end{tabular}
\end{table}

\subsection{Results}\label{sec:sim_results}

We first report the baseline Monte Carlo results without additional fixed effects. The results are organized around the two main designs described above. Experiment 1 is the iid Gaussian benchmark, while Experiment 2 allows for serial dependence, within-cluster CSD, and non-Gaussian errors. For each experiment, we report rejection frequencies for the predetermined-group (i.e., true group assignment) test, the naive post-clustering tests, and the proposed conditional tests, together with coverage and average length of confidence intervals for the scalar contrast in \(H_{0,2}\). Additional Monte Carlo results with unit fixed effects and grouped fixed effects are reported in Appendix~\ref{sec:additional_sim}.

Table~\ref{tab:exp1_case1_rejection} reports rejection frequencies for Experiment 1. The results demonstrate the key problem in post-clustering inferences. When there is no group separation, as in DGP1, the naive post-clustering tests severely over-reject. For \(H_{0,1}\), the naive TSK and PCR tests reject with frequency one at both values of \(T\), and the naive GFE test over-rejects less dramatically, at 0.63 and 0.49. Similar distortions appear for \(H_{0,2}\), where naive TSK rejects with frequency one and naive PCR at 0.86.

The conditional tests correct these distortions. Under DGP1, the conditional rejection frequencies are close to the nominal 5 percent level for all three clustering procedures, ranging between 0.06 and 0.07 for \(H_{0,1}\) and between 0.04 and 0.06 for \(H_{0,2}\). Hence, even in the most difficult case where the slope parameters are homogeneous, and the estimated groups are entirely driven by noise, the selective correction restores size control.

Under DGP2, where one slope differs across groups but \(H_{0,2}\) remains true, the naive tests are much less distorted than under DGP1, but still mildly oversized when \(T=20\). The conditional tests remain close to the nominal level. In particular, for \(T=50\), all procedures have rejection frequencies between 0.05 and 0.06. This confirms that the proposed correction does not rely on failure of separation; it also behaves well when partial separation is present.

Panel (b) of Table~\ref{tab:exp1_case1_rejection} reports power. The conditional tests retain high power across all alternatives. For \(H_{0,3}\) under DGP1, rejection frequencies are between 0.95 and 0.98 when \(T=20\), and between 0.98 and 0.99 when \(T=50\). Under DGP2 and DGP3, power is essentially one in most designs. Thus, the selective correction substantially improves size while preserving the ability to detect false restrictions.

\begin{table}
\centering
\scriptsize
\begin{threeparttable}
\caption{Rejection rates of naive tests and proposed tests under different null hypotheses \\
Experiment 1: iid Gaussian design
}
\label{tab:exp1_case1_rejection}
\begin{tabular*}{\textwidth}{@{\extracolsep{\fill}} ccc ccccccc}
\toprule
$T$ & Test & DGP & \shortstack{Predetermined} & \shortstack{Naive\\TSK} & \shortstack{Naive\\PCR} & \shortstack{Naive\\GFE} & \shortstack{Conditional\\TSK} & \shortstack{Conditional\\PCR} & \shortstack{Conditional\\GFE} \\
\midrule
\multicolumn{10}{l}{\textbf{Panel (a): Size}} \\
\midrule
\multicolumn{10}{l}{\emph{DGP1: No group separation}} \\
20 & $H_{0,1}$ & DGP1 & 0.05 & 1.00 & 1.00 & 0.63 & 0.06 & 0.07 & 0.07 \\
20 & $H_{0,2}$ & DGP1 & 0.06 & 1.00 & 0.86 & 0.42 & 0.06 & 0.06 & 0.06 \\
50 & $H_{0,1}$ & DGP1 & 0.05 & 1.00 & 1.00 & 0.49 & 0.06 & 0.07 & 0.07 \\
50 & $H_{0,2}$ & DGP1 & 0.05 & 1.00 & 0.86 & 0.36 & 0.06 & 0.06 & 0.04 \\
\midrule
\multicolumn{10}{l}{\emph{DGP2: Partial group separation}} \\
20 & $H_{0,2}$ & DGP2 & 0.07 & 0.10 & 0.09 & 0.08 & 0.09 & 0.08 & 0.07 \\
50 & $H_{0,2}$ & DGP2 & 0.04 & 0.06 & 0.05 & 0.05 & 0.06 & 0.05 & 0.05 \\
\midrule
\multicolumn{10}{l}{\textbf{Panel (b): Power}} \\
\midrule
\multicolumn{10}{l}{\emph{DGP1: No group separation}} \\
20 & $H_{0,3}$ & DGP1 & 1.00 & 1.00 & 1.00 & 1.00 & 0.95 & 0.97 & 0.98 \\
50 & $H_{0,3}$ & DGP1 & 1.00 & 1.00 & 1.00 & 1.00 & 0.98 & 0.98 & 0.99 \\
\midrule
\multicolumn{10}{l}{\emph{DGP2: Partial group separation}} \\
20 & $H_{0,1}$ & DGP2 & 1.00 & 1.00 & 1.00 & 1.00 & 0.98 & 0.99 & 0.96 \\
20 & $H_{0,3}$ & DGP2 & 1.00 & 1.00 & 1.00 & 1.00 & 1.00 & 1.00 & 0.99 \\
50 & $H_{0,1}$ & DGP2 & 1.00 & 1.00 & 1.00 & 1.00 & 1.00 & 1.00 & 0.99 \\
50 & $H_{0,3}$ & DGP2 & 1.00 & 1.00 & 1.00 & 1.00 & 1.00 & 1.00 & 1.00 \\
\midrule
\multicolumn{10}{l}{\emph{DGP3: Full group separation}} \\
20 & $H_{0,1}$ & DGP3 & 1.00 & 1.00 & 1.00 & 1.00 & 1.00 & 1.00 & 0.98 \\
20 & $H_{0,2}$ & DGP3 & 1.00 & 1.00 & 1.00 & 1.00 & 0.97 & 0.97 & 0.93 \\
20 & $H_{0,3}$ & DGP3 & 1.00 & 1.00 & 1.00 & 1.00 & 1.00 & 1.00 & 1.00 \\
50 & $H_{0,1}$ & DGP3 & 1.00 & 1.00 & 1.00 & 1.00 & 1.00 & 1.00 & 1.00 \\
50 & $H_{0,2}$ & DGP3 & 1.00 & 1.00 & 1.00 & 1.00 & 0.99 & 0.99 & 0.99 \\
50 & $H_{0,3}$ & DGP3 & 1.00 & 1.00 & 1.00 & 1.00 & 1.00 & 1.00 & 1.00 \\
\bottomrule
\end{tabular*}
\begin{tablenotes}[flushleft]
\footnotesize
\item \textit{Notes:} The table reports rejection frequencies. $H_{0,1}$ tests equality of all group-specific slopes, $H_{0,2}$ tests equality of a subset of slopes, and $H_{0,3}$ is false in the designs reported in Panel (b), so rejection frequencies in that panel measure power. ``Predetermined'' uses the true group structure. ``Naive'' tests condition on the estimated group structure but do not account for group selection uncertainty. ``Conditional'' tests apply the proposed selective inference correction. Since Experiment 1 is an iid Gaussian design, all test statistics are computed using the iid variance estimator with the corresponding finite-sample degrees of freedom correction.
\end{tablenotes}
\end{threeparttable}
\end{table}

Table~\ref{tab:exp1_case1_ci} reports coverage and average length for confidence intervals associated with \(H_{0,2}\). The coverage results mirror the rejection-frequency results. Under DGP1, with no group separation, the naive intervals after TSK have zero coverage for both \(T=20\) and \(T=50\). The naive PCR intervals also cover very poorly, at 0.13 and 0.14, and the naive GFE intervals, while better, still under-cover substantially, at 0.56 and 0.62.

The conditional confidence intervals restore coverage. Under DGP1, the conditional TSK, PCR, and GFE intervals all have coverage between 0.93 and 0.94, close to both the predetermined-group benchmark and the nominal 95 percent level. The price is an increase in length reflecting uncertainty due to estimating the groups: at \(T=20\), the conditional TSK and PCR intervals average 0.73 and 0.52, compared with 0.18 and 0.16 for their naive counterparts. The GFE intervals are an exception, remaining as short as the naive ones while achieving nominal coverage, because, under DGP1, the grouped fixed effects absorb the time variation that drives the truncation for the other two procedures.

Under DGP2, where the null \(H_{0,2}\) is true, but the groups are partially separated, both naive and conditional intervals have coverage close to the nominal level. Under DGP3, where the groups are fully separated, coverage remains high for the conditional intervals, but their average length can be substantially larger, especially for GFE. This occurs because the selective confidence set accounts for the truncation induced by the clustering event. In finite samples, this correction can produce wider intervals even when the selected groups are informative.

\begin{table}
\centering
\scriptsize
\begin{threeparttable}
\caption{Coverage and average length of confidence intervals for $H_{0,2}$ \\
Experiment 1: iid Gaussian design
}
\label{tab:exp1_case1_ci}
\begin{tabular*}{\textwidth}{@{\extracolsep{\fill}} cc ccccccc}
\toprule
$T$ & DGP & \shortstack{Predetermined} & \shortstack{Naive\\TSK} & \shortstack{Naive\\PCR} & \shortstack{Naive\\GFE} & \shortstack{Conditional\\TSK} & \shortstack{Conditional\\PCR} & \shortstack{Conditional\\GFE} \\
\midrule
\multicolumn{9}{l}{\textbf{Panel (a): Coverage}} \\
\midrule
\multicolumn{9}{l}{\emph{DGP1: No group separation}} \\
20 & DGP1 & 0.94 & 0.00 & 0.13 & 0.56 & 0.94 & 0.93 & 0.93 \\
50 & DGP1 & 0.95 & 0.00 & 0.14 & 0.62 & 0.93 & 0.93 & 0.94 \\
\midrule
\multicolumn{9}{l}{\emph{DGP2: Partial group separation}} \\
20 & DGP2 & 0.93 & 0.90 & 0.91 & 0.92 & 0.91 & 0.91 & 0.94 \\
50 & DGP2 & 0.95 & 0.94 & 0.95 & 0.95 & 0.94 & 0.94 & 0.95 \\
\midrule
\multicolumn{9}{l}{\emph{DGP3: Full group separation}} \\
20 & DGP3 & 0.95 & 0.93 & 0.93 & 0.94 & 0.95 & 0.96 & 0.99 \\
50 & DGP3 & 0.95 & 0.93 & 0.93 & 0.94 & 0.94 & 0.94 & 0.98 \\
\midrule
\multicolumn{9}{l}{\textbf{Panel (b): Average length}} \\
\midrule
\multicolumn{9}{l}{\emph{DGP1: No group separation}} \\
20 & DGP1 & 0.19 & 0.18 & 0.16 & 0.17 & 0.73 & 0.52 & 0.18 \\
50 & DGP1 & 0.12 & 0.11 & 0.11 & 0.11 & 0.44 & 0.35 & 0.10 \\
\midrule
\multicolumn{9}{l}{\emph{DGP2: Partial group separation}} \\
20 & DGP2 & 0.19 & 0.19 & 0.17 & 0.18 & 0.18 & 0.15 & 0.15 \\
50 & DGP2 & 0.12 & 0.12 & 0.11 & 0.12 & 0.12 & 0.11 & 0.10 \\
\midrule
\multicolumn{9}{l}{\emph{DGP3: Full group separation}} \\
20 & DGP3 & 0.19 & 0.19 & 0.17 & 0.18 & 0.70 & 0.75 & 1.54 \\
50 & DGP3 & 0.12 & 0.12 & 0.11 & 0.12 & 0.27 & 0.31 & 1.09 \\
\bottomrule
\end{tabular*}
\begin{tablenotes}[flushleft]
\footnotesize
\item \textit{Notes:} The table reports empirical coverage and average length of confidence intervals for the scalar contrast in $H_{0,2}$, evaluated relative to the appropriate group-label-aligned truth. Column definitions and the variance estimator are as in Table~\ref{tab:exp1_case1_rejection}.
\end{tablenotes}
\end{threeparttable}
\end{table}

Table~\ref{tab:exp2_case1_rejection} reports rejection frequencies for Experiment 2. This experiment is more challenging because it combines serial dependence, within-cluster CSD, and non-Gaussian errors. The predetermined-group test is no longer perfectly sized in small samples, especially when \(T=20\). For example, under DGP1 its rejection frequency is 0.23 for \(H_{0,1}\) and 0.15 for \(H_{0,2}\). This reflects the difficulty of estimating the long-run covariance matrix in short panels under dependence.

Despite this more demanding setting, the main comparison between naive and conditional post-clustering inference remains clear. Under DGP1, the naive TSK and PCR tests continue to reject almost always, with rejection frequencies equal to one for \(H_{0,1}\), and equal to one and 0.90 for \(H_{0,2}\) when \(T=20\). The naive GFE test also over-rejects, although less severely. By contrast, the conditional tests remain close to the nominal level. Under DGP1, the conditional rejection frequencies range from 0.05 to 0.07 across hypotheses, time dimensions, and estimators.

Under DGP2, the null \(H_{0,2}\) is true despite partial group separation. The conditional tests behave similarly to, or better than, the naive tests. Rejection frequencies are mildly above 5 percent when \(T=20\), ranging from 0.09 to 0.13, but improve when \(T=50\), where they range from 0.07 to 0.10. These results indicate that the proposed procedure remains effective under serial and CSD, although finite-sample covariance estimation can still lead to mild size distortions in short panels.

Panel (b) shows that the conditional tests maintain high power in the dependent design. Under DGP1 and \(H_{0,3}\), rejection frequencies are between 0.91 and 0.98 for \(T=20\), and between 0.97 and 0.98 for \(T=50\). Under DGP2 and DGP3, power is again close to one in most cases. Thus, even in the more realistic dependent design, the selective correction controls spurious rejections under true nulls without eliminating power against false restrictions.

\begin{table}
\centering
\scriptsize
\begin{threeparttable}
\caption{Rejection rates of naive tests and proposed tests under different null hypotheses \\
Experiment 2: serial and within-cluster dependence
}
\label{tab:exp2_case1_rejection}
\begin{tabular*}{\textwidth}{@{\extracolsep{\fill}} ccc ccccccc}
\toprule
$T$ & Test & DGP & \shortstack{Predetermined} & \shortstack{Naive\\TSK} & \shortstack{Naive\\PCR} & \shortstack{Naive\\GFE} & \shortstack{Conditional\\TSK} & \shortstack{Conditional\\PCR} & \shortstack{Conditional\\GFE} \\
\midrule
\multicolumn{10}{l}{\textbf{Panel (a): Size}} \\
\midrule
\multicolumn{10}{l}{\emph{DGP1: No group separation}} \\
20 & $H_{0,1}$ & DGP1 & 0.23 & 1.00 & 1.00 & 0.33 & 0.07 & 0.05 & 0.06 \\
20 & $H_{0,2}$ & DGP1 & 0.15 & 1.00 & 0.90 & 0.24 & 0.06 & 0.06 & 0.07 \\
50 & $H_{0,1}$ & DGP1 & 0.12 & 1.00 & 1.00 & 0.19 & 0.05 & 0.05 & 0.05 \\
50 & $H_{0,2}$ & DGP1 & 0.10 & 1.00 & 0.85 & 0.16 & 0.05 & 0.06 & 0.05 \\
\midrule
\multicolumn{10}{l}{\emph{DGP2: Partial group separation}} \\
20 & $H_{0,2}$ & DGP2 & 0.13 & 0.13 & 0.14 & 0.11 & 0.10 & 0.13 & 0.09 \\
50 & $H_{0,2}$ & DGP2 & 0.10 & 0.09 & 0.10 & 0.08 & 0.09 & 0.10 & 0.07 \\
\midrule
\multicolumn{10}{l}{\textbf{Panel (b): Power}} \\
\midrule
\multicolumn{10}{l}{\emph{DGP1: No group separation}} \\
20 & $H_{0,3}$ & DGP1 & 1.00 & 1.00 & 1.00 & 1.00 & 0.91 & 0.95 & 0.98 \\
50 & $H_{0,3}$ & DGP1 & 1.00 & 1.00 & 1.00 & 1.00 & 0.97 & 0.98 & 0.98 \\
\midrule
\multicolumn{10}{l}{\emph{DGP2: Partial group separation}} \\
20 & $H_{0,1}$ & DGP2 & 1.00 & 1.00 & 1.00 & 1.00 & 0.95 & 0.96 & 0.93 \\
20 & $H_{0,3}$ & DGP2 & 1.00 & 1.00 & 1.00 & 1.00 & 0.99 & 1.00 & 0.99 \\
50 & $H_{0,1}$ & DGP2 & 1.00 & 1.00 & 1.00 & 1.00 & 1.00 & 1.00 & 0.98 \\
50 & $H_{0,3}$ & DGP2 & 1.00 & 1.00 & 1.00 & 1.00 & 1.00 & 1.00 & 1.00 \\
\midrule
\multicolumn{10}{l}{\emph{DGP3: Full group separation}} \\
20 & $H_{0,1}$ & DGP3 & 1.00 & 1.00 & 1.00 & 1.00 & 0.98 & 0.98 & 0.97 \\
20 & $H_{0,2}$ & DGP3 & 1.00 & 1.00 & 1.00 & 1.00 & 0.94 & 0.96 & 0.92 \\
20 & $H_{0,3}$ & DGP3 & 1.00 & 1.00 & 1.00 & 1.00 & 0.99 & 1.00 & 1.00 \\
50 & $H_{0,1}$ & DGP3 & 1.00 & 1.00 & 1.00 & 1.00 & 1.00 & 1.00 & 0.99 \\
50 & $H_{0,2}$ & DGP3 & 1.00 & 1.00 & 1.00 & 1.00 & 0.99 & 0.98 & 0.98 \\
50 & $H_{0,3}$ & DGP3 & 1.00 & 1.00 & 1.00 & 1.00 & 1.00 & 1.00 & 1.00 \\
\bottomrule
\end{tabular*}
\begin{tablenotes}[flushleft]
\footnotesize
\item \textit{Notes:} The table reports rejection frequencies; hypotheses and column definitions are as in Table~\ref{tab:exp1_case1_rejection}. Since Experiment 2 allows serial dependence and within-cluster contemporaneous dependence, all test statistics use the dependence-robust variance estimator of Section~\ref{sec:variance}, applied to the true partition for the predetermined benchmark and to the estimated partition otherwise.
\end{tablenotes}
\end{threeparttable}
\end{table}

Table~\ref{tab:exp2_case1_ci} reports coverage and average length for \(H_{0,2}\) in Experiment 2. The confidence interval results reinforce the conclusions from the rejection frequencies. Under DGP1, the naive TSK intervals again have zero coverage, while the naive PCR intervals cover only 0.10 of the time for \(T=20\) and 0.16 for \(T=50\). The naive GFE intervals perform better, with coverage equal to 0.78 and 0.84, but still fall short of the nominal level.

The conditional intervals restore coverage under no separation. For DGP1, the conditional TSK, PCR, and GFE intervals have coverage between 0.94 and 0.95 for both values of \(T\). This is notable because the predetermined-group intervals under-cover in this dependent design, with coverage equal to 0.85 for \(T=20\) and 0.90 for \(T=50\). The selective intervals therefore correct the additional distortion caused by group selection, even when covariance estimation under dependence is itself challenging.

As in the iid design, the improvement in coverage comes with wider intervals: under DGP1 and \(T=20\), the conditional TSK and PCR intervals average 0.87 and 0.65, against 0.25 and 0.22 for their naive counterparts. This is the finite-sample cost of valid post-selection inference when the estimated groups are not well separated.

\begin{table}
\centering
\scriptsize
\begin{threeparttable}
\caption{Coverage and average length of confidence intervals for $H_{0,2}$ \\
Experiment 2: serial and within-cluster dependence
}
\label{tab:exp2_case1_ci}
\begin{tabular*}{\textwidth}{@{\extracolsep{\fill}} cc ccccccc}
\toprule
$T$ & DGP & \shortstack{Predetermined} & \shortstack{Naive\\TSK} & \shortstack{Naive\\PCR} & \shortstack{Naive\\GFE} & \shortstack{Conditional\\TSK} & \shortstack{Conditional\\PCR} & \shortstack{Conditional\\GFE} \\
\midrule
\multicolumn{9}{l}{\textbf{Panel (a): Coverage}} \\
\midrule
\multicolumn{9}{l}{\emph{DGP1: No group separation}} \\
20 & DGP1 & 0.85 & 0.00 & 0.10 & 0.78 & 0.95 & 0.94 & 0.94 \\
50 & DGP1 & 0.90 & 0.00 & 0.16 & 0.84 & 0.94 & 0.95 & 0.94 \\
\midrule
\multicolumn{9}{l}{\emph{DGP2: Partial group separation}} \\
20 & DGP2 & 0.87 & 0.87 & 0.86 & 0.89 & 0.88 & 0.88 & 0.93 \\
50 & DGP2 & 0.90 & 0.91 & 0.90 & 0.92 & 0.91 & 0.90 & 0.93 \\
\midrule
\multicolumn{9}{l}{\emph{DGP3: Full group separation}} \\
20 & DGP3 & 0.86 & 0.86 & 0.86 & 0.90 & 0.95 & 0.94 & 0.99 \\
50 & DGP3 & 0.90 & 0.90 & 0.90 & 0.92 & 0.91 & 0.91 & 0.98 \\
\midrule
\multicolumn{9}{l}{\textbf{Panel (b): Average length}} \\
\midrule
\multicolumn{9}{l}{\emph{DGP1: No group separation}} \\
20 & DGP1 & 0.21 & 0.25 & 0.22 & 0.19 & 0.87 & 0.65 & 0.14 \\
50 & DGP1 & 0.15 & 0.16 & 0.15 & 0.13 & 0.53 & 0.42 & 0.09 \\
\midrule
\multicolumn{9}{l}{\emph{DGP2: Partial group separation}} \\
20 & DGP2 & 0.21 & 0.23 & 0.21 & 0.21 & 0.21 & 0.19 & 0.17 \\
50 & DGP2 & 0.15 & 0.16 & 0.15 & 0.15 & 0.15 & 0.15 & 0.13 \\
\midrule
\multicolumn{9}{l}{\emph{DGP3: Full group separation}} \\
20 & DGP3 & 0.21 & 0.23 & 0.21 & 0.22 & 1.07 & 1.00 & 1.59 \\
50 & DGP3 & 0.15 & 0.16 & 0.15 & 0.15 & 0.37 & 0.38 & 1.22 \\
\bottomrule
\end{tabular*}
\begin{tablenotes}[flushleft]
\footnotesize
\item \textit{Notes:} The table reports empirical coverage and average length of confidence intervals for the scalar contrast in $H_{0,2}$, evaluated relative to the appropriate group-label-aligned truth. Column definitions and the variance estimator are as in Table~\ref{tab:exp2_case1_rejection}.
\end{tablenotes}
\end{threeparttable}
\end{table}

For DGP2 and DGP3, the conditional intervals continue to provide reasonable coverage. Under partial separation, coverage is slightly below 0.95 when \(T=20\), ranging from 0.88 to 0.93, and improves for \(T=50\). Under full separation, the conditional TSK and PCR intervals have coverage around 0.91 to 0.95, while the conditional GFE intervals are conservative, with coverage equal to 0.99 for \(T=20\) and 0.98 for \(T=50\). The corresponding GFE intervals are also substantially longer, reflecting the fact that the selected GFE clustering imposes a more complex conditioning event.

The Monte Carlo evidence leads to three main conclusions. First, naive post-clustering inference can be severely misleading when group separation is weak or absent. In the no-separation design, conventional post-clustering tests often reject with probability close to 1 even though the null hypothesis is true, and naive confidence intervals can have coverage close to zero. Second, the proposed conditional procedure restores size control and coverage in precisely these difficult cases. The correction is effective not only in the iid Gaussian benchmark, but also under serial dependence, within-cluster CSD, and non-Gaussian errors. Third, the selective correction preserves high power against false restrictions. Across the power designs, conditional rejection frequencies are generally close to one, especially as \(T\) increases.

Overall, the results show that accounting for group-selection uncertainty is essential for reliable inference in latent-group panel models. At the same time, when the groups are partially or fully separated, the proposed method remains competitive and retains strong power.

\section{Empirical Application: Growth Convergence}\label{sec:applications}

This section applies the proposed post-clustering inference procedure to the empirical study of growth convergence clubs. The purpose is twofold. First, we revisit the classical convergence club question using a grouped transition equation in the spirit of \citet{canova2004testing} and the convergence club evidence of \citet{phillips2007transition}. Second, we illustrate how conventional post-clustering tests can substantially overstate the evidence for heterogeneous club dynamics when the group structure is estimated from the same data used for inference.

\subsection{Empirical model and data}\label{subsec:growth_model_data}

The model we estimate comes from Example 1 of Section~\ref{sec:examples}. In the more general case, we model the transition dynamics of income as
\begin{equation}\label{eq:growth_convergence_model}
\begin{split}
      \Delta \log GDP_{it} &= \theta_{1,g_i} \log GDP_{i,t-1} + \theta_{2,g_i} HC_{i,t-1}+\theta_{3,g_i} INV_{i,t-1} + \theta_{4,g_i} GOV_{i,t-1} \\
      & + \theta_{5,g_i} \Delta \log POP_{it} + \theta_{6,g_i} + \varepsilon_{it},
\end{split}
\end{equation}
where $GDP$ is real GDP per capita, $HC$ is the human capital index, $INV$ is the investment share, $GOV$ is the government consumption share, and $POP$ is the population. The choice of conditioning variables follows the Barro-type cross-country growth regression literature \citep{barro1991economic}. Human capital, the investment share, and the government consumption share are lagged by one year, while population growth enters contemporaneously. The coefficient $\theta_{1,g_i}$ measures the club-specific speed of convergence: the response of income growth to its own lagged value. In addition, we estimate a version of \eqref{eq:growth_convergence_model} without controls, keeping only $\log GDP_{i,t-1}$ and the club-specific intercept $\theta_{6,g_i}$.

The null hypothesis of homogeneous transition dynamics across clubs is
$
    H_0:
    \theta_1=\theta_2=\cdots=\theta_G
$,
where $\theta_g$ collects all coefficients of club $g$. We also report coefficient-specific homogeneity tests. In particular, the null hypothesis of a common convergence coefficient is
$
    H_0:
    \theta_{1,1}=\theta_{1,2}=\cdots=\theta_{1,G}
$,
which tests whether the speed of convergence is the same across the estimated clubs.

We compare naive and selective tests of club heterogeneity for $G \in \{2,\dots,8\}$. For each specification and each value of $G$, we use 10,000 random initializations of the PCR algorithm and retain the solution with the smallest objective function. We focus our detailed discussion on $G=4$: this value lies within the range of $G$ for which the selective test rejects the null of homogeneous dynamics in the controlled specification, and offers a parsimonious partition into an economically interpretable set of clubs.

The data are taken from the Penn World Table 11.0. Real GDP per capita is constructed as real GDP divided by population. The raw sample is annual and covers 1950--2023. After losing the first year to the one-year lag in the transition equation and to construct population growth, the estimation sample covers 1951--2023. To ensure that the two specifications are directly comparable, and that any difference between them reflects the inclusion of controls rather than a difference in sample composition, we first construct the balanced panel for the specification with controls and then impose the same country-year observations on the specification without controls. This yields an identical balanced panel of $N=53$ countries observed over $T=73$ years for both the specification with controls and the specification without, so that naive and selective tests, and comparisons across the two specifications, are computed on the same data throughout. All inference is conducted both naively, treating the estimated clubs as fixed, and selectively, conditioning on the estimated club structure. The variance estimator allows for serial dependence and within-club contemporaneous dependence, with bandwidth $\lfloor T^{1/3}\rfloor = 4$.

\subsection{Results}\label{subsec:growth_results}

\begin{figure}[!htbp]
\centering
\includegraphics[scale=0.75]{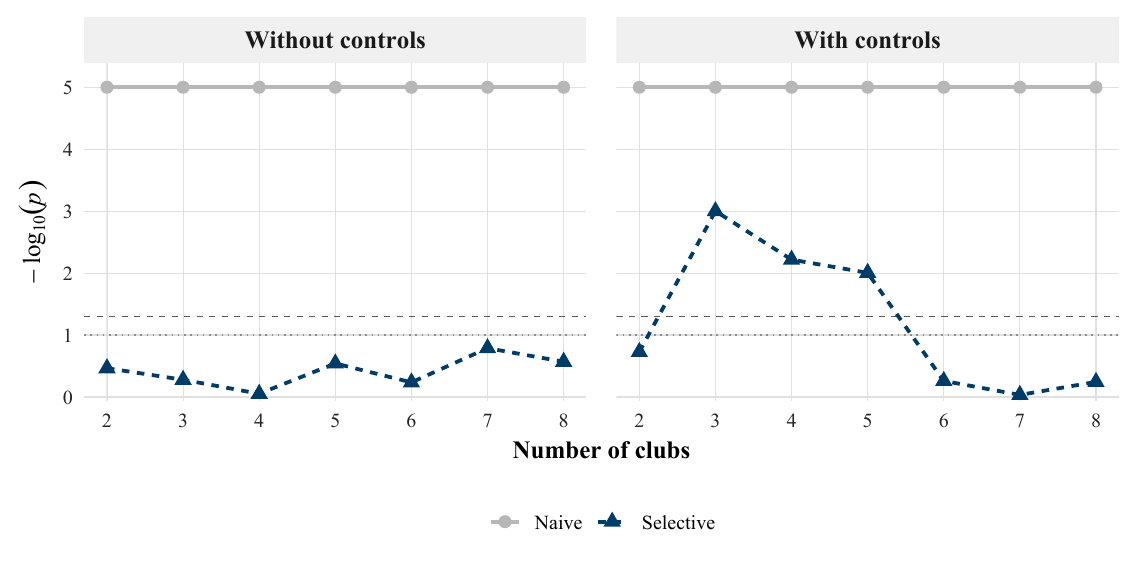}
\caption{Joint tests of club heterogeneity across the number of clubs}
\label{fig:pwt_pcr_joint_pvalues}
\footnotesize
\textit{Notes:} The figure reports joint homogeneity tests for the PCR estimator using PWT data. The vertical axis reports $-\log_{10}(p)$, so larger values indicate stronger evidence against homogeneity. Results are shown separately for the specification without controls and the specification with PWT controls. The solid grey line reports conventional post-clustering Wald $p$-values. The dashed blue line reports selective $p$-values obtained from the proposed conditional procedure. The horizontal reference lines correspond to the 10\% and 5\% significance levels.
\end{figure}

We start with the joint heterogeneity test results. Figure~\ref{fig:pwt_pcr_joint_pvalues} reports joint tests of club heterogeneity after estimating the PCR model for several possible numbers of clubs, $G \in \{2,\dots,8\}$. This exercise is useful because the empirical conclusion should not rely on a single arbitrary choice of \(G\). The vertical axis reports \(-\log_{10}(p)\), so larger values indicate stronger evidence against joint homogeneity. The horizontal reference lines correspond to the 10\% and 5\% significance levels.

The figure shows a sharp difference between naive and selective inference. The naive post-clustering test rejects joint homogeneity for essentially all values of \(G\), both with and without PWT controls: this is precisely the pattern one would expect if conventional inference treats the estimated clubs as fixed and ignores the fact that the same data were used to construct them.

The selective tests lead to a much more disciplined conclusion. Without controls, the selective test does not reject joint homogeneity at any value of \(G\) considered: once the estimated partition is properly conditioned on, the evidence for heterogeneous dynamics in a bare convergence equation disappears entirely, even though the naive test rejects at every \(G\) with $p<0.001$. With PWT controls, the selective test rejects joint homogeneity for \(G=3\), \(4\), and \(5\), with selective \(p\)-values of 0.001, 0.006, and 0.010, respectively, while the evidence is weak for \(G=2\) and for \(G=6\) through \(8\). Overall, the results indicate that controlling for post-clustering selection removes essentially all of the naive evidence of heterogeneity in the uncontrolled specification, and substantially attenuates it in the controlled specification, but does not eliminate it: robust evidence remains for economically relevant intermediate partitions of the country sample once a richer set of controls is included.

Table~\ref{tab:pwt_coefficient_homogeneity_controls} decomposes the joint evidence of club heterogeneity in the controlled PWT specification into coefficient-specific tests. The contrast between naive and selective inference is again substantial: the naive tests reject homogeneity for every coefficient with $p<0.001$. After conditioning on the estimated PCR club structure, the pattern is more informative.

For \(G=4\), the selective tests reject homogeneity for every coefficient except the convergence coefficient itself: human capital ($p=0.003$), the investment share ($p=0.001$), the government consumption share ($p=0.004$), population growth ($p=0.009$), and the intercept ($p=0.046$) all differ across clubs, while the speed of convergence does not ($p=0.945$). This is a substantively interesting split: at four clubs, the estimated heterogeneity is concentrated in the level and control coefficients rather than in the convergence mechanism.

The joint test remains significant under selective inference: the selective $p$-value is 0.006, confirming that the full coefficient vector differs across clubs even though no single coefficient drives the rejection.

\begin{table}[!htbp]
\centering
\scriptsize
\caption{Coefficient-specific homogeneity tests ($G=4$)}
\label{tab:pwt_coefficient_homogeneity_controls}
\begin{threeparttable}
\setlength{\tabcolsep}{6pt}
\begin{tabular*}{\textwidth}{@{\extracolsep{\fill}} lrrr}
\toprule
Coefficient & Wald & Naive $p$ & Selective $p$ \\
\midrule
$\log GDP_{i,t-1}$     & 41.50  & $<0.001$ & 0.945 \\
$HC_{i,t-1}$           & 30.43  & $<0.001$ & 0.003 \\
$INV_{i,t-1}$          & 18.99  & $<0.001$ & 0.001 \\
$GOV_{i,t-1}$          & 37.09  & $<0.001$ & 0.004 \\
$\Delta \log POP_{it}$ & 101.28 & $<0.001$ & 0.009 \\
Intercept              & 185.33 & $<0.001$ & 0.046 \\
\addlinespace
Joint                  & 392.12 & $<0.001$ & 0.006 \\
\bottomrule
\end{tabular*}
\begin{tablenotes}[flushleft]
\footnotesize
\item \textit{Notes:} The table reports coefficient-specific homogeneity tests for the PCR estimator in the specification with PWT controls, for the four-club partition. The degrees of freedom are $G-1=3$ for coefficient-specific tests and $6(G-1)=18$ for the joint test. The selective $p$-values condition on the estimated PCR club structure. $INV$ is the investment share and $GOV$ is the government consumption share in the PWT. The main convergence coefficient is the coefficient on lagged log GDP per capita.
\end{tablenotes}
\end{threeparttable}
\end{table}

Table~\ref{tab:pwt_club_coefficients_g4} reports the club-specific estimates underlying the coefficient-specific tests in Table~\ref{tab:pwt_coefficient_homogeneity_controls}, together with naive and selective tests of whether each individual coefficient is zero within each club. This view is complementary to the across-club homogeneity tests: Table~\ref{tab:pwt_coefficient_homogeneity_controls} asks whether a coefficient differs across clubs, while Table~\ref{tab:pwt_club_coefficients_g4} asks, club by club, whether that coefficient is distinguishable from zero at all. The two exercises need not agree, and in several cases here they diverge sharply.

The clearest illustration is the convergence coefficient. Naively, $\log GDP_{i,t-1}$ appears strongly significant in the two largest clubs (Club~3, $n=21$: naive $p<0.001$; Club~4, $n=12$: naive $p<0.001$) and only marginal or absent elsewhere. Once selection is accounted for, this pattern reverses almost entirely: the selective $p$-values for Club~3 and Club~4 rise to 0.800 and 0.958, respectively, while Club~1 and Club~2 retain evidence of genuine mean reversion (selective $p=0.024$ and $p=0.019$). The two largest clubs, which jointly account for 33 of the 53 countries in the sample, are precisely where the naive test's apparent evidence of convergence is weakest once the selection event is conditioned on. This is consistent with the coefficient-specific homogeneity test in Table~\ref{tab:pwt_coefficient_homogeneity_controls}, where the convergence coefficient is the only one for which cross-club homogeneity cannot be rejected selectively ($p=0.945$): the individual-club results show that this is not because every club exhibits the same degree of convergence, but because the two clubs with genuine convergence and the two without are each imprecisely enough estimated, once selection is accounted for, that the joint test cannot distinguish between them.

A second pattern worth noting is that the selective correction does not move uniformly in one direction. For Club~1's convergence coefficient (naive $p=0.814$, selective $p=0.024$), Club~3's intercept (naive $p=0.237$, selective $p=0.036$), and Club~4's government consumption share (naive $p=0.488$, selective $p=0.018$), the selective test is markedly \emph{more} decisive than the naive one. This runs counter to the common presumption that selective inference is simply a more conservative version of naive inference. Because the selective distribution is a truncation of the sampling distribution to the region consistent with the observed partition, rather than a uniformly wider reference distribution, conditioning can sharpen as well as dampen the evidence against a given null, depending on where the observed statistic falls within the truncated support.

The remaining coefficients display an economically coherent and, in places, cautionary pattern across clubs. The government consumption share is negative and selectively significant in both of the two largest clubs (Club~3, $p=0.025$; Club~4, $p=0.018$), consistent with the Barro-type prediction that government consumption crowds out growth. The investment share is positive in both clubs, in line with the same prediction, but is selectively significant only in Club~4 ($p=0.006$): in Club~3, the naive test suggests significance ($p=0.001$) that disappears once selection is accounted for ($p=0.879$), another instance of the naive-selective divergence noted above. Club~2 breaks the government-share pattern, with a positive and selectively significant coefficient ($p=0.006$); since this 17-country club spans nearly the full income range of the sample, from the poorest to some of the richest economies, its coefficients likely average over genuinely different fiscal-growth relationships rather than reflecting a single underlying mechanism. Club~1 is the most fragile of the four: with only three countries, its coefficients on population growth ($7.332$) and the investment share ($-0.158$) are an order of magnitude larger, and of the opposite sign from the other clubs in the case of investment, than anywhere else in the table. Given that this club is dominated by Venezuela's exceptional income collapse, these estimates should be read as describing that specific episode rather than as evidence of a distinct general growth regime, and we treat them with corresponding caution in the discussion that follows.

Overall, the club-specific estimates reinforce the main empirical message of this section: naive inference systematically overstates the number and location of individually significant relationships, most severely in the largest clubs, while selective inference isolates a smaller set of coefficients, concentrated in the smaller clubs and in the population-growth and distributional (intercept, government share) coefficients, for which the evidence survives conditioning on the estimated partition.

\begin{table}[!htbp]
\centering
\scriptsize
\caption{Club-specific coefficient estimates ($G=4$)}
\label{tab:pwt_club_coefficients_g4}
\begin{threeparttable}
\setlength{\tabcolsep}{6pt}
\begin{tabular*}{\textwidth}{@{\extracolsep{\fill}} l r r r}
\toprule
Coefficient & Estimate (SE) & Naive $p$ & Selective $p$ \\
\midrule
\multicolumn{4}{l}{\textbf{Club 1} \textit{(n=3, poorest)}} \\
\cmidrule(lr){1-4}
$\log GDP_{i,t-1}$ & $-0.0008$ (0.0032) & 0.814 & 0.024 \\
$HC_{i,t-1}$       & $\phantom{-}0.0734$ (0.0134) & $<0.001$ & 0.004 \\
$INV_{i,t-1}$      & $-0.1577$ (0.0478) & 0.001 & 0.001 \\
$GOV_{i,t-1}$      & $-0.1537$ (0.0686) & 0.025 & 0.020 \\
$\Delta \log POP_{it}$           & $\phantom{-}7.3320$ (0.8549) & $<0.001$ & 0.014 \\
Intercept          & $-0.2322$ (0.0252) & $<0.001$ & 0.013 \\
\addlinespace
\multicolumn{4}{l}{\textbf{Club 2} \textit{(n=17)}} \\
\cmidrule(lr){1-4}
$\log GDP_{i,t-1}$ & $-0.0080$ (0.0021) & $<0.001$ & 0.019 \\
$HC_{i,t-1}$       & $\phantom{-}0.0006$ (0.0026) & 0.813 & 0.972 \\
$INV_{i,t-1}$      & $\phantom{-}0.0373$ (0.0134) & 0.005 & 0.002 \\
$GOV_{i,t-1}$      & $\phantom{-}0.0573$ (0.0218) & 0.009 & 0.006 \\
$\Delta \log POP_{it}$           & $-1.1069$ (0.1692) & $<0.001$ & 0.001 \\
Intercept          & $\phantom{-}0.0974$ (0.0186) & $<0.001$ & 0.009 \\
\addlinespace
\multicolumn{4}{l}{\textbf{Club 3} \textit{(n=12)}} \\
\cmidrule(lr){1-4}
$\log GDP_{i,t-1}$ & $\phantom{-}0.0001$ (0.0038) & 0.978 & 0.958 \\
$HC_{i,t-1}$       & $\phantom{-}0.0050$ (0.0061) & 0.407 & 0.206 \\
$INV_{i,t-1}$      & $\phantom{-}0.0514$ (0.0151) & 0.001 & 0.879 \\
$GOV_{i,t-1}$      & $-0.1955$ (0.0385) & $<0.001$ & 0.025 \\
$\Delta \log POP_{it}$           & $-0.2745$ (0.1021) & 0.007 & 0.708 \\
Intercept          & $\phantom{-}0.0299$ (0.0253) & 0.237 & 0.036 \\
\addlinespace
\multicolumn{4}{l}{\textbf{Club 4} \textit{(n=21, richest)}} \\
\cmidrule(lr){1-4}
$\log GDP_{i,t-1}$ & $-0.0304$ (0.0040) & $<0.001$ & 0.800 \\
$HC_{i,t-1}$       & $\phantom{-}0.0098$ (0.0038) & 0.009 & 0.013 \\
$INV_{i,t-1}$      & $\phantom{-}0.0621$ (0.0198) & 0.002 & 0.006 \\
$GOV_{i,t-1}$      & $-0.0121$ (0.0175) & 0.488 & 0.018 \\
$\Delta \log POP_{it}$           & $-0.5195$ (0.1100) & $<0.001$ & 0.915 \\
Intercept          & $\phantom{-}0.2953$ (0.0332) & $<0.001$ & 0.830 \\
\bottomrule
\end{tabular*}
\begin{tablenotes}[flushleft]
\footnotesize
\item \textit{Notes:} The table reports club-specific PCR coefficient estimates for the four-club partition of the specification with PWT controls. Clubs are ordered by mean initial log GDP per capita, poorest to richest, and correspond to the club labels in Figure~\ref{fig:pwt_pcr_income_ladder}. Standard errors, in parentheses, are computed from the Driscoll--Kraay-type variance estimator with bandwidth 4. Naive $p$-values are conventional Wald $p$-values that treat the estimated club as fixed. Selective $p$-values condition on the estimated PCR partition. $INV$ is the investment share and $GOV$ is the government consumption share.
\end{tablenotes}
\end{threeparttable}
\end{table}

\begin{figure}[!htbp]
\centering
\includegraphics[scale=0.75]{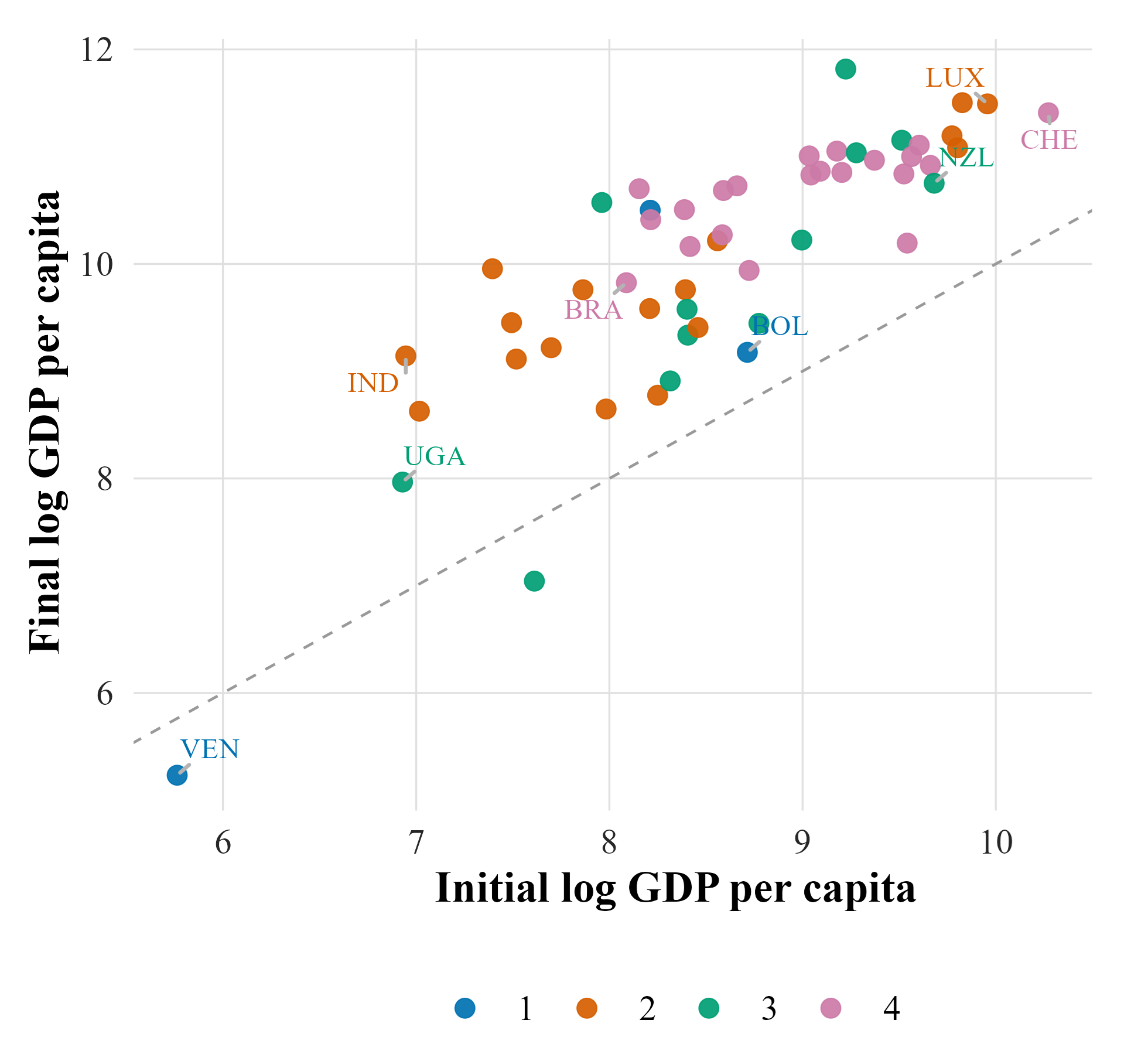}
\caption{Initial and final income by estimated club}
\label{fig:pwt_pcr_income_ladder}
\footnotesize
\textit{Notes:} The figure plots final log GDP per capita against initial log GDP per capita for the PCR clubs estimated from the specification with PWT controls. Results are shown for $G=4$. Each point is a country, and colors indicate estimated club membership. Clubs are ordered by mean initial income. The dashed 45-degree line indicates equal initial and final log GDP per capita. Selected country labels are displayed to highlight influential or economically informative observations.
\end{figure}

Figure~\ref{fig:pwt_pcr_income_ladder} provides a country-level view of the PCR club allocations in the controlled PWT specification. Each point represents a country, with initial log GDP per capita on the horizontal axis and final log GDP per capita on the vertical axis. The dashed line is the 45-degree line. Countries above this line experienced an increase in log GDP per capita over the sample period, while those below it experienced a decline. The figure shows the partition obtained for \(G=4\), the specification emphasized in the empirical analysis.

The figure shows that the estimated clubs are not simple rankings by initial income. At \(G=4\), the smallest club is dominated by Venezuela's marked income decline, alongside Bolivia and Panama. A large 17-country club spans nearly the full income range, from India, Kenya, and Pakistan to Australia, Norway, and the United States, suggesting that these countries share estimated transition dynamics despite very different starting points. The remaining 12-country club is the most heterogeneous of all: it contains both one of the sample's steepest decliners, the Democratic Republic of the Congo, and some of its strongest risers, Ireland, Iceland, and the Netherlands, which is consistent with its near-zero estimated convergence coefficient. The largest club, with 21 countries, is concentrated among Western European economies together with a set of middle-income Latin American and Mediterranean countries, and ends the sample as the richest group. Overall, the figure supports the interpretation that the PCR clubs identify economically meaningful convergence patterns, capturing both differences in initial income and sharply differing subsequent growth trajectories among countries with similar starting points.

This exercise illustrates the empirical relevance of accounting for the data-driven construction of convergence clubs. Naive post-clustering inference points to pervasive heterogeneity across every specification, every value of \(G\), and every coefficient. Selective inference delivers a markedly more disciplined conclusion: in the uncontrolled specification, the evidence for heterogeneity vanishes entirely once club-selection uncertainty is accounted for; in the controlled specification, meaningful evidence survives, but only for intermediate partitions (\(G=3\), \(4\), and \(5\)). The estimated clubs are also economically interpretable, though not through a simple ranking by initial income: several clubs group together countries with sharply different starting points but similar subsequent transition dynamics. Thus, the results support the presence of convergence-club heterogeneity in a sufficiently rich specification, while showing that conventional post-clustering tests dramatically overstate both its strength and its scope.

\section{Conclusion}\label{sec:otherquestions}

This paper presents methods for robust statistical inference in the presence of potential violations of group separation conditions. We propose a selective conditional inference approach that conditions on the estimated group structure and certain nuisance parameters. Our methods yield valid inferences even without group separation, as demonstrated by theoretical analysis and numerical simulations.
Specifically, our methods can be used to test whether two groups are identical and to determine whether a subset of coefficient parameters is equal across groups. It is important to note that existing asymptotic results can be used to test the homogeneity of a subset of coefficients. However, they require that the coefficients unrelated to the null hypothesis satisfy the group separation condition. Our method addresses this limitation by providing valid inferences even without group separation. Additionally, we allow for general linear hypotheses. Even when there is group separation, or when the null hypothesis is not directly related to the group separation condition, our procedure demonstrates superior performance by effectively accounting for the impact of group-structure estimation.

There are several avenues for future research. First, although our procedures accommodate serial and CSD via the score-based long-run variance estimators in Section~\ref{sec:variance}, the exact finite-sample truncation results are established under spherical Gaussian errors; a finite-sample theory for general dependence structures remains open. The work by \cite{gonzalez2023post}, which focuses on dependent observations, highlights the challenges that selective inference faces in this direction.
Second, while our simulation results indicate that our procedure has strong power, a theoretical examination of its power properties would be desirable.
Third, we have focused on linear panel data models; however, the clustering method could also be applied to nonlinear models, such as binary choice models. Exploring extensions to these models would expand the scope of selective inference.
Finally, we have treated the number of groups $G$ as prespecified; extending the conditional framework to accommodate a data-driven choice of $G$ is a natural next step.

\section*{Appendices}

\begin{appendices}
\counterwithin{lemma}{section}
\counterwithin{assumption}{section}
\counterwithin{proposition}{section}

\section{Calculation of the Truncation Sets}\label{sec:truncationset}

In this section, we develop the analytical formulas for the truncation sets $\mathcal{T}_{\textrm{PCR}}$ and $\mathcal{T}_{\textrm{TSK}}$ of Theorems~\ref{thm:pcr} and~\ref{thm:tsk}.

The truncation sets characterize the conditioning event that accommodates the entire algorithm iteration, from the initial values to the final estimated group structure.
The conditioning event in Definition \ref{definition:selectivepvalue} is equivalent to $\mathcal{C} = \left\lbrace \bigcap_{i=1}^N \lbrace g_{i,D} = g_{i,d} \rbrace \right\rbrace$, where $g_{i,D}$ is the $i$-th element of $\gamma_{D}$ and $g_{i,d}$ is its realization.
The truncation sets in \eqref{eq:truncationset_pcr} and \eqref{eq:truncationset_tsk} can be written as
\begin{gather}
    \mathcal{T}_{\textrm{PCR}} = \left\lbrace \phi^2 \in \mathbbm{R}_{\geq 0} : \bigcap_{m=0}^M \bigcap_{i=1}^N \lbrace \hat{g}_{i,d(\phi)}^{(m)} = \hat{g}_{i,d}^{(m)} \rbrace \right\rbrace, \label{eq:spcr_set_iterations} \\
    \mathcal{T}_{\textrm{TSK}} = \left\lbrace \phi^2 \in \mathbbm{R}_{\geq 0} : \bigcap_{m=0}^M \bigcap_{i=1}^N \lbrace \tilde{g}_{i,d(\phi)}^{(m)} = \tilde{g}_{i,d}^{(m)} \rbrace \right\rbrace, \label{eq:stsk_set_iterations}
\end{gather}
where for TSK, $d(\phi) = \{ \hat{b}_{i}(\phi); i=1, \dots, N \}$ with $\hat{b}_{i}(\phi) = [\hat{b}(\phi)]_i^K$ and $\hat{b}(\phi)$ as defined in \eqref{eq:perturbation_simple_KMeans}, and for PCR, $d(\phi) = \{ [y_{it}(\phi), X_{it}]; i=1, \dots, N, t=1,\dots, T \}$ with $y_{it}(\phi)$ the $(i,t)$-th element of the perturbed outcome vector $y(\phi)$ defined in \eqref{eq:perturbation_pcr}.

Both truncation sets decompose as
\begin{equation*}
\begin{split}
    \mathcal{T}_{\textrm{PCR}} &\equiv \mathcal{T}^{(0)}_{\textrm{PCR}} \bigcap \left( \bigcap_{m=1}^M \mathcal{T}^{(m)}_{\textrm{PCR}} \right), \\
    \mathcal{T}_{\textrm{TSK}} &\equiv \mathcal{T}^{(0)}_{\textrm{TSK}} \bigcap \left( \bigcap_{m=1}^M \mathcal{T}^{(m)}_{\textrm{TSK}} \right),
\end{split}
\end{equation*}
where $\mathcal{T}^{(m)}$ captures the constraint that the $m$-th iteration of the algorithm produces the same group assignment as in the observed data.

We initialize both algorithms randomly: the initial group label of each unit is drawn uniformly from $\{1,\dots, G\}$, and the initial parameter values are then computed by least squares given these labels. As is well known, Algorithms~\ref{algo:panel_KMeans} and~\ref{algo:simple_KMeans} may converge to local minima; we therefore use a large number of random starts and retain the solution with the smallest objective.

Random initialization has a convenient implication for the truncation sets. The conditioning event covers all iterations of the algorithm, including the initialization step, whose contribution is $\mathcal{T}^{(0)} = \{\phi \in \mathbbm{R}_{\geq 0} : \gamma^{(0)}_{d(\phi)} = \gamma^{(0)}_d\}$. Because the initial labels are drawn independently of the data, they do not vary with the perturbation, so $\gamma^{(0)}_{d(\phi)} = \gamma^{(0)}_d$ for every $\phi$ and $\mathcal{T}^{(0)}_{\textrm{TSK}} = \mathcal{T}^{(0)}_{\textrm{PCR}} = \mathbbm{R}_{\geq 0}$. The initialization step imposes no constraint and may be omitted; the derivations below therefore focus on $m \geq 1$. We condition on the realized random seed, so that the entire algorithmic path is a deterministic function of the perturbed data. A data-dependent initialization, such as KMeans++ \citep{arthur2007kmeansp} applied to $\{\hat{B}_i\}$, would instead make $\gamma^{(0)}_{d(\phi)}$ vary with $\phi$ and require its own quadratic truncation contribution; we leave the trade-off between better initialization and the cost of this additional conditioning for future work.

\subsection{Calculation of \texorpdfstring{$\mathcal{T}^{(m)}_{\mathrm{PCR}}$}{T(m) PCR} for \texorpdfstring{$m \geq 1$}{m >= 1}}
\label{sec:s_pcr_set}

We now derive the formulas for $\mathcal{T}^{(m)}_{\textrm{PCR}}$, $m \geq 1$. In the PCR algorithm, the group assignment at iteration $m$ is determined by the group whose fitted values are closest to the observed outcomes. Hence, the associated truncation set can be written as
\begin{equation}
\label{eq:pcr_p}
\begin{aligned}
\mathcal{T}^{(m)}_{\textrm{PCR}}
=
\bigcap_{i=1}^N
\bigcap_{g=1}^G
\Bigl\{
\phi^2 \in \mathbb{R}_{\geq 0}
:
\left\|
[y(\phi)]_i^T
-
X_i
\mathbb{A}_{\hat{g}_{i,d}^{(m)}}^{(m-1)}
y(\phi)
\right\|^2 \leq
\left\|
[y(\phi)]_i^T
-
X_i
\mathbb{A}_g^{(m-1)}
y(\phi)
\right\|^2
\Bigr\}.
\end{aligned}
\end{equation}
Here, $[y(\phi)]_i^T$ denotes the $T$-vector of perturbed outcomes for unit $i$, obtained by extracting the $((i-1)T+1)$-th to $(iT)$-th elements of $y(\phi)$ as defined in \eqref{eq:perturbation_pcr}.

Let $\widehat{H}^{(m-1)}$ be the $N \times G$ group-dummy matrix at iteration $m-1$, and let $\mathbb{X}(\widehat{H}^{(m-1)})$ denote the corresponding grouped design matrix. Define
\[
\mathbb{A}_g^{(m-1)}
=
\left(\mathbf{e}_g' \otimes I_K\right)
\left\{
\mathbb{X}(\widehat{H}^{(m-1)})'
\mathbb{X}(\widehat{H}^{(m-1)})
\right\}^{-1}
\mathbb{X}(\widehat{H}^{(m-1)})'.
\]
Thus, the group-$g$ coefficient estimate at iteration $m-1$ satisfies
$
\hat{\theta}_g^{(m-1)}(\phi)
=
\mathbb{A}_g^{(m-1)}y(\phi)
$.

From \eqref{eq:perturbation_pcr}, we have
$
y(\phi)
=
\phi \cdot \mathbb{Q}_{\textrm{PCR}}\hat{j}
+
\hat{v},
$
so that
$
[y(\phi)]_i^T
=
\phi \cdot
[\mathbb{Q}_{\textrm{PCR}}\hat{j}]_i^T
+
[\hat{v}]_i^T.
$
Define
$
\hat{q}_i
=
[\mathbb{Q}_{\textrm{PCR}}\hat{j}]_i^T
$
and
$
\hat{u}_i
=
[\hat{v}]_i^T
$,
so that
$
[y(\phi)]_i^T
=
\phi \cdot \hat{q}_i+\hat{u}_i
$.

\begin{proposition}\normalfont
For all $i \in \{1,\dots,N\}$, $g \in \{1,\dots,G\}$, and
$m \in \{1,\dots,M\}$,
\[
\left\|
[y(\phi)]_i^T
-
X_i \mathbb{A}_g^{(m-1)}y(\phi)
\right\|^2
=
\psi^{(m)}_{i,g,1}\phi^2
+
\psi^{(m)}_{i,g,2}\phi
+
\psi^{(m)}_{i,g,3},
\]
where, letting
$
\hat{c}_{i,g}^{(m)}
=
\hat{q}_i
-
X_i \mathbb{A}_{g}^{(m-1)}
\mathbb{Q}_{\textrm{PCR}}\hat{j}
$
and
$
\hat{d}_{i,g}^{(m)}
=
\hat{u}_i
-
X_i \mathbb{A}_g^{(m-1)}
\hat{v}
$,
we have
\begin{align*}
\psi^{(m)}_{i,g,1}
=
\left\|
\hat{c}_{i,g}^{(m)}
\right\|^2, \quad
\psi^{(m)}_{i,g,2}
=
2
\left(\hat{c}_{i,g}^{(m)}\right)'
\hat{d}_{i,g}^{(m)}, \quad
\psi^{(m)}_{i,g,3} =
\left\|
\hat{d}_{i,g}^{(m)}
\right\|^2.
\end{align*}
\end{proposition}

Under random initialization,
\[
\mathcal{T}_{\textrm{PCR}}
=
\bigcap_{m=1}^M
\bigcap_{i=1}^N
\bigcap_{g=1}^G
\left\{
\phi^2 \in \mathbb{R}_{\geq 0}
:
\Psi_{i,g}^{(m)}(\phi)
\leq 0
\right\},
\]
where
\[
\begin{aligned}
\Psi_{i,g}^{(m)}(\phi)
=
\left[
\psi^{(m)}_{i,\hat{g}_{i,d}^{(m)},1}
-
\psi^{(m)}_{i,g,1}
\right]\phi^2 +
\left[
\psi^{(m)}_{i,\hat{g}_{i,d}^{(m)},2}
-
\psi^{(m)}_{i,g,2}
\right]\phi +
\left[
\psi^{(m)}_{i,\hat{g}_{i,d}^{(m)},3}
-
\psi^{(m)}_{i,g,3}
\right],
\end{aligned}
\]
and all coefficients are computable from the data.

\subsection{Calculation of \texorpdfstring{$\mathcal{T}^{(m)}_{\mathrm{TSK}}$}{T(m) TSK} for \texorpdfstring{$m \geq 1$}{m >= 1}}
\label{sec:s_tsk_set}

We derive analytical formulas for $\mathcal{T}^{(m)}_{\textrm{TSK}}$, $m \geq 1$.
Following Proposition~2 of \cite{chen23}, the $m$-th iteration of Algorithm~\ref{algo:simple_KMeans} assigns unit $i$ to group $\tilde{g}_{i,d}^{(m)}$ if and only if
\begin{equation}
\label{eq:tsk_nu}
\mathcal{T}^{(m)}_{\textrm{TSK}}
=
\bigcap_{i=1}^N
\bigcap_{g=1}^G
\left\{
\phi^2 \in \mathbb{R}_{\geq 0}
:
\left\|
\hat{b}_{i}(\phi)
-
\mathbb{B}_{\tilde{g}_{i,d}^{(m)}}^{(m-1)}
\hat{b}(\phi)
\right\|^2
\leq
\left\|
\hat{b}_{i}(\phi)
-
\mathbb{B}_{g}^{(m-1)}
\hat{b}(\phi)
\right\|^2
\right\}.
\end{equation}
Here, letting $\tilde{H}^{(m-1)}$ be the $N \times G$ group-dummy matrix at iteration $m-1$,
\[
\mathbb{B}_{g}^{(m-1)}
=
(\mathbf{e}_g' \otimes I_K)
(
[
(\tilde{H}^{(m-1)})'
\tilde{H}^{(m-1)}
]^{-1}
\otimes I_K
)
(
(\tilde{H}^{(m-1)})'
\otimes I_K
).
\]
From \eqref{eq:perturbation_simple_KMeans}, we have
\begin{equation}
\label{eq:tsk_perturbation_expanded}
\hat{b}(\phi)
=
\phi \cdot \mathbb{Q}_{\textrm{TSK}}\tilde{j}
+
\tilde{u},
\end{equation}
so that
$
\hat{b}_{i}(\phi)
=
[\hat{b}(\phi)]_i^K
=
\phi \cdot
\left[
\mathbb{Q}_{\textrm{TSK}}\tilde{j}
\right]_i^K
+
[\tilde{u}]_i^K
$.
For conciseness, define
$
\tilde{q}_i
=
\left[
\mathbb{Q}_{\textrm{TSK}}\tilde{j}
\right]_i^K
$
and
$
\tilde{u}_i
=
[\tilde{u}]_i^K
$,
so that
$
\hat{b}_i(\phi)
=
\phi \cdot \tilde{q}_i+\tilde{u}_i
$.

\begin{proposition}\normalfont
For all $i \in \{1,\dots,N\}$, $g \in \{1,\dots,G\}$, and
$m \in \{1,\dots,M\}$,
\[
\left\|
\hat{b}_{i}(\phi)
-
\mathbb{B}_{g}^{(m-1)}
\hat{b}(\phi)
\right\|^2
=
\lambda^{(m)}_{i,g,1}\phi^2
+
\lambda^{(m)}_{i,g,2}\phi
+
\lambda^{(m)}_{i,g,3},
\]
where, letting
$
\tilde{c}_{i,g}^{(m)}
=
\tilde{q}_i
-
\mathbb{B}_g^{(m-1)}
\mathbb{Q}_{\textrm{TSK}}\tilde{j}
$
and
$
\tilde{d}_{i,g}^{(m)}
=
\tilde{u}_i
-
\mathbb{B}_g^{(m-1)}\tilde{u}
$,
we have
\begin{align*}
\lambda^{(m)}_{i,g,1}
=
\left\|
\tilde{c}_{i,g}^{(m)}
\right\|^2, \quad
\lambda^{(m)}_{i,g,2}
=
2
(\tilde{c}_{i,g}^{(m)})'
\tilde{d}_{i,g}^{(m)}, \quad
\lambda^{(m)}_{i,g,3}
=
\left\|
\tilde{d}_{i,g}^{(m)}
\right\|^2.
\end{align*}
\end{proposition}

Under random initialization,
\[
\mathcal{T}_{\textrm{TSK}}
=
\bigcap_{m=1}^M
\bigcap_{i=1}^N
\bigcap_{g=1}^G
\left\{
\phi^2 \in \mathbb{R}_{\geq 0}
:
\Lambda_{i,g}^{(m)}(\phi)
\leq 0
\right\},
\]
where
$$
\Lambda_{i,g}^{(m)}(\phi)
=
\left[
\lambda^{(m)}_{i,\tilde{g}_{i,d}^{(m)},1}
-
\lambda^{(m)}_{i,g,1}
\right]\phi^2 +
\left[
\lambda^{(m)}_{i,\tilde{g}_{i,d}^{(m)},2}
-
\lambda^{(m)}_{i,g,2}
\right]\phi +
\left[
\lambda^{(m)}_{i,\tilde{g}_{i,d}^{(m)},3}
-
\lambda^{(m)}_{i,g,3}
\right],
$$
and all coefficients are computable from the data.

\section{Technical appendix for PCR}

This appendix contains the proof of Theorem \ref{thm:pcr} and related technical details. We first discuss various theoretical properties of the random variable underlying the test statistics. The proof follows similar steps to those of Theorem \ref{thm:tsk}.

\subsection{Lemmas}

We now establish several lemmas which hold for fixed $\gamma$. Eventually, we consider the case with $\gamma = \hat{\gamma}_d $. 

\begin{lemma}
	\label{lemma-wind-pcr}
    Suppose that $S \sim N(\mathbb{X}'\mathbb{X}B, \sigma^2 \mathbb{X}'\mathbb{X})$ and $\sum_{t=1}^T X_{it} X_{it}' = \Sigma$ for any $i$. Then, it holds that 
	\begin{align*}
		 R \hat{\theta}_{D} \;\bot\; \hat{U}.
	\end{align*}
\end{lemma}

\begin{proof}
Note that 
$
     R \hat{\theta}_{D}
     = R (\widehat{\mathbb{H}}'\mathbb{X}' \mathbb{X}\widehat{\mathbb{H}})^{-1} \widehat{\mathbb{H}}' S
$
and 
\begin{align*}
    \hat{U}
    &= 
    {\mathbb{X}'X_{\hat{\gamma}}}\hat{\theta}_{R,D} 
 + \mathbb{X}' \hat{\varepsilon} \\
&= {\mathbb{X}'X_{\hat{\gamma}}}
 \hat{\theta}_D - {\mathbb{X}'X_{\hat{\gamma}}} (X_{\hat{\gamma}}'X_{\hat{\gamma}})^{-1} R' \{R(X_{\hat{\gamma}}'X_{\hat{\gamma}})^{-1} R'\}^{-1} [R \hat{\theta}_D - r]  +\mathbb{X}'Y - {\mathbb{X}'X_{\hat{\gamma}}} \hat{\theta}_D \\
&= -{\mathbb{X}'X_{\hat{\gamma}}} (X_{\hat{\gamma}}'X_{\hat{\gamma}})^{-1} R' \{R(X_{\hat{\gamma}}'X_{\hat{\gamma}})^{-1} R'\}^{-1} R \hat{\theta}_D + \mathbb{X}'Y + {\mathbb{X}'X_{\hat{\gamma}}} (X_{\hat{\gamma}}'X_{\hat{\gamma}})^{-1} R' \{R(X_{\hat{\gamma}}'X_{\hat{\gamma}})^{-1} R'\}^{-1} r \\
&= -{\mathbb{X}'X_{\hat{\gamma}}} (X_{\hat{\gamma}}'X_{\hat{\gamma}})^{-1} R' \{R(X_{\hat{\gamma}}'X_{\hat{\gamma}})^{-1} R'\}^{-1} R (X_{\hat{\gamma}}'X_{\hat{\gamma}})^{-1} \widehat{\mathbb{H}}' S \\
&\quad + S + {\mathbb{X}'X_{\hat{\gamma}}} (X_{\hat{\gamma}}'X_{\hat{\gamma}})^{-1} R' \{R(X_{\hat{\gamma}}'X_{\hat{\gamma}})^{-1} R'\}^{-1} r .
\end{align*}
The last term, ${\mathbb{X}'X_{\hat{\gamma}}} (X_{\hat{\gamma}}'X_{\hat{\gamma}})^{-1} R' \{R(X_{\hat{\gamma}}'X_{\hat{\gamma}})^{-1} R'\}^{-1} r$ is a constant and thus does not affect the independence. 
Because $S \sim N(\mathbb{X}'\mathbb{X}B, \sigma^2 \mathbb{X}'\mathbb{X})$, by the properties of the multivariate normal distributions, it is sufficient to show that 
\begin{align*}
    ( {\mathbb{X}'X_{\hat{\gamma}}} (X_{\hat{\gamma}}'X_{\hat{\gamma}})^{-1} R' \{R(X_{\hat{\gamma}}'X_{\hat{\gamma}})^{-1} R'\}^{-1} R (X_{\hat{\gamma}}'X_{\hat{\gamma}})^{-1} \widehat{\mathbb{H}}' - \mathbf{I})
    (\mathbb{X}'\mathbb{X})
    \widehat{\mathbb{H}} (X_{\hat{\gamma}}'X_{\hat{\gamma}})^{-1}R' =0.
\end{align*}
To show this, we use the normalization that 
$
    \sum_{t=1}^T X_{it} X_{it}' = \Sigma
$
for any $i$.
Under this normalization, {$\mathbb{X}'\mathbb{X} = I_N \otimes \Sigma$, so that $\widehat{\mathbb{H}}'\mathbb{X}'\mathbb{X}\widehat{\mathbb{H}} = \hat{\mathcal{N}}_{\Sigma} \equiv \mathrm{diag}(\hat{n}_1, \dots, \hat{n}_G) \otimes \Sigma$ and $\mathbb{X}'\mathbb{X}\widehat{\mathbb{H}} = (I_N \otimes \Sigma)\widehat{\mathbb{H}}$.} Thus, we have 
\begin{align*}
    & ( \mathbb{X}'X_{\hat{\gamma}} (X_{\hat{\gamma}}'X_{\hat{\gamma}})^{-1} R' \{R(X_{\hat{\gamma}}'X_{\hat{\gamma}})^{-1} R'\}^{-1} R (X_{\hat{\gamma}}'X_{\hat{\gamma}})^{-1} \widehat{\mathbb{H}}' - \mathbf{I}) (\mathbb{X}' \mathbb{X}) \widehat{\mathbb{H}} (X_{\hat{\gamma}}'X_{\hat{\gamma}})^{-1}R' \\
	&= (I_N\otimes \Sigma ) \widehat{\mathbb{H}} \hat{\mathcal{N}}_{\Sigma}^{-1} R' \{R \hat{\mathcal{N}}_{\Sigma}^{-1} R'\}^{-1} R \hat{\mathcal{N}}_{\Sigma}^{-1} \widehat{\mathbb{H}}' (I_N \otimes \Sigma) \widehat{\mathbb{H}} \hat{\mathcal{N}}_{\Sigma}^{-1} R' \\
    & \quad - (I_N \otimes \Sigma) \widehat{\mathbb{H}} \hat{\mathcal{N}}_{\Sigma}^{-1} R' \\
	&=
	(I_N\otimes \Sigma ) \widehat{\mathbb{H}} \hat{\mathcal{N}}_{\Sigma}^{-1} R' \{R \hat{\mathcal{N}}_{\Sigma}^{-1} R'\}^{-1}  R \hat{\mathcal{N}}_{\Sigma}^{-1} \hat{\mathcal{N}}_{\Sigma} \hat{\mathcal{N}}_{\Sigma}^{-1} R' - (I_N \otimes \Sigma) \widehat{\mathbb{H}} \hat{\mathcal{N}}_{\Sigma}^{-1} R' \\
	&=
	(I_N\otimes \Sigma ) \widehat{\mathbb{H}} \hat{\mathcal{N}}_{\Sigma}^{-1} R' - (I_N \otimes \Sigma) \widehat{\mathbb{H}} \hat{\mathcal{N}}_{\Sigma}^{-1} R'
	= 0,
\end{align*}
where we use the fact that $ \widehat{\mathbb{H}}'(I_N \otimes \Sigma)\widehat{\mathbb{H}} = \hat{\mathcal{N}}_{\Sigma}$.
\end{proof}

Note that this lemma holds regardless of whether the null hypothesis holds. The following, in contrast, requires that the null hypothesis be true.

\begin{lemma}\label{lemma-dirind-pcr}
Under $H_0$,
$
\mathcal{L}_D(\gamma)'\,\mathcal{L}_D(\gamma)
\;\bot\;
\operatorname{dir}\!\left\{\mathcal{L}_D(\gamma)\right\}
$.
\end{lemma}

\begin{proof}
	Because 
	$
		 R\hat{\theta}_{D} - r \sim N(0,  \sigma^2R (X_{\hat{\gamma}}'X_{\hat{\gamma}})^{-1} R')
	$
	under $H_0$, it holds that 
	$
		 \mathcal{L}_D (\gamma)  \sim N (0, I_q )
	$.
	Therefore, its norm and the direction are independent by Proposition 4.11 and Corollary 4.3 of \cite{bilodeau1999theory}.
\end{proof}

\subsection{Proof of Theorem \ref{thm:pcr}}

Recall that 
$
s(\phi)
= \phi \cdot \mathbb{P}_{\textrm{PCR}} \mathrm{dir} (\mathcal{L}_{d} ) + \hat{u}
$.
Consider an event $W_{\textrm{PCR}}(\hat{\gamma}_D) \in E$ for some $E$. We observe that
\begin{align*}
	&\Pr_{H_0} \left( W_{\textrm{PCR}}(\hat{\gamma}_D) \in E \; \middle | \; \mathcal{C}_{\textrm{PCR}} \right) \\
	&= \Pr_{H_0} \left( W_{\textrm{PCR}}(\hat{\gamma}_D) \in E \; \middle | \; 
	    \bigcap_{m=0}^M \left\lbrace 
        \hat{\gamma}^{(m)}_D = 
        \hat{\gamma}^{(m)}_d 
    \right\rbrace, \, \mathrm{dir} \left\lbrace 
       \mathcal{L}_D (\hat{\gamma}_{D})
    \right\rbrace 
    = \mathrm{dir} \left\lbrace 
        \mathcal{L}_{d}
    \right\rbrace, 
     \, \hat{U} (\hat{\gamma}_{D}) = 
    {\hat{u}}  \right) \\
	&= \Pr_{H_0} \left( W_{\textrm{PCR}}(\hat{\gamma}_d) \in E \; \middle | \; 
	    \bigcap_{m=0}^M \left\lbrace 
        \hat{\gamma}^{(m)}_D = 
        \hat{\gamma}^{(m)}_d 
    \right\rbrace, \, \mathrm{dir} \left\lbrace 
       \mathcal{L}_{D} (\hat{\gamma}_d)
    \right\rbrace 
    = \mathrm{dir} \left\lbrace 
        \mathcal{L}_{d} 
    \right\rbrace, 
     \, \hat{U} (\hat{\gamma}_{d}) = 
    {\hat{u}}  \right).
\end{align*}
The first equality follows from the definition of $\mathcal{C}_{\textrm{PCR}}$. The second equality uses the fact that $\hat{\gamma}_D =\hat{\gamma}_d$ in this conditioning set.

We write 
\begin{align*}
	\hat{\gamma}_D^{(m)}
	&= \hat{\gamma}^{(m)} ( S ) \\
	 &= \hat{\gamma}^{(m)}  \left( \mathbb{P}_{\textrm{PCR}}(\hat{\gamma}_{d}) \mathcal{L}_{d}  + {\hat{U}(\hat{\gamma}_{d})} \right) \\
 	 &= \hat{\gamma}^{(m)}  \left( \sqrt{ W_{\textrm{PCR}}(\hat{\gamma}_d) } \mathbb{P}_{\textrm{PCR}}(\hat{\gamma}_{d}) \mathrm{dir} ( \mathcal{L}_{d})  + {\hat{U}(\hat{\gamma}_{d})} \right) .
\end{align*}

Using this decomposition, we write
\begin{align*}
&\Pr_{H_0} \left( W_{\textrm{PCR}}(\hat{\gamma}_D) \in E \mid \mathcal{C}_{\textrm{PCR}} \right) \\
	&= \Pr_{H_0} \bigg( W_{\textrm{PCR}}(\hat{\gamma}_d) \in E \mid \\
& \quad  \bigcap_{m=0}^M  \left\lbrace\   \hat{\gamma}^{(m)} \left(  \sqrt{ W_{\textrm{PCR}}(\hat{\gamma}_d) } \mathbb{P}_{\textrm{PCR}}(\hat{\gamma}_{d}) \mathrm{dir} ( \mathcal{L}_{d})  + {\hat{U}(\hat{\gamma}_{d})} \right)= 
        \hat{\gamma}^{(m)}_d 
    \right\rbrace, \\
    & \quad \mathrm{dir} \left\lbrace 
       \mathcal{L}_{d}
    \right\rbrace 
    = \mathrm{dir} \left\lbrace 
      \mathcal{L}_{d}
    \right\rbrace, 
     \, \hat{U} = 
    {\hat{u}}  \bigg).
\end{align*}
We insert the latter two conditions into the argument of $\hat{\gamma}^{(m)}$, and the probability becomes
\begin{align*}
	&= \Pr_{H_0} \bigg( W_{\textrm{PCR}}(\hat{\gamma}_d) \in E \mid \\
& \quad  \bigcap_{m=0}^M  \left\lbrace\   \hat{\gamma}^{(m)} \left(  \sqrt{ W_{\textrm{PCR}}(\hat{\gamma}_d) } \mathbb{P}_{\textrm{PCR}}(\hat{\gamma}_{d}) \mathrm{dir} ( \mathcal{L}_{d})  + {\hat{u}} \right)= 
        \hat{\gamma}^{(m)}_d 
    \right\rbrace, \\
    & \quad \mathrm{dir} \left\lbrace 
       \mathcal{L}_{d}
    \right\rbrace 
    = \mathrm{dir} \left\lbrace 
      \mathcal{L}_{d}
    \right\rbrace, 
     \, \hat{U} = 
    {\hat{u}}  \bigg).
\end{align*}
The latter two conditions can be dropped from the conditioning set because of the independence properties established in Lemmas \ref{lemma-wind-pcr} and \ref{lemma-dirind-pcr}. So we obtain
\begin{align*}
	&=  \Pr_{H_0} \bigg( W_{\textrm{PCR}}(\hat{\gamma}_d) \in E \mid  \bigcap_{m=0}^M  \left\lbrace\   \hat{\gamma}^{(m)} \left(  \sqrt{ W_{\textrm{PCR}}(\hat{\gamma}_d) } \mathbb{P}_{\textrm{PCR}}(\hat{\gamma}_{d}) \mathrm{dir} ( \mathcal{L}_{d})  + {\hat{u}} \right)= 
        \hat{\gamma}^{(m)}_d 
    \right\rbrace   \bigg) \\
	&= \Pr_{H_0} \bigg( \phi^2 \in E \mid  \left\lbrace\   \hat{\gamma}^{(m)} \left(   {s(\phi)} \right)= 
        \hat{\gamma}^{(m)}_d 
    \right\rbrace  \bigg) \\
	&= \Pr_{H_0} \bigg( \phi^2 \in E \mid \mathcal{T}_{\textrm{PCR}} \bigg)
\end{align*}
where $\phi^2 \sim \chi^2_{q}$.

\subsection{Proof of Theorem \ref{thm:pcr_asymptotic}}

We first introduce various notations. 
Let $\Gamma = \{ s \mid \bigcap_{m=0}^M \lbrace \hat{\gamma}^{(m)} (s) = \hat{\gamma}^{(m)}_d  \rbrace \}$. We also note that $\mathcal{T}_{\textrm{PCR}}$ can be written as a union of intervals whose upper and lower bounds depend on $S$ so that we write $\mathcal{T}_{\textrm{PCR}} = \cup_{j=1}^{J} [\phi_{jl} (S), \phi_{ju} (S)]$. Let $S^* \sim N( \mathbb{X}'\mathbb{X}B, \sigma^2 \mathbb{X}'\mathbb{X})$. We also write $W_{\textrm{PCR}}$ as $W_{\textrm{PCR}} (S)$ to clarify the dependence of $W_{\textrm{PCR}}$ on $S$.

Using this notation, we write
\begin{align*}
    \Pr\left(F_{\textrm{PCR}} (W_{\textrm{PCR}}(S)) \geq 1- \alpha 
    \; \middle | \;
    \bigcap_{m=0}^M \lbrace \hat{\gamma}^{(m)}_D = \hat{\gamma}^{(m)}_d  \rbrace \right)
    &= \Pr (F_{\textrm{PCR}} (W_{\textrm{PCR}}(S)) \geq 1- \alpha  \mid S \in \Gamma ) \\
    &= \frac{\Pr (F_{\textrm{PCR}} (W_{\textrm{PCR}}(S)) \geq 1- \alpha ,  S \in \Gamma ) }{\Pr ( S \in \Gamma )}.
\end{align*}

We first show that $\Pr (S \in \Gamma ) \to \Pr ( S^* \in \Gamma ) $. This holds by the assumption that $S \to_d S^*$ and the observation from Section \ref{sec:s_pcr_set} that the probability measure of $S^*$ on $ \partial \Gamma$ in the space of $\mathbb{R}^{NK}$ is zero. 

Next, we examine $\Pr (F_{\textrm{PCR}} (W_{\textrm{PCR}}(S)) \geq 1- \alpha ,  S \in \Gamma ) $. Let $\chi^2 (\cdot) $ be the cumulative distribution function of $\chi^2_{q}$.
We observe that 
\begin{align*}
    F_{\textrm{PCR}} (W_{\textrm{PCR}}(S))  = \frac{\sum_{j=1}^{J} \max (0, \min (\chi^2 (W_{\textrm{PCR}}(S)), \chi^2 (\phi_{ju}^2 ({S})) ) - \chi^2 (\phi_{jl}^2 ({S}) )) }{ \sum_{j=1}^{J}( \chi^2 (\phi_{ju}^2 ({S}) ) - \chi^2 (\phi_{jl}^2 ({S}) )) }.
\end{align*}
We observe that the max and min functions and $\chi^2$ are continuous. The observation from Section \ref{sec:s_pcr_set} implies that $\phi_{jl}$ and $\phi_{ju}$ are continuous. Thus, we can apply the continuous mapping theorem that 
\begin{align*}
    \Pr (F_{\textrm{PCR}} (W_{\textrm{PCR}}(S)) \geq 1- \alpha ,  S \in \Gamma ) \to 
    \Pr (F_{\textrm{PCR}} (W_{\textrm{PCR}} (S^*)) \geq 1- \alpha ,  S^* \in \Gamma ) .
\end{align*}

It thus holds that
\begin{align*}
        \Pr\left(F_{\textrm{PCR}} (W_{\textrm{PCR}}(S)) \geq 1- \alpha 
    \; \middle | \;
    \bigcap_{m=0}^M \lbrace \hat{\gamma}^{(m)}_D = \hat{\gamma}^{(m)}_d  \rbrace \right)
\to & \frac{ \Pr (F_{\textrm{PCR}} (W_{\textrm{PCR}} (S^*)) \geq 1- \alpha ,  S^* \in \Gamma )}{\Pr ( S^* \in \Gamma )} \\
    &= \Pr (F_{\textrm{PCR}} (W_{\textrm{PCR}} (S^*)) \geq 1- \alpha \mid  S^* \in \Gamma ) \\
    &= \alpha,
\end{align*}
where the last equality follows from Theorem \ref{thm:pcr}.

\section{Technical appendix for TSK}

This appendix provides the proof of Theorem \ref{thm:tsk}. We first establish several lemmas. Then, we present the proof of Theorem \ref{thm:tsk}.

\subsection{Lemmas}

The following lemmas establish the independence conditions needed to prove the main theorem. We fix $\gamma$, and the results below hold. Note that we eventually set $\gamma$ to $\tilde{\gamma}_{d}$.

\begin{lemma}
	\label{lemma-wind}
 Suppose that $\hat{B} \sim N(B, \sigma^2 I_N \otimes \Sigma^{-1})$ and $\gamma$ is fixed. Then, it holds that 
	$
		 R \tilde{\theta}_{D} \;\bot\; \tilde{U}.
	$.
\end{lemma}

\begin{proof}
	Note that
$
R\tilde{\theta}_{D}
=
R(\widetilde{\mathbb{H}}'\widetilde{\mathbb{H}})^{-1}\widetilde{\mathbb{H}}'\hat{B}
$,
and
\begin{align*}
\tilde{U}
&=
\widetilde{\mathbb{H}}\left[
\tilde{\theta}_{D}
-
\widetilde{\mathcal{N}}_{\Sigma}^{-1}R'
\left(
R\widetilde{\mathcal{N}}_{\Sigma}^{-1}R'
\right)^{-1}
\left(
R\tilde{\theta}_{D}-r
\right)
\right]
\\
&\quad
+
\left[
I_{NK}
-
\widetilde{\mathbb{H}}(\widetilde{\mathbb{H}}'\widetilde{\mathbb{H}})^{-1}
\widetilde{\mathbb{H}}'
\right]\hat{B}
\\
&=
\Bigg[
I_{NK}
-
\widetilde{\mathbb{H}}\Bigg(
I_{GK}
-
\widetilde{\mathcal{N}}_{\Sigma}^{-1}R'
\left(
R\widetilde{\mathcal{N}}_{\Sigma}^{-1}R'
\right)^{-1}R
\Bigg)
(\widetilde{\mathbb{H}}'\widetilde{\mathbb{H}})^{-1}\widetilde{\mathbb{H}}'
\widetilde{\mathbb{H}}(\widetilde{\mathbb{H}}'\widetilde{\mathbb{H}})^{-1}
\widetilde{\mathbb{H}}'
\Bigg]\hat{B}
\\
&\quad
+
\widetilde{\mathbb{H}}\widetilde{\mathcal{N}}_{\Sigma}^{-1}R'
\left(
R\widetilde{\mathcal{N}}_{\Sigma}^{-1}R'
\right)^{-1}r.
\end{align*}
The last term, $\widetilde{\mathbb{H}}  \widetilde{\mathcal{N}}_{\Sigma}^{-1} R' (R \widetilde{\mathcal{N}}_{\Sigma}^{-1} R')^{-1} r$, is a constant and thus does not affect the independence. 

Because $ \hat{B} \sim N(B, \sigma^2 I_N \otimes \Sigma^{-1})$, by the properties of the multivariate normal distributions, it is sufficient to show that 
\begin{align*}
&\bigg(  I_{NK} - \widetilde{\mathbb{H}} (I_{GK} - \widetilde{\mathcal{N}}_{\Sigma}^{-1} R' (R \widetilde{\mathcal{N}}_{\Sigma}^{-1} R')^{-1} R)  (\widetilde{\mathbb{H}}'\widetilde{\mathbb{H}})^{-1}\widetilde{\mathbb{H}}')  
\widetilde{\mathbb{H}}  (\widetilde{\mathbb{H}}'\widetilde{\mathbb{H}})^{-1} 
    \widetilde{\mathbb{H}}' \bigg) \\
    & \times 
    (I_N \otimes \Sigma^{-1})
    \widetilde{\mathbb{H}} (\widetilde{\mathbb{H}}'\widetilde{\mathbb{H}})^{-1} R' =0.
\end{align*}
	We observe that 
 	\begin{align*}
&\bigg(  I_{NK} - \widetilde{\mathbb{H}} (I_{GK} - \widetilde{\mathcal{N}}_{\Sigma}^{-1} R' (R \widetilde{\mathcal{N}}_{\Sigma}^{-1} R')^{-1} R)  (\widetilde{\mathbb{H}}'\widetilde{\mathbb{H}})^{-1}\widetilde{\mathbb{H}}')  
\widetilde{\mathbb{H}}  (\widetilde{\mathbb{H}}'\widetilde{\mathbb{H}})^{-1} 
    \widetilde{\mathbb{H}}' \bigg) \\
    & \quad \times 
    (I_N \otimes \Sigma^{-1})
    \widetilde{\mathbb{H}} (\widetilde{\mathbb{H}}'\widetilde{\mathbb{H}})^{-1} R' \\
    &= (I_N \otimes \Sigma^{-1})\widetilde{\mathbb{H}} (\widetilde{\mathbb{H}}'\widetilde{\mathbb{H}})^{-1} R'  \\
    & \quad - \widetilde{\mathbb{H}}  (\widetilde{\mathbb{H}}'\widetilde{\mathbb{H}})^{-1}\widetilde{\mathbb{H}}' (I_N \otimes \Sigma^{-1})\widetilde{\mathbb{H}} (\widetilde{\mathbb{H}}'\widetilde{\mathbb{H}})^{-1} R' \\
    & \quad + \widetilde{\mathbb{H}} \widetilde{\mathcal{N}}_{\Sigma}^{-1} R' (R \widetilde{\mathcal{N}}_{\Sigma}^{-1} R')^{-1} R  (\widetilde{\mathbb{H}}' \widetilde{\mathbb{H}} )^{-1} \widetilde{\mathbb{H}}'(I_N \otimes \Sigma^{-1}) \widetilde{\mathbb{H}}(\widetilde{\mathbb{H}}' \widetilde{\mathbb{H}} )^{-1}  R' \\
    & \quad -  
    \widetilde{\mathbb{H}}     (\widetilde{\mathbb{H}}'\widetilde{\mathbb{H}})^{-1} 
    \widetilde{\mathbb{H}}' (I_N \otimes \Sigma^{-1})\widetilde{\mathbb{H}} (\widetilde{\mathbb{H}}'\widetilde{\mathbb{H}})^{-1} R'\\
  &= (I_N \otimes \Sigma^{-1})\widetilde{\mathbb{H}} (\widetilde{\mathbb{H}}'\widetilde{\mathbb{H}})^{-1} R'  
  - \widetilde{\mathbb{H}}  \widetilde{\mathcal{N}}_{\Sigma}^{-1} R' \\
    & \quad + \widetilde{\mathbb{H}} \widetilde{\mathcal{N}}_{\Sigma}^{-1} R' (R \widetilde{\mathcal{N}}_{\Sigma}^{-1} R')^{-1} R  \widetilde{\mathcal{N}}_{\Sigma}^{-1} R'
  - \widetilde{\mathbb{H}}  \widetilde{\mathcal{N}}_{\Sigma}^{-1} R' \\
    & = \widetilde{\mathbb{H}}  \widetilde{\mathcal{N}}_{\Sigma}^{-1} R' -\widetilde{\mathbb{H}}  \widetilde{\mathcal{N}}_{\Sigma}^{-1} R'  +\widetilde{\mathbb{H}}  \widetilde{\mathcal{N}}_{\Sigma}^{-1} R' - \widetilde{\mathbb{H}}  \widetilde{\mathcal{N}}_{\Sigma}^{-1} R' = 0,   
	\end{align*}
	where the second equality is based on the fact that $(I_N \otimes \Sigma^{-1})\widetilde{\mathbb{H}} (\widetilde{\mathbb{H}}'\widetilde{\mathbb{H}})^{-1} = \widetilde{\mathbb{H}}  \widetilde{\mathcal{N}}_{\Sigma}^{-1}$ and 
    \begin{align*}
      (\widetilde{\mathbb{H}}'\widetilde{\mathbb{H}})^{-1} \widetilde{\mathbb{H}}'(I_N \otimes \Sigma^{-1}) \widetilde{\mathbb{H}} (\widetilde{\mathbb{H}}'\widetilde{\mathbb{H}})^{-1} = \widetilde{\mathcal{N}}_{\Sigma}^{-1}.  
    \end{align*}
\end{proof}
Note that this lemma holds regardless of whether the null hypothesis holds.

The following lemma requires the null hypothesis to be true, unlike the above lemma.
\begin{lemma} 
	\label{lemma-dirind}
	Suppose that $\hat{B} \sim N(B, \sigma^2 I_N \otimes \Sigma^{-1})$ and $\gamma$ is fixed. Under $H_0$,
	\begin{align*}
		(R\tilde{\theta}_{D} - r)' (R\widetilde{\mathcal{N}}_{\Sigma}^{-1}R')^{-1} (R\tilde{\theta}_{D} - r)
		\;\bot\;
		\mathrm{dir} [
        (R\widetilde{\mathcal{N}}_{\Sigma}^{-1}R')^{-1/2} ( R\tilde{\theta}_{D} - r )
    ]
	\end{align*}
\end{lemma}

\begin{proof}
	Because 
	$
		 R\tilde{\theta}_{D} - r \sim N(0,  \sigma^2 R\widetilde{\mathcal{N}}_{\Sigma}^{-1}R')
	$
	under $H_0$, it holds that 
	$
		 (R\widetilde{\mathcal{N}}_{\Sigma}^{-1}R')^{-1/2} (R\tilde{\theta}_{D} - r) \sim N (0, \sigma^2 I_q )
	$.
Thus, by Proposition 4.11 and Corollary 4.3 of \cite{bilodeau1999theory}, its norm and direction are independent.
\end{proof}

\subsection{Proof of Theorem \ref{thm:tsk}}
Let 
$
	\mathcal{J}_{D} (\gamma) = [\sigma^2 R \widetilde{\mathcal{N}}_{\Sigma}^{-1} R']^{-1/2} (R \tilde{\theta}_{D} - r)
$.
Consider an event $W_{\textrm{TSK}} \in E$ for some $E$. We observe that 
\begin{align*}
	&\Pr_{H_0} \left( W_{\textrm{TSK}} \in E \mid \mathcal{C}_{\textrm{TSK}} \right) \\
	&=\Pr_{H_0} \bigg( W_{\textrm{TSK}} \in E \mid 
	    \bigcap_{m=0}^M \left\lbrace 
        \tilde{\gamma}^{(m)}_D = 
        \tilde{\gamma}^{(m)}_d 
    \right\rbrace,  \, \mathrm{dir} \left\lbrace 
       \mathcal{J}_{D} (\tilde{\gamma}_{D})
    \right\rbrace 
    = \mathrm{dir} \left\lbrace 
        \mathcal{J}_{d} (\tilde{\gamma}_{d}) 
    \right\rbrace, 
     \, \tilde{U} = 
    \tilde{u}  \bigg) \\
	&= \Pr_{H_0} \bigg( W_{\textrm{TSK}} \in E \mid 
	    \bigcap_{m=0}^M \left\lbrace 
        \tilde{\gamma}^{(m)}_D = 
        \tilde{\gamma}^{(m)}_d 
    \right\rbrace,  \, \mathrm{dir} \left\lbrace 
       \mathcal{J}_{D} (\tilde{\gamma}_{d})
    \right\rbrace 
    = \mathrm{dir} \left\lbrace 
        \mathcal{J}_{d} (\tilde{\gamma}_{d}) 
    \right\rbrace, 
     \, \tilde{U} = 
    \tilde{u}  \bigg).
\end{align*}
The first equality follows from the definition of $\mathcal{C}_{\textrm{TSK}}$. The second equality uses the fact that $\tilde{\gamma}_D =\tilde{\gamma}_d$ in this conditioning set.

We use the decomposition:
$
	 \hat{B} = \mathbb{Q}_{\textrm{TSK}} (\gamma) \mathcal{J}_{D} (\gamma)
	 + {\tilde{U}(\gamma)}
$,
where 
$
	\mathbb{Q}_{\textrm{TSK}} (\gamma) =  \widetilde{\mathbb{H}} \widetilde{\mathcal{N}}_{\Sigma}^{-1} R' [R \widetilde{\mathcal{N}}_{\Sigma}^{-1} R']^{-1/2} \sigma
$.
Note that this decomposition holds for any $\gamma$ so is valid for $\gamma =  \tilde{\gamma}_d$. We write 
\begin{align*}
	\tilde{\gamma}_D^{(m)}
	= \tilde{\gamma}^{(m)} ( \hat{B} )
	 = \tilde{\gamma}^{(m)}  \left(  \mathbb{Q}_{\textrm{TSK}}(\tilde{\gamma}_{d}) \mathcal{J}_{D} (\tilde{\gamma}_{d})  + {\tilde{U}(\tilde{\gamma}_{d})} \right)
 	 = \tilde{\gamma}^{(m)}  \left( W_{\textrm{TSK}}^{1/2}  \mathbb{Q}_{\textrm{TSK}}(\tilde{\gamma}_{d}) \mathrm{dir} ( \mathcal{J}_{D} (\tilde{\gamma}_{d}))  + {\tilde{U}(\tilde{\gamma}_{d})} \right) .
\end{align*}

Using this decomposition, we write
\begin{align*}
&\Pr_{H_0} \left( W_{\textrm{TSK}} \in E \mid \mathcal{C}_{\textrm{TSK}} \right) \\
	&= \Pr_{H_0} \bigg( W_{\textrm{TSK}} \in E \mid \\
& \quad  \bigcap_{m=0}^M  \left\lbrace\   \tilde{\gamma}^{(m)} \left(  W_{\textrm{TSK}}^{1/2}  \mathbb{Q}_{\textrm{TSK}}(\tilde{\gamma}_{d}) \mathrm{dir} ( \mathcal{J}_{D} (\tilde{\gamma}_{d}))  + {\tilde{U}(\tilde{\gamma}_{d})} \right)= 
        \tilde{\gamma}^{(m)}_d 
    \right\rbrace, \mathrm{dir} \left\lbrace 
       \mathcal{J}_{D} (\tilde{\gamma}_{d})
    \right\rbrace 
    = \mathrm{dir} \left\lbrace 
      \mathcal{J}_{d} (\tilde{\gamma}_{d})
    \right\rbrace, 
     \, \tilde{U} = 
    \tilde{u}  \bigg).
\end{align*}
We insert the latter two conditions into the argument of $\tilde{\gamma}^{(m)}$, and the probability becomes
\begin{align*}
	&= \Pr_{H_0} \bigg( W_{\textrm{TSK}} \in E \mid \\
& \quad  \bigcap_{m=0}^M  \left\lbrace\   \tilde{\gamma}^{(m)} \left(  W_{\textrm{TSK}}^{1/2}  \mathbb{Q}_{\textrm{TSK}}(\tilde{\gamma}_{d}) \mathrm{dir} ( \mathcal{J}_{d} (\tilde{\gamma}_{d}))  +               \tilde{u} \right)= 
        \tilde{\gamma}^{(m)}_d 
    \right\rbrace, \mathrm{dir} \left\lbrace 
       \mathcal{J}_{D} (\tilde{\gamma}_{d})
    \right\rbrace 
    = \mathrm{dir} \left\lbrace 
      \mathcal{J}_{d} (\tilde{\gamma}_{d})
    \right\rbrace, 
     \, \tilde{U} = 
    \tilde{u}  \bigg).
\end{align*}
The latter two conditions in the conditioning set can be dropped because of the independence properties established in Lemmas \ref{lemma-wind} and \ref{lemma-dirind}. So we obtain
\begin{align*}
	&=  \Pr_{H_0} \bigg( W_{\textrm{TSK}} \in E \mid \bigcap_{m=0}^M  \left\lbrace\   \tilde{\gamma}^{(m)} \left(  W_{\textrm{TSK}}^{1/2}  \mathbb{Q}_{\textrm{TSK}}(\tilde{\gamma}_{d}) \mathrm{dir} ( \mathcal{J}_{d} (\tilde{\gamma}_{d}))  + {\tilde{u}} \right)= 
        \tilde{\gamma}^{(m)}_d 
    \right\rbrace   \bigg) \\
	&= \Pr_{H_0} \bigg( \phi^2 \in E \mid  \left\lbrace\   \tilde{\gamma}^{(m)} \left(   \hat{b}{(\phi)} \right)= 
        \tilde{\gamma}^{(m)}_d 
    \right\rbrace  \bigg) \\
	&= \Pr_{H_0} \bigg( \phi^2 \in E \mid \mathcal{T}_{\textrm{TSK}} \bigg)
\end{align*}
where $\phi^2 \sim \chi^2_{q}$.

\subsection{Proof of Theorem \ref{thm:tsk_asymptotic}}

The proof is almost identical to that of Theorem \ref{thm:pcr_asymptotic} and is thus omitted.

\subsection{Weighted KMeans under heteroskedasticity}\label{sec:weighted-exact}

Remark~\ref{rem:wkmeans} claims that, under inverse-variance weighting, the conditional distribution of the Wald statistic remains exactly truncated $\chi^2_q$ even when the unit-specific estimators are heteroskedastic with unequal designs. The exact theory of Section~\ref{sec:simple_KMeans_tests} was established for the homoskedastic case $\operatorname{Var}(\hat{B}) = \sigma^2 (I_N \otimes \Sigma^{-1})$, which underlies Lemma~\ref{lemma-wind}. This section establishes the independence result in the general form required by the remark.

Let
$
  \hat{B} \sim N(B, V)
  $,
  $
  V = \operatorname{bdiag}(V_1, \dots, V_N)
$,
where each $V_i$ is an $K \times K$ symmetric positive definite matrix. Let $\Phi = \operatorname{bdiag}(\Phi_1, \dots, \Phi_N)$ be a block-diagonal weighting matrix with each $\Phi_i$ symmetric positive definite. The weighted TSK estimator of the group-specific coefficients is
\begin{equation}
  \tilde{\theta}_D = (\widetilde{\mathbb{H}}'\Phi \widetilde{\mathbb{H}})^{-1} \widetilde{\mathbb{H}}'\Phi \hat{B},
\end{equation}
with variance
\begin{equation}
  \operatorname{Var}(R\tilde{\theta})
  = R (\widetilde{\mathbb{H}}'\Phi \widetilde{\mathbb{H}})^{-1} \widetilde{\mathbb{H}}'\Phi \, V \, \Phi \widetilde{\mathbb{H}} (\widetilde{\mathbb{H}}'\Phi \widetilde{\mathbb{H}})^{-1} R'.
\end{equation}
Write $A_\Phi = (\widetilde{\mathbb{H}}'\Phi \widetilde{\mathbb{H}})^{-1} \widetilde{\mathbb{H}}'\Phi$ for the weighted projector, so that
$\tilde{\theta}_D = A_\Phi \hat{B}$ and $R\tilde{\theta}_D = R A_\Phi \hat{B}$. The constrained estimator and the nuisance component are defined exactly as in Section~\ref{sec:simple_KMeans_tests}, with $N_\Sigma(\gamma)^{-1}$ replaced throughout by
$(\widetilde{\mathbb{H}}'\Phi \widetilde{\mathbb{H}})^{-1}$:
\begin{align}
  \tilde{\theta}_{R,D}
  &= \tilde{\theta}_D
     - (\widetilde{\mathbb{H}}'\Phi \widetilde{\mathbb{H}})^{-1} R' \bigl( R (\widetilde{\mathbb{H}}'\Phi \widetilde{\mathbb{H}})^{-1} R' \bigr)^{-1}
       (R\tilde{\theta}_D - r), \\
  \tilde{U}
  &= \widetilde{\mathbb{H}} \tilde{\theta}_{R,D}
     + \bigl( I_{NK} - \widetilde{\mathbb{H}} (\widetilde{\mathbb{H}}'\Phi \widetilde{\mathbb{H}})^{-1} \widetilde{\mathbb{H}}'\Phi \bigr) \hat{B}
   = \widetilde{\mathbb{H}} \tilde{\theta}_{R,D} + (I_{NK} - \widetilde{\mathbb{H}} A_\Phi) \hat{B}.
\end{align}

The following lemma is the weighted analog of Lemma~\ref{lemma-wind}.

\begin{lemma}\label{lem:tsk-weighted-independence}
  Suppose $\hat{B} \sim N(B, V)$ with $V = \operatorname{bdiag}(V_1, \dots, V_N)$, and
  $\gamma$ is fixed. If the weighting matrix is chosen as the inverse variance,
  $\Phi = V^{-1}$, then
  \begin{equation*}
    R\tilde{\theta}_D \perp \tilde{U}.
  \end{equation*}
\end{lemma}

\begin{proof}
  Both $R\tilde{\theta}_D$ and $\tilde{U}$ are affine functions of $\hat{B}$:
  \begin{equation*}
    R\tilde{\theta}_D = R A_\Phi \hat{B}, \quad \tilde{U} = C \hat{B} + c,
  \end{equation*}
  where, collecting the terms in the definition of $\tilde{U}$,
  \begin{equation*}
    C = (I_{NK} - \widetilde{\mathbb{H}} A_\Phi)
        + \widetilde{\mathbb{H}} \Bigl( I_{GK}
          - (\widetilde{\mathbb{H}}'\Phi \widetilde{\mathbb{H}})^{-1} R' \bigl( R (\widetilde{\mathbb{H}}'\Phi \widetilde{\mathbb{H}})^{-1} R' \bigr)^{-1} R \Bigr) A_\Phi,
  \end{equation*}
  and $c = \widetilde{\mathbb{H}} (\widetilde{\mathbb{H}}'\Phi \widetilde{\mathbb{H}})^{-1} R' \bigl( R (\widetilde{\mathbb{H}}'\Phi \widetilde{\mathbb{H}})^{-1} R' \bigr)^{-1} r$ is a constant that does not affect independence. Simplifying $C$,
  \begin{equation*}
    C = I_{NK}
        - \widetilde{\mathbb{H}} (\widetilde{\mathbb{H}}'\Phi \widetilde{\mathbb{H}})^{-1} R' \bigl( R (\widetilde{\mathbb{H}}'\Phi \widetilde{\mathbb{H}})^{-1} R' \bigr)^{-1} R A_\Phi.
  \end{equation*}

  Since $\hat{B}$ is Gaussian, $R\tilde{\theta}_D$ and $\tilde{U}$ are jointly Gaussian, so the independence is equivalent to the vanishing of their cross-covariance:
  \begin{equation*}
    \operatorname{Cov}(R\tilde{\theta}_D, \tilde{U}) = R A_\Phi \, V \, C' = 0.
  \end{equation*}
  It therefore suffices to show $R A_\Phi V C' = 0$, or equivalently   $C V A_\Phi' R' = 0$. We compute $C V A_\Phi'$:
  \begin{equation*}
    C V A_\Phi'
    = \Bigl( I_{NK}
        - \widetilde{\mathbb{H}} (\widetilde{\mathbb{H}}'\Phi \widetilde{\mathbb{H}})^{-1} R' \bigl( R (\widetilde{\mathbb{H}}'\Phi \widetilde{\mathbb{H}})^{-1} R' \bigr)^{-1} R A_\Phi
      \Bigr) V A_\Phi'.
  \end{equation*}

  The key simplification uses the inverse-variance choice $\Phi = V^{-1}$. With this choice, the weighted projector collapses, because
  \begin{equation*}
    A_\Phi V A_\Phi'
    = (\widetilde{\mathbb{H}}'V^{-1}\widetilde{\mathbb{H}})^{-1} \widetilde{\mathbb{H}}'V^{-1} \, V \, V^{-1} \widetilde{\mathbb{H}} (\widetilde{\mathbb{H}}'V^{-1}\widetilde{\mathbb{H}})^{-1}
    = (\widetilde{\mathbb{H}}'V^{-1}\widetilde{\mathbb{H}})^{-1},
  \end{equation*}
  which is exactly $\operatorname{Var}(\tilde{\theta}_D)$, and
  \begin{equation*}
    V A_\Phi' = V \, V^{-1} \widetilde{\mathbb{H}} (\widetilde{\mathbb{H}}'V^{-1}\widetilde{\mathbb{H}})^{-1} = \widetilde{\mathbb{H}} (\widetilde{\mathbb{H}}'V^{-1}\widetilde{\mathbb{H}})^{-1}.
  \end{equation*}
  Substituting the latter,
  \begin{align*}
    C V A_\Phi'
    &= V A_\Phi'
       - \widetilde{\mathbb{H}} (\widetilde{\mathbb{H}}'V^{-1}\widetilde{\mathbb{H}})^{-1} R' \bigl( R (\widetilde{\mathbb{H}}'V^{-1}\widetilde{\mathbb{H}})^{-1} R' \bigr)^{-1}
         R \, A_\Phi V A_\Phi' \\
    &= \widetilde{\mathbb{H}} (\widetilde{\mathbb{H}}'V^{-1}\widetilde{\mathbb{H}})^{-1}
       - \widetilde{\mathbb{H}} (\widetilde{\mathbb{H}}'V^{-1}\widetilde{\mathbb{H}})^{-1} R' \bigl( R (\widetilde{\mathbb{H}}'V^{-1}\widetilde{\mathbb{H}})^{-1} R' \bigr)^{-1}
         R (\widetilde{\mathbb{H}}'V^{-1}\widetilde{\mathbb{H}})^{-1},
  \end{align*}
  where the second equality follows from $A_\Phi V A_\Phi' = (\widetilde{\mathbb{H}}'V^{-1}\widetilde{\mathbb{H}})^{-1}$. Postmultiplying by
  $R'$,
  \begin{align*}
    C V A_\Phi' R'
    &= \widetilde{\mathbb{H}} (\widetilde{\mathbb{H}}'V^{-1}\widetilde{\mathbb{H}})^{-1} R' \\
    &\quad - \widetilde{\mathbb{H}} (\widetilde{\mathbb{H}}'V^{-1}\widetilde{\mathbb{H}})^{-1} R'
       \bigl( R (\widetilde{\mathbb{H}}'V^{-1}\widetilde{\mathbb{H}})^{-1} R' \bigr)^{-1} R (\widetilde{\mathbb{H}}'V^{-1}\widetilde{\mathbb{H}})^{-1} R' \\
    &= \widetilde{\mathbb{H}} (\widetilde{\mathbb{H}}'V^{-1}\widetilde{\mathbb{H}})^{-1} R' - \widetilde{\mathbb{H}} (\widetilde{\mathbb{H}}'V^{-1}\widetilde{\mathbb{H}})^{-1} R' \\
    &= 0.
  \end{align*}
 Hence, $\operatorname{Cov}(R\tilde{\theta}_D, \tilde{U}) = R A_\Phi V C'  = (C V A_\Phi' R')' = 0$, and the two jointly Gaussian quantities are independent.
\end{proof}

This lemma reduces to Lemma~\ref{lemma-wind} when $V = \sigma^2 (I_N \otimes \Sigma^{-1})$ and $\Phi = V^{-1} = \sigma^{-2} (I_N \otimes \Sigma)$: in that case $\widetilde{\mathbb{H}}'\Phi \widetilde{\mathbb{H}} = \sigma^{-2} N_\Sigma(\gamma)$ and the weighted projector coincides with $(\widetilde{\mathbb{H}}'\widetilde{\mathbb{H}})^{-1} \widetilde{\mathbb{H}}'$ up to the cancellation of $\Sigma$, recovering the identity $(\widetilde{\mathbb{H}}'\widetilde{\mathbb{H}})^{-1} \widetilde{\mathbb{H}}' (I_N \otimes \Sigma^{-1}) \widetilde{\mathbb{H}} (\widetilde{\mathbb{H}}'\widetilde{\mathbb{H}})^{-1} = N_\Sigma(\gamma)^{-1}$ used there.

The independence in Lemma~\ref{lem:tsk-weighted-independence} is the only place where the homoskedastic structure entered the proof of Theorem~\ref{thm:tsk}; the norm-direction decomposition (the analog of Lemma~\ref{lemma-dirind}) requires only that $R\tilde{\theta}_D - r$ be Gaussian with covariance $R (\widetilde{\mathbb{H}}' V^{-1}\widetilde{\mathbb{H}})^{-1} R'$ under $H_0$, which continues to hold. Consequently, with $\Phi = V^{-1}$, the conditional distribution of the weighted Wald statistic is exactly truncated $\chi^2_q$, without requiring a common error variance or a common design across units, as claimed in Remark~\ref{rem:wkmeans}.

For a generic positive definite $\Phi \neq V^{-1}$, the collapse $A_\Phi V A_\Phi' = (\widetilde{\mathbb{H}}'\Phi \widetilde{\mathbb{H}})^{-1}$ fails: an extra factor $\Phi V \Phi$ survives in $A_\Phi V A_\Phi'$, the cross-covariance does not vanish, and the exact independence is lost. In that case, the procedure is justified only asymptotically, under consistency of the estimated weights, which is the fallback stated in Remark~\ref{rem:wkmeans}.

\section{Additional Monte Carlo Results}\label{sec:additional_sim}

This section reports additional Monte Carlo evidence for specifications with fixed effects and for the GFE estimator. The design is the same as in Section~\ref{sec:mc_design}: \(N=120\), \(T\in\{20,50\}\), 1000 replications, two true groups with \(g_i=1\) for \(i\leq 40\) and \(g_i=2\) for \(i>40\), and the same three slope DGPs, corresponding to no separation, partial separation, and full separation. The model is
\[
Y_{it}=X_{it}'\theta_{g_i}+\xi_{it}+\varepsilon_{it},
\quad i=1,\dots,N,\quad t=1,\dots,T.
\]

We consider three cases for the intercept component \(\xi_{it}\). Case 1 is the baseline of the main text, with \(\xi_{it}=0\). Case 2 introduces unit-specific heterogeneity, \(\xi_{it}=\mu_i\) with \(\mu_i\sim N(0,0.5^2)\). Case 3 adds grouped time effects, \(\xi_{it}=\mu_i+\eta_{g_i t}\) with
\[
\eta_{1t}=0.8\sin(2\pi t/T),
\quad
\eta_{2t}=2+\sin(2\pi t/T+\pi/4).
\]
The first two cases preserve the slope-based group-separation structure. In contrast,  Case 3 creates systematic time-varying intercept differences across groups, so that the groups may remain separated even when the slope parameters are identical.

The regressor and error processes, the null hypotheses \(H_{0,1}\), \(H_{0,2}\), and \(H_{0,3}\), and the comparison between the predetermined-group test, the naive post-clustering tests, and the proposed conditional tests are all as described in Section~\ref{sec:mc_design}. TSK, PCR, and GFE are implemented with \(G=2\); for TSK and PCR, the fixed-effect cases are handled by applying the relevant within-transformations before clustering and estimation, while GFE directly estimates grouped time effects and is therefore particularly relevant in Case 3. The tables report rejection frequencies for all combinations of experiment, case, DGP, hypothesis, and estimator, and coverage and average length of confidence intervals for the scalar null \(H_{0,2}\).

Tables~\ref{tab:exp1_case2_rejection} and~\ref{tab:exp1_case3_rejection} report rejection frequencies for the iid Gaussian design. Case 2 results are very close to the baseline of the main text: under DGP1, the naive TSK and PCR tests reject almost always, naive GFE over-rejects with frequencies between 0.35 and 0.67, and the conditional tests range from 0.05 to 0.07. Case 3 is more demanding, since grouped time effects create systematic intercept differences across groups even when the slopes are homogeneous, and the predetermined-group benchmark is itself oversized under DGP1, with rejection frequencies of 0.17 and 0.15 for \(H_{0,1}\). The conditional TSK and PCR tests reduce the over-rejection substantially but remain mildly oversized, at around 0.10 to 0.14, whereas the conditional GFE test stays close to 0.05 under true nulls. This is expected, since GFE directly models grouped time effects and is therefore better aligned with the Case 3 data-generating process. Power remains high throughout, especially as \(T\) increases.

\begin{table}
\centering
\scriptsize
\begin{threeparttable}
\caption{Rejection rates of naive tests and proposed tests under different null hypotheses \\
Experiment 1: iid Gaussian design \\
Case 2: Unit fixed effects}
\label{tab:exp1_case2_rejection}
\begin{tabular*}{\textwidth}{@{\extracolsep{\fill}} ccc ccccccc}
\toprule
$T$ & Test & DGP & \shortstack{Predetermined} & \shortstack{Naive\\TSK} & \shortstack{Naive\\PCR} & \shortstack{Naive\\GFE} & \shortstack{Conditional\\TSK} & \shortstack{Conditional\\PCR} & \shortstack{Conditional\\GFE} \\
\midrule
\multicolumn{10}{l}{\textbf{Panel (a): Size}} \\
\midrule
\multicolumn{10}{l}{\emph{DGP1: No group separation}} \\
20 & $H_{0,1}$ & DGP1 & 0.06 & 1.00 & 1.00 & 0.67 & 0.07 & 0.07 & 0.05 \\
20 & $H_{0,2}$ & DGP1 & 0.05 & 1.00 & 0.90 & 0.49 & 0.05 & 0.07 & 0.05 \\
50 & $H_{0,1}$ & DGP1 & 0.05 & 1.00 & 1.00 & 0.51 & 0.06 & 0.07 & 0.07 \\
50 & $H_{0,2}$ & DGP1 & 0.05 & 1.00 & 0.88 & 0.35 & 0.07 & 0.06 & 0.06 \\
\midrule
\multicolumn{10}{l}{\emph{DGP2: Partial group separation}} \\
20 & $H_{0,2}$ & DGP2 & 0.06 & 0.10 & 0.09 & 0.07 & 0.07 & 0.07 & 0.06 \\
50 & $H_{0,2}$ & DGP2 & 0.06 & 0.06 & 0.06 & 0.06 & 0.06 & 0.06 & 0.05 \\
\midrule
\multicolumn{10}{l}{\textbf{Panel (b): Power}} \\
\midrule
\multicolumn{10}{l}{\emph{DGP1: No group separation}} \\
20 & $H_{0,3}$ & DGP1 & 1.00 & 1.00 & 1.00 & 1.00 & 0.94 & 0.95 & 0.97 \\
50 & $H_{0,3}$ & DGP1 & 1.00 & 1.00 & 1.00 & 1.00 & 0.97 & 0.98 & 0.99 \\
\midrule
\multicolumn{10}{l}{\emph{DGP2: Partial group separation}} \\
20 & $H_{0,1}$ & DGP2 & 1.00 & 1.00 & 1.00 & 1.00 & 0.95 & 0.97 & 0.97 \\
20 & $H_{0,3}$ & DGP2 & 1.00 & 1.00 & 1.00 & 1.00 & 0.99 & 1.00 & 1.00 \\
50 & $H_{0,1}$ & DGP2 & 1.00 & 1.00 & 1.00 & 1.00 & 1.00 & 1.00 & 0.99 \\
50 & $H_{0,3}$ & DGP2 & 1.00 & 1.00 & 1.00 & 1.00 & 1.00 & 1.00 & 1.00 \\
\midrule
\multicolumn{10}{l}{\emph{DGP3: Full group separation}} \\
20 & $H_{0,1}$ & DGP3 & 1.00 & 1.00 & 1.00 & 1.00 & 0.98 & 0.99 & 0.98 \\
20 & $H_{0,2}$ & DGP3 & 1.00 & 1.00 & 1.00 & 1.00 & 0.94 & 0.95 & 0.94 \\
20 & $H_{0,3}$ & DGP3 & 1.00 & 1.00 & 1.00 & 1.00 & 1.00 & 1.00 & 1.00 \\
50 & $H_{0,1}$ & DGP3 & 1.00 & 1.00 & 1.00 & 1.00 & 1.00 & 1.00 & 1.00 \\
50 & $H_{0,2}$ & DGP3 & 1.00 & 1.00 & 1.00 & 1.00 & 0.99 & 0.99 & 0.98 \\
50 & $H_{0,3}$ & DGP3 & 1.00 & 1.00 & 1.00 & 1.00 & 1.00 & 1.00 & 1.00 \\
\bottomrule
\end{tabular*}
\begin{tablenotes}[flushleft]
\footnotesize
\item \textit{Notes:} The table reports rejection frequencies. $H_{0,1}$ tests equality of all group-specific slopes, $H_{0,2}$ tests equality of a subset of slopes, and $H_{0,3}$ is false in the designs reported in Panel (b), so rejection frequencies in that panel measure power. ``Predetermined'' uses the true group structure. ``Naive'' tests condition on the estimated group structure but do not account for group selection uncertainty. ``Conditional'' tests apply the proposed selective inference correction. Case 2 includes additive unit effects, which are removed before estimation according to the transformations described in the Monte Carlo design. Since Experiment 1 is an iid Gaussian design, all test statistics are computed using the iid variance estimator with the corresponding finite-sample degrees of freedom correction.
\end{tablenotes}
\end{threeparttable}
\end{table}

\begin{table}
\centering
\scriptsize
\begin{threeparttable}
\caption{Rejection rates of naive tests and proposed tests under different null hypotheses \\
Experiment 1: iid Gaussian design \\
Case 3: Unit fixed effects and grouped time effects}
\label{tab:exp1_case3_rejection}
\begin{tabular*}{\textwidth}{@{\extracolsep{\fill}} ccc ccccccc}
\toprule
$T$ & Test & DGP & \shortstack{Predetermined} & \shortstack{Naive\\TSK} & \shortstack{Naive\\PCR} & \shortstack{Naive\\GFE} & \shortstack{Conditional\\TSK} & \shortstack{Conditional\\PCR} & \shortstack{Conditional\\GFE} \\
\midrule
\multicolumn{10}{l}{\textbf{Panel (a): Size}} \\
\midrule
\multicolumn{10}{l}{\emph{DGP1: No group separation}} \\
20 & $H_{0,1}$ & DGP1 & 0.17 & 1.00 & 1.00 & 0.40 & 0.09 & 0.14 & 0.06 \\
20 & $H_{0,2}$ & DGP1 & 0.12 & 1.00 & 0.91 & 0.27 & 0.10 & 0.12 & 0.06 \\
50 & $H_{0,1}$ & DGP1 & 0.15 & 1.00 & 1.00 & 0.12 & 0.10 & 0.13 & 0.04 \\
50 & $H_{0,2}$ & DGP1 & 0.14 & 1.00 & 0.91 & 0.12 & 0.10 & 0.10 & 0.05 \\
\midrule
\multicolumn{10}{l}{\emph{DGP2: Partial group separation}} \\
20 & $H_{0,2}$ & DGP2 & 0.11 & 0.17 & 0.17 & 0.06 & 0.09 & 0.13 & 0.05 \\
50 & $H_{0,2}$ & DGP2 & 0.12 & 0.12 & 0.14 & 0.06 & 0.11 & 0.13 & 0.06 \\
\midrule
\multicolumn{10}{l}{\textbf{Panel (b): Power}} \\
\midrule
\multicolumn{10}{l}{\emph{DGP1: No group separation}} \\
20 & $H_{0,3}$ & DGP1 & 1.00 & 1.00 & 1.00 & 1.00 & 0.93 & 0.95 & 0.97 \\
50 & $H_{0,3}$ & DGP1 & 1.00 & 1.00 & 1.00 & 1.00 & 0.97 & 0.98 & 0.99 \\
\midrule
\multicolumn{10}{l}{\emph{DGP2: Partial group separation}} \\
20 & $H_{0,1}$ & DGP2 & 1.00 & 1.00 & 1.00 & 1.00 & 0.88 & 0.95 & 0.96 \\
20 & $H_{0,3}$ & DGP2 & 1.00 & 1.00 & 1.00 & 1.00 & 0.99 & 1.00 & 1.00 \\
50 & $H_{0,1}$ & DGP2 & 1.00 & 1.00 & 1.00 & 1.00 & 0.99 & 0.99 & 0.99 \\
50 & $H_{0,3}$ & DGP2 & 1.00 & 1.00 & 1.00 & 1.00 & 1.00 & 1.00 & 1.00 \\
\midrule
\multicolumn{10}{l}{\emph{DGP3: Full group separation}} \\
20 & $H_{0,1}$ & DGP3 & 1.00 & 1.00 & 1.00 & 1.00 & 0.96 & 0.98 & 0.99 \\
20 & $H_{0,2}$ & DGP3 & 1.00 & 1.00 & 1.00 & 1.00 & 0.91 & 0.92 & 0.93 \\
20 & $H_{0,3}$ & DGP3 & 1.00 & 1.00 & 1.00 & 1.00 & 1.00 & 1.00 & 1.00 \\
50 & $H_{0,1}$ & DGP3 & 1.00 & 1.00 & 1.00 & 1.00 & 1.00 & 1.00 & 0.99 \\
50 & $H_{0,2}$ & DGP3 & 1.00 & 1.00 & 1.00 & 1.00 & 0.99 & 0.99 & 0.98 \\
50 & $H_{0,3}$ & DGP3 & 1.00 & 1.00 & 1.00 & 1.00 & 1.00 & 1.00 & 1.00 \\
\bottomrule
\end{tabular*}
\begin{tablenotes}[flushleft]
\footnotesize
\item \textit{Notes:} Hypotheses and column definitions are as in Table~\ref{tab:exp1_case2_rejection}. Case 3 includes unit effects and grouped time effects; unit effects are removed before estimation, and the GFE procedure additionally estimates the grouped time effects.
\end{tablenotes}
\end{threeparttable}
\end{table}

Tables~\ref{tab:exp2_case2_rejection} and~\ref{tab:exp2_case3_rejection} report the corresponding results under serial dependence, within-cluster CSD, and non-Gaussian errors. These designs are substantially more challenging, and the predetermined-group benchmark is itself oversized in short panels: under DGP1 and \(T=20\), its rejection frequency for \(H_{0,1}\) is 0.24 in Case 2 and 0.44 in Case 3. The naive TSK and PCR tests remain extremely distorted throughout, rejecting almost always under the no-separation null. The conditional tests nevertheless deliver rejection frequencies between 0.04 and 0.08 under DGP1 in both cases. Under DGP2, the conditional TSK and PCR tests are somewhat oversized when \(T=20\) but improve at \(T=50\), whereas the conditional GFE test remains the most stable procedure in the grouped-time-effect design. Power remains high across alternatives.

\begin{table}
\centering
\scriptsize
\begin{threeparttable}
\caption{Rejection rates of naive tests and proposed tests under different null hypotheses \\
Experiment 2: serial and within-cluster dependence \\
Case 2: Unit fixed effects}
\label{tab:exp2_case2_rejection}
\begin{tabular*}{\textwidth}{@{\extracolsep{\fill}} ccc ccccccc}
\toprule
$T$ & Test & DGP & \shortstack{Predetermined} & \shortstack{Naive\\TSK} & \shortstack{Naive\\PCR} & \shortstack{Naive\\GFE} & \shortstack{Conditional\\TSK} & \shortstack{Conditional\\PCR} & \shortstack{Conditional\\GFE} \\
\midrule
\multicolumn{10}{l}{\textbf{Panel (a): Size}} \\
\midrule
\multicolumn{10}{l}{\emph{DGP1: No group separation}} \\
20 & $H_{0,1}$ & DGP1 & 0.24 & 1.00 & 1.00 & 0.35 & 0.06 & 0.08 & 0.07 \\
20 & $H_{0,2}$ & DGP1 & 0.17 & 1.00 & 0.86 & 0.26 & 0.06 & 0.06 & 0.05 \\
50 & $H_{0,1}$ & DGP1 & 0.12 & 1.00 & 1.00 & 0.24 & 0.05 & 0.05 & 0.05 \\
50 & $H_{0,2}$ & DGP1 & 0.10 & 1.00 & 0.84 & 0.17 & 0.06 & 0.04 & 0.07 \\
\midrule
\multicolumn{10}{l}{\emph{DGP2: Partial group separation}} \\
20 & $H_{0,2}$ & DGP2 & 0.16 & 0.17 & 0.17 & 0.12 & 0.12 & 0.12 & 0.07 \\
50 & $H_{0,2}$ & DGP2 & 0.11 & 0.12 & 0.11 & 0.08 & 0.11 & 0.11 & 0.08 \\
\midrule
\multicolumn{10}{l}{\textbf{Panel (b): Power}} \\
\midrule
\multicolumn{10}{l}{\emph{DGP1: No group separation}} \\
20 & $H_{0,3}$ & DGP1 & 1.00 & 1.00 & 1.00 & 1.00 & 0.88 & 0.92 & 0.98 \\
50 & $H_{0,3}$ & DGP1 & 1.00 & 1.00 & 1.00 & 1.00 & 0.95 & 0.97 & 0.98 \\
\midrule
\multicolumn{10}{l}{\emph{DGP2: Partial group separation}} \\
20 & $H_{0,1}$ & DGP2 & 1.00 & 1.00 & 1.00 & 1.00 & 0.87 & 0.93 & 0.93 \\
20 & $H_{0,3}$ & DGP2 & 1.00 & 1.00 & 1.00 & 1.00 & 0.99 & 0.99 & 0.99 \\
50 & $H_{0,1}$ & DGP2 & 1.00 & 1.00 & 1.00 & 1.00 & 0.99 & 0.99 & 0.98 \\
50 & $H_{0,3}$ & DGP2 & 1.00 & 1.00 & 1.00 & 1.00 & 1.00 & 1.00 & 1.00 \\
\midrule
\multicolumn{10}{l}{\emph{DGP3: Full group separation}} \\
20 & $H_{0,1}$ & DGP3 & 1.00 & 1.00 & 1.00 & 1.00 & 0.96 & 0.98 & 0.97 \\
20 & $H_{0,2}$ & DGP3 & 1.00 & 1.00 & 1.00 & 1.00 & 0.91 & 0.91 & 0.92 \\
20 & $H_{0,3}$ & DGP3 & 1.00 & 1.00 & 1.00 & 1.00 & 0.99 & 0.99 & 1.00 \\
50 & $H_{0,1}$ & DGP3 & 1.00 & 1.00 & 1.00 & 1.00 & 1.00 & 0.99 & 0.99 \\
50 & $H_{0,2}$ & DGP3 & 1.00 & 1.00 & 1.00 & 1.00 & 0.97 & 0.96 & 0.97 \\
50 & $H_{0,3}$ & DGP3 & 1.00 & 1.00 & 1.00 & 1.00 & 1.00 & 1.00 & 1.00 \\
\bottomrule
\end{tabular*}
\begin{tablenotes}[flushleft]
\footnotesize
\item \textit{Notes:} Hypotheses and column definitions are as in Table~\ref{tab:exp1_case2_rejection}. Case 2 includes additive unit effects, which are removed before estimation according to the transformations described in the Monte Carlo design. Since Experiment 2 allows serial dependence and within-cluster contemporaneous dependence, all test statistics use the dependence-robust variance estimator of Section~\ref{sec:variance}, applied to the true partition for the predetermined benchmark and to the estimated partition otherwise.
\end{tablenotes}
\end{threeparttable}
\end{table}

\begin{table}
\centering
\scriptsize
\begin{threeparttable}
\caption{Rejection rates of naive tests and proposed tests under different null hypotheses \\
Experiment 2: serial and within-cluster dependence \\
Case 3: Unit fixed effects and grouped time effects}
\label{tab:exp2_case3_rejection}
\begin{tabular*}{\textwidth}{@{\extracolsep{\fill}} ccc ccccccc}
\toprule
$T$ & Test & DGP & \shortstack{Predetermined} & \shortstack{Naive\\TSK} & \shortstack{Naive\\PCR} & \shortstack{Naive\\GFE} & \shortstack{Conditional\\TSK} & \shortstack{Conditional\\PCR} & \shortstack{Conditional\\GFE} \\
\midrule
\multicolumn{10}{l}{\textbf{Panel (a): Size}} \\
\midrule
\multicolumn{10}{l}{\emph{DGP1: No group separation}} \\
20 & $H_{0,1}$ & DGP1 & 0.44 & 1.00 & 1.00 & 0.31 & 0.06 & 0.07 & 0.07 \\
20 & $H_{0,2}$ & DGP1 & 0.29 & 1.00 & 0.88 & 0.21 & 0.06 & 0.07 & 0.05 \\
50 & $H_{0,1}$ & DGP1 & 0.26 & 1.00 & 1.00 & 0.16 & 0.04 & 0.05 & 0.06 \\
50 & $H_{0,2}$ & DGP1 & 0.17 & 1.00 & 0.85 & 0.09 & 0.05 & 0.05 & 0.06 \\
\midrule
\multicolumn{10}{l}{\emph{DGP2: Partial group separation}} \\
20 & $H_{0,2}$ & DGP2 & 0.29 & 0.27 & 0.30 & 0.13 & 0.13 & 0.16 & 0.09 \\
50 & $H_{0,2}$ & DGP2 & 0.16 & 0.15 & 0.16 & 0.10 & 0.14 & 0.14 & 0.08 \\
\midrule
\multicolumn{10}{l}{\textbf{Panel (b): Power}} \\
\midrule
\multicolumn{10}{l}{\emph{DGP1: No group separation}} \\
20 & $H_{0,3}$ & DGP1 & 1.00 & 1.00 & 1.00 & 1.00 & 0.88 & 0.89 & 0.98 \\
50 & $H_{0,3}$ & DGP1 & 1.00 & 1.00 & 1.00 & 1.00 & 0.93 & 0.94 & 0.99 \\
\midrule
\multicolumn{10}{l}{\emph{DGP2: Partial group separation}} \\
20 & $H_{0,1}$ & DGP2 & 1.00 & 1.00 & 1.00 & 1.00 & 0.79 & 0.88 & 0.94 \\
20 & $H_{0,3}$ & DGP2 & 1.00 & 1.00 & 1.00 & 1.00 & 0.98 & 0.98 & 0.99 \\
50 & $H_{0,1}$ & DGP2 & 1.00 & 1.00 & 1.00 & 1.00 & 0.97 & 0.97 & 0.98 \\
50 & $H_{0,3}$ & DGP2 & 1.00 & 1.00 & 1.00 & 1.00 & 1.00 & 1.00 & 1.00 \\
\midrule
\multicolumn{10}{l}{\emph{DGP3: Full group separation}} \\
20 & $H_{0,1}$ & DGP3 & 1.00 & 1.00 & 1.00 & 1.00 & 0.93 & 0.96 & 0.98 \\
20 & $H_{0,2}$ & DGP3 & 1.00 & 1.00 & 1.00 & 1.00 & 0.83 & 0.87 & 0.92 \\
20 & $H_{0,3}$ & DGP3 & 1.00 & 1.00 & 1.00 & 1.00 & 0.98 & 0.99 & 1.00 \\
50 & $H_{0,1}$ & DGP3 & 1.00 & 1.00 & 1.00 & 1.00 & 0.99 & 0.99 & 0.99 \\
50 & $H_{0,2}$ & DGP3 & 1.00 & 1.00 & 1.00 & 1.00 & 0.95 & 0.96 & 0.98 \\
50 & $H_{0,3}$ & DGP3 & 1.00 & 1.00 & 1.00 & 1.00 & 1.00 & 1.00 & 1.00 \\
\bottomrule
\end{tabular*}
\begin{tablenotes}[flushleft]
\footnotesize
\item \textit{Notes:} Hypotheses, column definitions, and the variance estimator are as in Table~\ref{tab:exp2_case2_rejection}. Case 3 includes unit effects and grouped time effects; unit effects are removed before estimation, and the GFE procedure additionally estimates the grouped time effects.
\end{tablenotes}
\end{threeparttable}
\end{table}

Tables~\ref{tab:exp1_case2_ci} and~\ref{tab:exp1_case3_ci} report coverage and average length for \(H_{0,2}\) in the iid Gaussian design, and mirror the rejection-frequency evidence. Under DGP1, the naive TSK intervals have zero coverage and the naive PCR intervals cover only 0.09 to 0.12 of the time, while the conditional intervals restore coverage to between 0.89 and 0.96. As in the main text, the improvement comes with longer intervals for TSK and PCR: at \(T=20\) in Case 2, the conditional TSK interval averages 0.89, compared with 0.21 for its naive counterpart. In Case 3, GFE coverage is close to or above the nominal level in most designs, reflecting its ability to absorb grouped time effects, and the conditional intervals are again longest under DGP3.

\begin{table}
\centering
\scriptsize
\begin{threeparttable}
\caption{Coverage and average length of confidence intervals for $H_{0,2}$ \\
Experiment 1: iid Gaussian design \\
Case 2: Unit fixed effects}
\label{tab:exp1_case2_ci}
\begin{tabular*}{\textwidth}{@{\extracolsep{\fill}} cc ccccccc}
\toprule
$T$ & DGP & \shortstack{Predetermined} & \shortstack{Naive\\TSK} & \shortstack{Naive\\PCR} & \shortstack{Naive\\GFE} & \shortstack{Conditional\\TSK} & \shortstack{Conditional\\PCR} & \shortstack{Conditional\\GFE} \\
\midrule
\multicolumn{9}{l}{\textbf{Panel (a): Coverage}} \\
\midrule
\multicolumn{9}{l}{\emph{DGP1: No group separation}} \\
20 & DGP1 & 0.95 & 0.00 & 0.11 & 0.52 & 0.95 & 0.91 & 0.96 \\
50 & DGP1 & 0.95 & 0.00 & 0.12 & 0.63 & 0.92 & 0.95 & 0.94 \\
\midrule
\multicolumn{9}{l}{\emph{DGP2: Partial group separation}} \\
20 & DGP2 & 0.94 & 0.90 & 0.92 & 0.94 & 0.92 & 0.93 & 0.93 \\
50 & DGP2 & 0.94 & 0.94 & 0.94 & 0.94 & 0.94 & 0.94 & 0.94 \\
\midrule
\multicolumn{9}{l}{\emph{DGP3: Full group separation}} \\
20 & DGP3 & 0.95 & 0.92 & 0.92 & 0.93 & 0.98 & 0.96 & 0.99 \\
50 & DGP3 & 0.95 & 0.94 & 0.94 & 0.96 & 0.95 & 0.95 & 0.98 \\
\midrule
\multicolumn{9}{l}{\textbf{Panel (b): Average length}} \\
\midrule
\multicolumn{9}{l}{\emph{DGP1: No group separation}} \\
20 & DGP1 & 0.22 & 0.21 & 0.19 & 0.17 & 0.89 & 0.65 & 0.20 \\
50 & DGP1 & 0.13 & 0.13 & 0.12 & 0.11 & 0.50 & 0.39 & 0.11 \\
\midrule
\multicolumn{9}{l}{\emph{DGP2: Partial group separation}} \\
20 & DGP2 & 0.22 & 0.22 & 0.20 & 0.19 & 0.20 & 0.18 & 0.15 \\
50 & DGP2 & 0.13 & 0.13 & 0.13 & 0.12 & 0.13 & 0.12 & 0.11 \\
\midrule
\multicolumn{9}{l}{\emph{DGP3: Full group separation}} \\
20 & DGP3 & 0.22 & 0.22 & 0.20 & 0.19 & 1.08 & 0.95 & 1.57 \\
50 & DGP3 & 0.13 & 0.13 & 0.13 & 0.12 & 0.31 & 0.37 & 1.11 \\
\bottomrule
\end{tabular*}
\begin{tablenotes}[flushleft]
\footnotesize
\item \textit{Notes:} The table reports empirical coverage and average length of confidence intervals for the scalar contrast in $H_{0,2}$, evaluated relative to the appropriate group-label-aligned truth. Column definitions and the variance estimator are as in Table~\ref{tab:exp1_case2_rejection}. Case 2 includes additive unit effects, which are removed before estimation by the transformations described in the Monte Carlo design.
\end{tablenotes}
\end{threeparttable}
\end{table}

\begin{table}
\centering
\scriptsize
\begin{threeparttable}
\caption{Coverage and average length of confidence intervals for $H_{0,2}$ \\
Experiment 1: iid Gaussian design \\
Case 3: Unit fixed effects and grouped time effects}
\label{tab:exp1_case3_ci}
\begin{tabular*}{\textwidth}{@{\extracolsep{\fill}} cc ccccccc}
\toprule
$T$ & DGP & \shortstack{Predetermined} & \shortstack{Naive\\TSK} & \shortstack{Naive\\PCR} & \shortstack{Naive\\GFE} & \shortstack{Conditional\\TSK} & \shortstack{Conditional\\PCR} & \shortstack{Conditional\\GFE} \\
\midrule
\multicolumn{9}{l}{\textbf{Panel (a): Coverage}} \\
\midrule
\multicolumn{9}{l}{\emph{DGP1: No group separation}} \\
20 & DGP1 & 0.88 & 0.00 & 0.09 & 0.71 & 0.90 & 0.91 & 0.93 \\
50 & DGP1 & 0.86 & 0.00 & 0.09 & 0.90 & 0.90 & 0.89 & 0.95 \\
\midrule
\multicolumn{9}{l}{\emph{DGP2: Partial group separation}} \\
20 & DGP2 & 0.89 & 0.83 & 0.83 & 0.94 & 0.89 & 0.88 & 0.95 \\
50 & DGP2 & 0.88 & 0.88 & 0.86 & 0.94 & 0.88 & 0.85 & 0.94 \\
\midrule
\multicolumn{9}{l}{\emph{DGP3: Full group separation}} \\
20 & DGP3 & 0.87 & 0.85 & 0.81 & 0.92 & 0.99 & 0.97 & 0.99 \\
50 & DGP3 & 0.89 & 0.88 & 0.86 & 0.94 & 0.91 & 0.88 & 0.97 \\
\midrule
\multicolumn{9}{l}{\textbf{Panel (b): Average length}} \\
\midrule
\multicolumn{9}{l}{\emph{DGP1: No group separation}} \\
20 & DGP1 & 0.23 & 0.22 & 0.19 & 0.18 & 1.10 & 0.84 & 0.14 \\
50 & DGP1 & 0.14 & 0.14 & 0.13 & 0.12 & 0.66 & 0.53 & 0.07 \\
\midrule
\multicolumn{9}{l}{\emph{DGP2: Partial group separation}} \\
20 & DGP2 & 0.23 & 0.23 & 0.20 & 0.19 & 0.20 & 0.18 & 0.15 \\
50 & DGP2 & 0.14 & 0.14 & 0.13 & 0.12 & 0.13 & 0.13 & 0.11 \\
\midrule
\multicolumn{9}{l}{\emph{DGP3: Full group separation}} \\
20 & DGP3 & 0.23 & 0.23 & 0.20 & 0.19 & 1.44 & 1.29 & 1.57 \\
50 & DGP3 & 0.14 & 0.14 & 0.13 & 0.12 & 0.50 & 0.45 & 1.06 \\
\bottomrule
\end{tabular*}
\begin{tablenotes}[flushleft]
\footnotesize
\item \textit{Notes:} The table reports empirical coverage and average length of confidence intervals for the scalar contrast in $H_{0,2}$, evaluated relative to the appropriate group-label-aligned truth. Column definitions and the variance estimator are as in Table~\ref{tab:exp1_case2_rejection}. Case 3 includes unit effects and grouped time effects; unit effects are removed before estimation, and the GFE procedure additionally estimates the grouped time effects.
\end{tablenotes}
\end{threeparttable}
\end{table}

Tables~\ref{tab:exp2_case2_ci} and~\ref{tab:exp2_case3_ci} report the corresponding results for the dependent design. Under DGP1, the naive TSK intervals again have zero coverage and the naive PCR intervals cover only 0.13 to 0.16 of the time, while the conditional intervals attain 0.93 to 0.96. This is especially important because the predetermined-group intervals under-cover here, with coverage as low as 0.71 in Case 3 at \(T=20\): even when covariance estimation under dependence is challenging, the selective intervals correct the additional distortion caused by estimating the group structure. The price is again wider intervals, most visibly under DGP1 in Case 3, where the conditional TSK and PCR intervals average 1.19 and 0.91 at \(T=20\) against 0.33 and 0.30 for their naive counterparts. When the groups are partially or fully separated, and \(T=50\), the gap between naive and conditional lengths becomes much smaller for TSK and PCR, while GFE remains more conservative in some designs.

\begin{table}
\centering
\scriptsize
\begin{threeparttable}
\caption{Coverage and average length of confidence intervals for $H_{0,2}$ \\
Experiment 2: serial and within-cluster dependence \\
Case 2: Unit fixed effects}
\label{tab:exp2_case2_ci}
\begin{tabular*}{\textwidth}{@{\extracolsep{\fill}} cc ccccccc}
\toprule
$T$ & DGP & \shortstack{Predetermined} & \shortstack{Naive\\TSK} & \shortstack{Naive\\PCR} & \shortstack{Naive\\GFE} & \shortstack{Conditional\\TSK} & \shortstack{Conditional\\PCR} & \shortstack{Conditional\\GFE} \\
\midrule
\multicolumn{9}{l}{\textbf{Panel (a): Coverage}} \\
\midrule
\multicolumn{9}{l}{\emph{DGP1: No group separation}} \\
20 & DGP1 & 0.83 & 0.00 & 0.14 & 0.75 & 0.94 & 0.95 & 0.93 \\
50 & DGP1 & 0.90 & 0.00 & 0.16 & 0.83 & 0.96 & 0.96 & 0.94 \\
\midrule
\multicolumn{9}{l}{\emph{DGP2: Partial group separation}} \\
20 & DGP2 & 0.84 & 0.82 & 0.83 & 0.89 & 0.90 & 0.88 & 0.92 \\
50 & DGP2 & 0.89 & 0.88 & 0.89 & 0.92 & 0.89 & 0.90 & 0.93 \\
\midrule
\multicolumn{9}{l}{\emph{DGP3: Full group separation}} \\
20 & DGP3 & 0.84 & 0.85 & 0.84 & 0.89 & 0.96 & 0.96 & 0.99 \\
50 & DGP3 & 0.89 & 0.89 & 0.89 & 0.92 & 0.91 & 0.91 & 0.98 \\
\midrule
\multicolumn{9}{l}{\textbf{Panel (b): Average length}} \\
\midrule
\multicolumn{9}{l}{\emph{DGP1: No group separation}} \\
20 & DGP1 & 0.27 & 0.29 & 0.26 & 0.19 & 1.01 & 0.73 & 0.15 \\
50 & DGP1 & 0.19 & 0.19 & 0.18 & 0.13 & 0.59 & 0.47 & 0.09 \\
\midrule
\multicolumn{9}{l}{\emph{DGP2: Partial group separation}} \\
20 & DGP2 & 0.27 & 0.29 & 0.27 & 0.21 & 0.26 & 0.23 & 0.18 \\
50 & DGP2 & 0.20 & 0.20 & 0.20 & 0.14 & 0.19 & 0.19 & 0.13 \\
\midrule
\multicolumn{9}{l}{\emph{DGP3: Full group separation}} \\
20 & DGP3 & 0.27 & 0.29 & 0.27 & 0.21 & 1.26 & 1.20 & 1.62 \\
50 & DGP3 & 0.19 & 0.20 & 0.19 & 0.14 & 0.45 & 0.44 & 1.24 \\
\bottomrule
\end{tabular*}
\begin{tablenotes}[flushleft]
\footnotesize
\item \textit{Notes:} The table reports empirical coverage and average length of confidence intervals for the scalar contrast in $H_{0,2}$, evaluated relative to the appropriate group-label-aligned truth. Column definitions and the variance estimator are as in Table~\ref{tab:exp2_case2_rejection}. Case 2 includes additive unit effects, which are removed before estimation by the transformations described in the Monte Carlo design.
\end{tablenotes}
\end{threeparttable}
\end{table}

\begin{table}
\centering
\scriptsize
\begin{threeparttable}
\caption{Coverage and average length of confidence intervals for $H_{0,2}$ \\
Experiment 2: serial and within-cluster dependence \\
Case 3: Unit fixed effects and grouped time effects}
\label{tab:exp2_case3_ci}
\begin{tabular*}{\textwidth}{@{\extracolsep{\fill}} cc ccccccc}
\toprule
$T$ & DGP & \shortstack{Predetermined} & \shortstack{Naive\\TSK} & \shortstack{Naive\\PCR} & \shortstack{Naive\\GFE} & \shortstack{Conditional\\TSK} & \shortstack{Conditional\\PCR} & \shortstack{Conditional\\GFE} \\
\midrule
\multicolumn{9}{l}{\textbf{Panel (a): Coverage}} \\
\midrule
\multicolumn{9}{l}{\emph{DGP1: No group separation}} \\
20 & DGP1 & 0.71 & 0.00 & 0.13 & 0.79 & 0.94 & 0.93 & 0.93 \\
50 & DGP1 & 0.83 & 0.00 & 0.15 & 0.91 & 0.96 & 0.95 & 0.94 \\
\midrule
\multicolumn{9}{l}{\emph{DGP2: Partial group separation}} \\
20 & DGP2 & 0.71 & 0.73 & 0.70 & 0.87 & 0.86 & 0.81 & 0.91 \\
50 & DGP2 & 0.84 & 0.84 & 0.84 & 0.91 & 0.87 & 0.87 & 0.92 \\
\midrule
\multicolumn{9}{l}{\emph{DGP3: Full group separation}} \\
20 & DGP3 & 0.71 & 0.75 & 0.70 & 0.86 & 0.97 & 0.95 & 0.98 \\
50 & DGP3 & 0.82 & 0.84 & 0.82 & 0.92 & 0.89 & 0.88 & 0.97 \\
\midrule
\multicolumn{9}{l}{\textbf{Panel (b): Average length}} \\
\midrule
\multicolumn{9}{l}{\emph{DGP1: No group separation}} \\
20 & DGP1 & 0.30 & 0.33 & 0.30 & 0.20 & 1.19 & 0.91 & 0.14 \\
50 & DGP1 & 0.24 & 0.23 & 0.22 & 0.14 & 0.72 & 0.58 & 0.09 \\
\midrule
\multicolumn{9}{l}{\emph{DGP2: Partial group separation}} \\
20 & DGP2 & 0.30 & 0.32 & 0.30 & 0.21 & 0.30 & 0.29 & 0.18 \\
50 & DGP2 & 0.24 & 0.24 & 0.24 & 0.14 & 0.23 & 0.23 & 0.13 \\
\midrule
\multicolumn{9}{l}{\emph{DGP3: Full group separation}} \\
20 & DGP3 & 0.30 & 0.32 & 0.30 & 0.21 & 1.56 & 1.43 & 1.60 \\
50 & DGP3 & 0.24 & 0.25 & 0.24 & 0.14 & 0.65 & 0.59 & 1.20 \\
\bottomrule
\end{tabular*}
\begin{tablenotes}[flushleft]
\footnotesize
\item \textit{Notes:} The table reports empirical coverage and average length of confidence intervals for the scalar contrast in $H_{0,2}$, evaluated relative to the appropriate group-label-aligned truth. Column definitions and the variance estimator are as in Table~\ref{tab:exp2_case2_rejection}. Case 3 includes unit effects and grouped time effects; unit effects are removed before estimation, and the GFE procedure additionally estimates the grouped time effects.
\end{tablenotes}
\end{threeparttable}
\end{table}

Overall, the additional results reinforce the main message of the paper. Naive post-clustering inference can be highly misleading whenever the estimated groups are not strongly separated, and this problem persists in the presence of fixed effects, grouped time effects, serial dependence, CSD, and non-Gaussian errors. The proposed conditional procedure greatly reduces these distortions and restores reliable inference in the most problematic cases. The results also show that the choice of clustering estimator matters: TSK and PCR perform well when slope separation is the main source of heterogeneity, while GFE is particularly useful when the data contain grouped time effects.

\section{Additional Empirical Applications}\label{sec:add_applications}

This section presents two additional empirical applications to illustrate our procedures. The first focuses on R\&D dynamics, the second on the relationship between income and democracy. In both cases, the initial analyses reveal clear heterogeneity, but this evidence becomes more nuanced once we account for the estimation of the group structure.

\subsection{Firm level R\&D investment and the business cycle}

The model of \citet{aguilar-loyo_grouped_2024} is particularly relevant for illustrating the challenges of inference under latent grouping: firm-level R\&D responses to industry output are modeled with group-specific coefficients, and while their estimation strategy incorporates heteroskedasticity across groups, their analysis leaves open whether the apparent heterogeneity in slopes is genuine or spurious. The data, originally compiled by \citet{fabrizio2014empirical} and made available by \citet{van2019cyclicality}, consist of yearly information on U.S. firms over the period 1975--2002. We use the balanced panel without balance-sheet controls, which spans 28 years and includes 291 firms, and retain only industry-level output growth as a regressor:
\begin{equation*}
    \Delta \log RD_{it}
    =
    \theta_{g_i}\Delta \log X_{st}
    +
    \xi_{it}
    +
    \varepsilon_{it},
    \quad
    N=291,
    \quad
    T=28.
\end{equation*}
Here \(\Delta \log RD_{it}\) is the growth of firm \(i\)'s R\&D expenditure and \(\Delta \log X_{st}\) is industry-level output growth, so the group-specific slope \(\theta_{g_i}\) measures the cyclicality of R\&D investment. For TSK and PCR, we remove additive firm and year fixed effects by a two-way within transformation before clustering and estimation. For GFE, we use the raw outcome and regressor and let the intercept component \(\xi_{it}\) be represented by group-specific time effects, which allows a direct comparison between slope-based clustering after standard fixed-effect residualization and a specification that absorbs time-varying latent group intercepts.

Following \citet{aguilar-loyo_grouped_2024}, we set \(G=3\). Table~\ref{tab:rd_coef_ci_new} reports the group-specific estimates of the cyclicality coefficient with individual selective \(p\)-values and fixed-truncation confidence intervals. The point estimates display substantial heterogeneity across groups and estimators, but little of it survives the selective adjustment. For PCR, the coefficient is positive for Groups 1 and 3 and negative for Group 2, ranging from \(-0.963\) to \(1.137\), and the individual selective tests reject zero for Groups 2 and 3 (\(p=0.004\) and \(p=0.002\)). For TSK, Group 2 has a very large negative coefficient, \(-4.108\), but contains only 8 firms and the associated selective uncertainty is correspondingly large; no TSK coefficient is selectively significant. The GFE partition is highly unbalanced, with sizes \((230,58,3)\), and again none of the individual selective \(p\)-values is close to rejection. This last result is particularly informative, since GFE is estimated on the raw data and identifies the slopes after allowing for time-varying latent group intercepts: under this more flexible specification, the evidence for statistically significant group-specific cyclicality disappears.

\begin{table}[!htbp]
\centering
\scriptsize
\begin{threeparttable}
\caption{Firm-Level R\&D Investment: Group-Specific Estimates and Selective Confidence Intervals}
\label{tab:rd_coef_ci_new}
\setlength{\tabcolsep}{6.0pt}
\begin{tabular*}{\textwidth}{@{\extracolsep{\fill}} l c l r c c}
\toprule
Estimator & Group & Coefficient & Estimate & \shortstack{Individual\\selective $p$-val} & \shortstack{Fixed\\95\% CI} \\
\midrule
\multicolumn{6}{l}{PCR, group sizes: (65,63,163)} \\
\cmidrule(lr){1-6}
PCR & 1 & $\Delta\log X_{st}$ & 1.137 & 0.844 & [1.002, 1.272] \\
PCR & 2 & $\Delta\log X_{st}$ & -0.963 & 0.004$^{**}$ & [-1.146, -0.780] \\
PCR & 3 & $\Delta\log X_{st}$ & 0.180 & 0.002$^{**}$ & [0.139, 0.220] \\
\addlinespace[0.35em]
\multicolumn{6}{l}{TSK, group sizes: (95,8,188)} \\
\cmidrule(lr){1-6}
TSK & 1 & $\Delta\log X_{st}$ & 1.048 & 0.560 & [0.923, 1.173] \\
TSK & 2 & $\Delta\log X_{st}$ & -4.108 & 0.407 & [-5.485, -2.731] \\
TSK & 3 & $\Delta\log X_{st}$ & -0.173 & 0.561 & [-0.250, -0.096] \\
\addlinespace[0.35em]
\multicolumn{6}{l}{GFE, group sizes: (230,58,3)} \\
\cmidrule(lr){1-6}
GFE & 1 & $\Delta\log X_{st}$ & 0.313 & 0.595 & [0.281, 0.345] \\
GFE & 2 & $\Delta\log X_{st}$ & -0.137 & 0.608 & [-0.193, -0.081] \\
GFE & 3 & $\Delta\log X_{st}$ & -0.677 & 0.918 & [-1.978, 0.623] \\
\bottomrule
\end{tabular*}
\begin{tablenotes}[flushleft]
\footnotesize
\item \textit{Notes:} The table reports group-specific estimates for the coefficient on $\Delta\log X_{st}$ in the firm-level R\&D investment application. For each estimator, the group partition is estimated with $G=3$; group sizes are reported in parentheses in the order of the displayed group labels. PCR and TSK are implemented on the two-way transformed data. GFE is implemented on the raw data and includes grouped time effects. Individual selective $p$-values test the null hypothesis that the corresponding group-specific coefficient equals zero and account for the estimated group partition. Fixed 95\% confidence intervals are obtained by inverting the selective test with the truncation set held fixed at its observed value. Variance matrices are computed using the dependence-robust estimator described in Section~\ref{sec:variance}. Stars are based on individual selective $p$-values: $^{*}$ and $^{**}$ denote significance at the 10\% and 5\% levels.
\end{tablenotes}
\end{threeparttable}
\end{table}

Table~\ref{tab:rd_tests_new} reports homogeneity and pairwise tests. Because the specification contains a single slope coefficient, the joint slope homogeneity test coincides with the coefficient-specific test for \(\Delta\log X_{st}\). The contrast between naive and selective inference is substantial: across all three estimators, the conventional Wald tests reject slope homogeneity, with \(p\)-values below 0.001 in every case except two GFE pairwise comparisons, whereas the selective tests are much less likely to do so.

For PCR, the selective \(p\)-value for homogeneity is 0.052, providing only marginal evidence at the 10 percent level, and the pairwise results show that this evidence is driven entirely by the comparison between Groups 2 and 3 (\(p=0.004\)); the comparison between Groups 1 and 3 is far from significant (\(p=0.952\)). The most robust difference is therefore between the negative-cyclicality group and the group with a small positive output-growth response. For TSK and GFE, no selective test rejects: the homogeneity \(p\)-values are 0.560 and 0.605, and no pairwise selective \(p\)-value is close to conventional levels. For TSK, this is consistent with the coefficient table, where one group is very small and carries large selective uncertainty; for GFE, it means that once grouped time effects are allowed for, there is no selective evidence of slope heterogeneity in R\&D cyclicality at all.

\begin{table}[!htbp]
\centering
\scriptsize
\begin{threeparttable}
\caption{Firm-Level R\&D Investment: Homogeneity and Pairwise Tests}
\label{tab:rd_tests_new}
\setlength{\tabcolsep}{6.0pt}
\begin{tabular*}{\textwidth}{@{\extracolsep{\fill}} l l r c c}
\toprule
Estimator & Test & Statistic & Naive $p$-val & Selective $p$-val \\
\midrule
\multicolumn{5}{l}{PCR, group sizes: (65,63,163)} \\
\cmidrule(lr){1-5}
PCR & $\Delta\log X_{st}$ homogeneity & 89.206 & $<0.001$ & 0.052$^{*}$ \\
PCR & Group 1 vs Group 2 & 85.124 & $<0.001$ & 0.055$^{*}$ \\
PCR & Group 1 vs Group 3 & 48.544 & $<0.001$ & 0.952 \\
PCR & Group 2 vs Group 3 & 37.682 & $<0.001$ & 0.004$^{**}$ \\
\addlinespace[0.35em]
\multicolumn{5}{l}{TSK, group sizes: (95,8,188)} \\
\cmidrule(lr){1-5}
TSK & $\Delta\log X_{st}$ homogeneity & 78.972 & $<0.001$ & 0.560 \\
TSK & Group 1 vs Group 2 & 14.030 & $<0.001$ & 0.561 \\
TSK & Group 1 vs Group 3 & 69.290 & $<0.001$ & 0.560 \\
TSK & Group 2 vs Group 3 & 8.214 & 0.004 & 0.438 \\
\addlinespace[0.35em]
\multicolumn{5}{l}{GFE, group sizes: (230,58,3)} \\
\cmidrule(lr){1-5}
GFE & $\Delta\log X_{st}$ homogeneity & 50.214 & $<0.001$ & 0.605 \\
GFE & Group 1 vs Group 2 & 49.639 & $<0.001$ & 0.605 \\
GFE & Group 1 vs Group 3 & 0.729 & 0.393 & 0.713 \\
GFE & Group 2 vs Group 3 & 0.217 & 0.642 & 0.388 \\
\bottomrule
\end{tabular*}
\begin{tablenotes}[flushleft]
\footnotesize
\item \textit{Notes:} The table reports homogeneity and pairwise tests for the coefficient on $\Delta\log X_{st}$ in the firm-level R\&D investment application. For each estimator, the group partition is estimated with $G=3$; group sizes are reported in parentheses in the order of the displayed group labels. PCR and TSK are implemented on the two-way transformed data. GFE is implemented on the raw data and includes grouped time effects. Naive $p$-values are conventional Wald $p$-values computed after estimating the group partition and therefore ignore group selection uncertainty. Selective $p$-values are obtained from the proposed conditional procedure and account for the estimated group partition. Variance matrices are computed using the dependence-robust estimator described in Section~\ref{sec:variance}. Stars are based on selective $p$-values: $^{*}$ and $^{**}$ denote significance at the 10\% and 5\% levels.
\end{tablenotes}
\end{threeparttable}
\end{table}

Taken together, the two tables show that naive post-clustering tests systematically overstate the evidence for heterogeneity in this application. After the selective correction, the only clear rejection is the PCR pairwise comparison between Groups 2 and 3, the global PCR homogeneity test is marginal, and neither TSK nor GFE yields significant slope heterogeneity. The evidence for heterogeneous R\&D cyclicality is thus sensitive to the clustering method and becomes much weaker once group selection uncertainty is taken into account.

\subsection{Income and democracy}

We next consider the classical application of grouped fixed effects models to the relationship between income and democracy. Following \citet{bonhomme_grouped_2015}, we use their original balanced democracy panel, containing \(N=90\) countries observed over \(T=7\) periods, and estimate
\begin{equation*}
    DEM_{it}
    =
    \theta_{1,g_i} DEM_{i,t-1}
    +
    \theta_{2,g_i} \log GDP_{i,t-1}
    +
    \xi_{it}
    +
    \varepsilon_{it},
    \quad
    N=90,
    \quad
    T=7,
\end{equation*}
where \(DEM_{it}\) is the Freedom House democracy indicator and \(GDP_{it}\) is per capita income, so that the group-specific coefficients \((\theta_{1,g_i},\theta_{2,g_i})\) allow both the persistence of democracy and the effect of lagged income to vary across latent clusters of countries. As above, we apply a two-way within transformation for PCR and TSK, while for GFE we use the raw data and model \(\xi_{it}\) through grouped time effects, as in \citet{bonhomme_grouped_2015}. We set \(G=4\) following their analysis, and since the original application is the classical grouped fixed effects application, we focus the interpretation on the GFE estimator.

Table~\ref{tab:dem_coef_ci_g4_new} reports the GFE group-specific estimates. The four groups are relatively balanced, with sizes \((16,17,22,35)\), and the coefficient on lagged democracy is positive throughout and increases substantially across groups, from 0.132 in Group 1 to 0.689 in Group 4, suggesting important heterogeneity in democratic persistence. The selective evidence is strongest for Group 4, whose individual selective \(p\)-value is below 0.001; Group 3 is marginally significant (\(p=0.094\)), while for Groups 1 and 2 the selective \(p\)-values are 0.263 and 0.405. The income coefficient is positive in all four groups but small, ranging from 0.040 to 0.106, and no individual selective test rejects the null that it is zero, the smallest \(p\)-value being 0.117. Hence the main robust source of heterogeneity in this application is not the direct income effect, but the persistence of democracy.

\begin{table}[!htbp]
\centering
\scriptsize
\begin{threeparttable}
\caption{Income and Democracy: Group-Specific Estimates and Selective Confidence Intervals, $G=4$}
\label{tab:dem_coef_ci_g4_new}
\setlength{\tabcolsep}{6.0pt}
\begin{tabular*}{\textwidth}{@{\extracolsep{\fill}} l c l r c c}
\toprule
Estimator & Group & Coefficient & Estimate & \shortstack{Individual\\selective $p$-val} & \shortstack{Fixed\\95\% CI} \\
\midrule
\multicolumn{6}{l}{GFE, group sizes: (16,17,22,35)} \\
\cmidrule(lr){1-6}
GFE & 1 & $DEM_{i,t-1}$ & 0.132 & 0.263 & [-0.004, 0.268] \\
GFE & 1 & $\log GDP_{i,t-1}$ & 0.106 & 0.117 & [0.084, 0.128] \\
GFE & 2 & $DEM_{i,t-1}$ & 0.166 & 0.405 & [0.105, 0.227] \\
GFE & 2 & $\log GDP_{i,t-1}$ & 0.103 & 0.213 & [0.088, 0.119] \\
GFE & 3 & $DEM_{i,t-1}$ & 0.427 & 0.094$^{*}$ & [0.375, 0.479] \\
GFE & 3 & $\log GDP_{i,t-1}$ & 0.040 & 0.361 & [0.024, 0.056] \\
GFE & 4 & $DEM_{i,t-1}$ & 0.689 & $<0.001^{**}$ & [0.630, 0.748] \\
GFE & 4 & $\log GDP_{i,t-1}$ & 0.059 & 0.184 & [0.043, 0.075] \\
\bottomrule
\end{tabular*}
\begin{tablenotes}[flushleft]
\footnotesize
\item \textit{Notes:} The table reports group-specific estimates in the income and democracy application with $G=4$. Group sizes are reported in parentheses in the order of the displayed group labels. GFE is implemented on the raw data and includes grouped time effects. Individual selective $p$-values test whether the corresponding group-specific coefficient is equal to zero and account for the estimated group partition. Fixed 95\% confidence intervals are obtained by inverting the selective test with the truncation set held fixed at its observed value. Variance matrices are computed using the dependence-robust estimator described in Section~\ref{sec:variance}. Stars are based on individual selective $p$-values: $^{*}$ and $^{**}$ denote significance at the 10\% and 5\% levels.
\end{tablenotes}
\end{threeparttable}
\end{table}

Table~\ref{tab:dem_tests_new} reports homogeneity and pairwise vector tests. The same qualitative pattern as in the R\&D application holds for the PCR and TSK estimators, which we do not report in detail: the naive Wald tests reject almost all homogeneity restrictions, whereas none of the corresponding selective tests is significant at the 5 percent level. The GFE results differ. The joint slope homogeneity test has a selective \(p\)-value of 0.035, indicating statistically significant heterogeneity across the four groups after accounting for group selection, and this rejection is driven by heterogeneity in the coefficient on lagged democracy: the selective \(p\)-value for homogeneity in \(DEM_{i,t-1}\) is below 0.001, against 0.350 for \(\log GDP_{i,t-1}\). The pairwise tests localize the difference, with the strongest rejection between Groups 2 and 4 (\(p<0.001\)) and a marginal rejection between Groups 1 and 4 (\(p=0.076\)); all remaining comparisons are insignificant. The fourth group is therefore the main source of heterogeneity, consistent with its being the group with the largest persistence coefficient in Table~\ref{tab:dem_coef_ci_g4_new}.

\begin{table}[!htbp]
\centering
\scriptsize
\begin{threeparttable}
\caption{Income and Democracy: Homogeneity and Pairwise Vector Tests}
\label{tab:dem_tests_new}
\setlength{\tabcolsep}{6.0pt}
\begin{tabular*}{\textwidth}{@{\extracolsep{\fill}} l l r c c}
\toprule
Estimator & Test & Statistic & Naive $p$-val & Selective $p$-val \\
\midrule
\multicolumn{5}{l}{GFE, group sizes: (16,17,22,35)} \\
\cmidrule(lr){1-5}
GFE & Joint slope homogeneity & 75.380 & $<0.001$ & 0.035$^{**}$ \\
GFE & $DEM_{i,t-1}$ homogeneity & 72.226 & $<0.001$ & $<0.001^{**}$ \\
GFE & $\log GDP_{i,t-1}$ homogeneity & 14.657 & 0.002 & 0.350 \\
GFE & Group 1 vs Group 2 & 0.102 & 0.950 & 0.786 \\
GFE & Group 1 vs Group 3 & 11.032 & 0.004 & 0.419 \\
GFE & Group 1 vs Group 4 & 28.345 & $<0.001$ & 0.076$^{*}$ \\
GFE & Group 2 vs Group 3 & 20.617 & $<0.001$ & 0.370 \\
GFE & Group 2 vs Group 4 & 60.564 & $<0.001$ & $<0.001^{**}$ \\
GFE & Group 3 vs Group 4 & 19.891 & $<0.001$ & 0.559 \\
\bottomrule
\end{tabular*}
\begin{tablenotes}[flushleft]
\footnotesize
\item \textit{Notes:} The table reports homogeneity and pairwise tests in the income and democracy application with $G=4$. Group sizes are reported in parentheses in the order of the displayed group labels. GFE is implemented on the raw data and includes grouped time effects. Joint slope homogeneity tests equality of the full slope vector across all groups. Coefficient-specific homogeneity tests equality of the indicated coefficient across all groups. Pairwise tests impose equality of the full slope vector between the two listed groups. Naive $p$-values are conventional Wald $p$-values computed after estimating the group partition and therefore ignore group selection uncertainty. Selective $p$-values are obtained from the proposed conditional procedure and account for the estimated group partition. Variance matrices are computed using the dependence-robust estimator described in Section~\ref{sec:variance}. Stars are based on selective $p$-values: $^{*}$ and $^{**}$ denote significance at the 10\% and 5\% levels.
\end{tablenotes}
\end{threeparttable}
\end{table}

Table~\ref{tab:dem_pairwise_scalar_new} decomposes the pairwise vector tests into scalar contrasts and confirms that the heterogeneity is concentrated in democratic persistence. For \(DEM_{i,t-1}\), the Group 2 versus Group 4 contrast is significant at the 5 percent level after selective adjustment (\(p=0.018\)) and the Group 1 versus Group 4 contrast is marginally significant (\(p=0.066\)), while all other contrasts are insignificant. For the income coefficient, no pairwise scalar contrast is significant after selection adjustment.

\begin{table}[!htbp]
\centering
\scriptsize
\begin{threeparttable}
\caption{Income and Democracy: Pairwise Scalar Coefficient Tests}
\label{tab:dem_pairwise_scalar_new}
\setlength{\tabcolsep}{6.0pt}
\begin{tabular*}{\textwidth}{@{\extracolsep{\fill}} l l l r c c}
\toprule
Estimator & Pair & Coefficient & Estimate & Naive $p$-val & Selective $p$-val \\
\midrule
\multicolumn{6}{l}{GFE, group sizes: (16,17,22,35)} \\
\cmidrule(lr){1-6}
GFE & Group 1 vs Group 2 & $DEM_{i,t-1}$ & -0.034 & 0.757 & 0.573 \\
GFE & Group 1 vs Group 2 & $\log GDP_{i,t-1}$ & 0.003 & 0.903 & 0.402 \\
GFE & Group 1 vs Group 3 & $DEM_{i,t-1}$ & -0.295 & 0.005 & 0.454 \\
GFE & Group 1 vs Group 3 & $\log GDP_{i,t-1}$ & 0.066 & 0.004 & 0.284 \\
GFE & Group 1 vs Group 4 & $DEM_{i,t-1}$ & -0.557 & $<0.001$ & 0.066$^{*}$ \\
GFE & Group 1 vs Group 4 & $\log GDP_{i,t-1}$ & 0.047 & 0.035 & 0.371 \\
GFE & Group 2 vs Group 3 & $DEM_{i,t-1}$ & -0.261 & $<0.001$ & 0.306 \\
GFE & Group 2 vs Group 3 & $\log GDP_{i,t-1}$ & 0.064 & 0.001 & 0.393 \\
GFE & Group 2 vs Group 4 & $DEM_{i,t-1}$ & -0.523 & $<0.001$ & 0.018$^{**}$ \\
GFE & Group 2 vs Group 4 & $\log GDP_{i,t-1}$ & 0.044 & 0.020 & 0.451 \\
GFE & Group 3 vs Group 4 & $DEM_{i,t-1}$ & -0.262 & $<0.001$ & 0.288 \\
GFE & Group 3 vs Group 4 & $\log GDP_{i,t-1}$ & -0.020 & 0.320 & 0.395 \\
\bottomrule
\end{tabular*}
\begin{tablenotes}[flushleft]
\footnotesize
\item \textit{Notes:} The table reports pairwise scalar coefficient tests in the income and democracy application with $G=4$. Group sizes are reported in parentheses in the order of the displayed group labels. The estimate is the coefficient in the first listed group minus the coefficient in the second listed group. GFE is implemented on the raw data and includes grouped time effects. Naive $p$-values are conventional Wald $p$-values computed after estimating the group partition and therefore ignore group selection uncertainty. Selective $p$-values are obtained from the proposed conditional procedure and account for the estimated group partition. Variance matrices are computed using the dependence-robust estimator described in Section~\ref{sec:variance}. Stars are based on selective $p$-values: $^{*}$ and $^{**}$ denote significance at the 10\% and 5\% levels.
\end{tablenotes}
\end{threeparttable}
\end{table}

Overall, the democracy application yields a coherent picture. In the GFE specification with \(G=4\), the data contain statistically significant latent-group heterogeneity, but it is driven by differences in democratic persistence rather than by differences in the effect of lagged income. This is precisely the type of distinction obscured by naive post-clustering inference: conventional Wald tests reject many restrictions, whereas the selective procedure identifies a narrower and more interpretable source of heterogeneity.

\end{appendices}

\bibliography{refs}

\end{document}